\documentclass[a4paper,11pt]{article}
\usepackage{jheppub}
\usepackage{slashed}
\usepackage{lineno}
\usepackage{comment}
\usepackage{braket}
\usepackage{cancel}

\title{\boldmath Bubble wall velocity with out-of-equilibrium corrections}

\author{Carlo Branchina$^{1,2}$, Angela Conaci$^{1,2}$, Stefania De Curtis$^{3}$, Luigi Delle Rose$^{1,2}$}
\affiliation{$^{1}$Dipartimento di Fisica, Università della Calabria, I-87036 Arcavacata di Rende, Cosenza, Italy}
\affiliation{$^{2}$INFN, Gruppo Collegato di Cosenza, Arcavacata di Rende, I-87036, Cosenza, Italy}
\affiliation{$^{3}$INFN, Sezione di Firenze, Via Sansone, 1, I-50019 Sesto Fiorentino (FI), Italy}

\emailAdd{carlo.branchina@unical.it}
\emailAdd{angela.conaci@unical.it}
\emailAdd{stefania.decurtis@fi.infn.it}
\emailAdd{luigi.dellerose@unical.it}

\abstract{
We study how out-of-equilibrium effects modify the steady-state propagation of bubble walls during a cosmological first-order electroweak phase transition. Going beyond the local thermal equilibrium approximation, we numerically solve the coupled system of scalar field, hydrodynamic and Boltzmann equations using a spectral algorithm that allows a first-principle treatment of the collision integral. This approach enables a quantitative assessment of non-equilibrium perturbations in the plasma and their backreaction on the wall motion. Focusing on the singlet extension of the Standard Model as a minimal benchmark scenario, we find that out-of-equilibrium corrections substantially enhance the effective friction on the expanding front, leading to slower wall velocities and broader wall profiles compared to the equilibrium case. These modifications have significant implications for cosmological observables. For instance, they enhance the efficiency of electroweak baryogenesis, thus improving the viability of baryon asymmetry generation within realistic parameter regions that can also be probed by future gravitational wave interferometers.
}

\makeatletter
\def\@fpheader{\relax}
\makeatother

\begin{document}
\maketitle
\flushbottom

\section{Introduction}

The dynamics of cosmological phase transitions is governed by the interplay between scalar field evolution and the surrounding hot plasma. In particular, the interactions between expanding scalar field configurations and the plasma can significantly influence the propagation of the transition front, leading to friction effects that determine whether the bubble wall reaches a stationary velocity or experiences acceleration. Understanding these processes is crucial to describing how the electroweak (EW) vacuum emerged during the thermal history of the universe and to assessing the associated cosmological relics such as the matter–antimatter asymmetry and backgrounds of gravitational waves.

Within the Standard Model (SM), the electroweak phase transition (EWPhT) is known to be a smooth crossover \cite{Kajantie:1995kf,Kajantie:1996mn,Csikor:1998eu}. However, beyond the SM (BSM) extensions with enlarged scalar sectors can accommodate a strongly first-order phase transition, thereby opening the possibility of producing observable imprints such as gravitational waves, primordial baryon asymmetry, or other remnants of new physics \cite{Hindmarsh:2020hop,Athron:2023xlk}. In these scenarios, accurately describing the wall dynamics, and in particular, the bubble wall velocity $v_w$ and width, is essential to quantify the expected signals and to connect particle physics models with cosmology.

A full and self-consistent treatment of the wall motion requires solving the coupled system of scalar field equations, hydrodynamic conservation laws for the plasma, and the Boltzmann equations for the distribution functions of all relevant species. Since the seminal works \cite{Moore:1995si,Moore:1995ua}, a huge effort has been put forward to develop methods allowing to study the electroweak phase transition dynamics~\cite{Moore:2000wx,John:2000zq,Cline:2000nw,Bodeker:2009qy,Megevand:2009gh,Espinosa:2010hh,Leitao:2010yw,Huber:2013kj,Megevand:2013hwa,Megevand:2013yua,Megevand:2014yua,Konstandin:2014zta,Leitao:2014pda,Megevand:2014dua,Kozaczuk:2015owa,Bodeker:2017cim,Basler:2018cwe,Dorsch:2018pat,DeCurtis:2019rxl,Cline:2020jre,BarrosoMancha:2020fay,Basler:2020nrq,Hoche:2020ysm,Laurent:2020gpg,Friedlander:2020tnq,Azatov:2020ufh,Balaji:2020yrx,Cai:2020djd,Wang:2020zlf,Cline:2021iff,Bigazzi:2021ucw,Dorsch:2021ubz,Cline:2021dkf,Ai:2021kak,Lewicki:2021pgr,Gouttenoire:2021kjv,Dorsch:2021nje,DeCurtis:2022hlx,Laurent:2022jrs,Lewicki:2022nba,Janik:2022wsx,DeCurtis:2022llw,Ellis:2022lft,DeCurtis:2022djw,Jiang:2022btc,DeCurtis:2023hil,Ai:2023see,Krajewski:2023clt,Baldes:2023cih,Azatov:2023xem,Dorsch:2023tss,Sanchez-Garitaonandia:2023zqz,Ai:2024shx,DeCurtis:2024hvh,Krajewski:2024gma,Basler:2024aaf,Wang:2024wcs,Azatov:2024auq,Barni:2024lkj,Yuwen:2024hme,Branchina:2024rva,Ekstedt:2024fyq,Ai:2024btx,Krajewski:2024xuz,Krajewski:2024zxg,Dorsch:2024jjl,Ramsey-Musolf:2025jyk,Ai:2025bjw,Carena:2025flp,Branchina:2025jou,Biekotter:2025fjx,Chala:2025oul,Si:2025vdt,Lee:2025hgb,Searle:2025cnj,Bhatnagar:2025jhh,Eriksson:2025owh,Roy:2025zvo,Braathen:2025svl}. In several previous studies, the set of equations describing bubble dynamics have been solved under the assumption of local thermal equilibrium (LTE), where the deviations of the plasma from equilibrium are neglected, and the friction acting on the wall is effectively captured by temperature gradients. Although the LTE approximation provides valuable insights, it fails to account for the out-of-equilibrium (OOE) dynamics that arise as the moving wall disturbs the plasma distribution. These non-equilibrium effects are expected to play a central role in determining the dynamics of the phase transition and the efficiency of baryon asymmetry generation.

A quantitative treatment of these effects has been recently proposed by developing a new numerical algorithm capable of solving the Boltzmann equation without resorting to simplified ansätze for the distribution functions \cite{DeCurtis:2022hlx,DeCurtis:2023hil,DeCurtis:2024hvh}. In this approach, the collision integral, traditionally the most challenging component of the problem, is evaluated iteratively using a spectral decomposition in terms of eigenmodes of the collision operator, greatly reducing its complexity and allowing to control the accuracy of its evaluation by exploiting the hierarchy of the eigenvalues. This method allows one to compute the non-equilibrium perturbations $\delta f_i$ of the plasma species directly from first principles and to assess their backreaction on the scalar field dynamics through the corresponding friction terms.

In the present work, we use this framework to study the quantitative impact of out-of-equilibrium corrections on the bubble wall dynamics during the electroweak phase transition. We focus on the $\mathbb{Z}_2$-symmetric singlet extension of the Standard Model (SSM), which constitutes a minimal benchmark allowing for a two-step symmetry breaking pattern and a strongly first-order transition. 
In a previous analysis \cite{Branchina:2025jou} we presented a detailed study of bubble dynamics in the LTE approximation. By comparing the results obtained with and without non-equilibrium corrections, we systematically evaluate here how deviations from LTE modify the stationary solutions, including the wall velocity, wall widths, and plasma temperature and velocity profiles.

We conduct a comprehensive numerical investigation across the parameter space of the model, employing the optimised algorithm introduced in \cite{DeCurtis:2022hlx,DeCurtis:2023hil,DeCurtis:2024hvh} to iteratively solve the Boltzmann equation for the distribution functions of species together with the scalar and hydrodynamic equations. 

Our analysis reveals that out-of-equilibrium effects significantly enhance the friction acting on the wall, leading to a systematic reduction of the stationary wall velocity and an increase in the wall widths. These corrections are particularly pronounced in regions of the parameter space corresponding to relatively weak phase transitions, where the LTE approximation tends to overestimate the temperature gradients required to balance the driving pressure. Importantly, we find that the inclusion of OOE contributions not only shifts the quantitative predictions but also qualitatively modifies the phase transition dynamics, allowing for stationary deflagration solutions in regimes where LTE would predict ultrarelativistic detonations.

In this study we consider only the OOE contributions arising from the top quark, the one with the largest coupling to the Higgs field. As shown in \cite{DeCurtis:2024hvh}, out-of-equilibrium effects from additional species, particularly the $W$ gauge bosons, can yield further corrections to the phase transition parameters, which could be comparable in magnitude to those arising from the top quark. Further investigations are currently ongoing.

Beyond their intrinsic theoretical interest, the OOE corrections have direct implications for cosmological observables. The wall velocity and the wall widths (mainly the one of the SM-like scalar field) strongly influence both the gravitational wave spectrum produced during the transition and the efficiency of electroweak baryogenesis (EWBG). To illustrate these implications, we apply our results to compute the resulting gravitational wave signals and the baryon asymmetry (with the latter generated in an augmented version of the SSM that includes a CP-violating dimension-five operator coupling the singlet to the top quark). We show that accounting for non-equilibrium dynamics leads to a reduction of the predicted gravitational wave amplitude and, conversely, an enhancement of the baryon asymmetry, improving the viability of successful EWBG within this framework. We also show that, in a certain region of the parameter space where the phase transition is strong, an efficient EWBG can be compatible with a potentially observable gravitational wave spectrum at future interferometers.

The remainder of this paper is structured as follows. In Section \ref{sec: set-up}, we introduce the theoretical framework, outlining the hydrodynamic equations, scalar field dynamics, and the Boltzmann equation governing the non-equilibrium distribution functions. Section \ref{sec: numerical results} describes the numerical method and presents our results for the wall profiles and velocities across the SSM parameter space, comparing them with their LTE counterparts. In Section \ref{sec- applications}, we apply these results to compute the gravitational wave spectra and baryon asymmetry generated by the transition. We conclude in Section \ref{sec- conclusions} with a discussion of the implications of our findings and future extensions. 
In Appendix \ref{sec: appendix baryo} we present the explicit derivation of the transport equations for EWBG, and in Appendix \ref{sec- lambdas=2} we show some numerical results for a different choice of the self-interaction singlet coupling of the SSM.

\section{Theoretical set-up}
\label{sec: set-up}

The dynamics of the bubble wall is determined by its interaction with the surrounding plasma. As the bubble nucleates and starts expanding in an initially homogeneous false vacuum background due to the pressure difference between the exterior and its interior, the plasma is driven out of equilibrium. The presence of the moving phase transition front triggers a response from the plasma, that develops non-trivial temperature and velocity profiles, $T(z)$ and $v_p(z)$ respectively, to ensure energy and momentum conservation. Together with the decelerating effect determined by interactions with particles impinging on the wall, this generates a friction that tends to slow down the front. If a balance between the outward pressure and the friction thus produced is reached during the expansion of the bubble, the acceleration vanishes and a steady-state regime sets in, where the plasma and field profiles freeze (in a frame comoving with the wall) and the wall keeps expanding at a constant velocity $v_w$, that we typically refer to as the ``wall velocity". In the following, we assume that such a stationary state is reached, and solve a set of fluid equations to fully determine the related dynamics. 

It is convenient to think of the system as being made up of three components: (i) the scalar fields driving the transition, (ii) species strongly coupled to the scalars, and whose out-of-equilibrium (OOE) contributions can have an important impact on the dynamics, and (iii) species weakly coupled to the scalars, that can be described as being in local thermal equilibrium (LTE), to a good approximation.   

The basic ingredient to describe the dynamics of the plasma is the stress-energy tensor
\begin{equation}
\label{eq: EMT}
    T_{\mu\nu}^{pl}=\sum_i\int\frac{d^3k}{(2\pi)^3 2 E_k}k_\mu k_\nu f_i(k,x), 
\end{equation}
where the sum extends to all the species the plasma is made of, and $f_i(k,x)$ are the particle distribution functions. For each species, we conveniently parametrise them as 
\begin{equation}
 f_i(k,x)=f_{0,i}(k,x)+\delta f_i(k,x), 
\end{equation}
where $f_{0,i}$ is the LTE distribution function with non-trivial temperature and velocity profiles (the $\pm$ sign is for the Bose-Einstein or Fermi-Dirac distribution, while $u_\mu \equiv \gamma(1, \vec{v_p})$ is the plasma four-velocity), 
\begin{equation}
\label{eq: LTE distribution function}
f_{0,i}=\frac{1}{e^{p^\mu u_{\mu}(x)/T(x)}\pm 1},
\end{equation}
and we refer to $\delta f_i$ as the OOE contributions.

Using Lorentz invariance, the equilibrium part of the stress-energy tensor can be written in the form 
\begin{equation}
	T_{\mu \nu}^{\rm LTE} = w\, u_\mu u_\nu - g_{\mu \nu}\, p,
\end{equation}
where $w = e + p = T\,\partial_{_T} p $ is the enthalpy, $e$ the energy density and $p$ the pressure. The latter is given by the thermal contribution to the effective potential 
\begin{equation}p=-V_{_T}=T\sum_i\mp\,  n_i \int \frac{d^3p}{(2\pi)^3} \ln \left(1\mp e^{-E_i/T}\right) ,
\end{equation}
where the upper (lower) sign refers to bosons (fermions) and $n_i$ denotes the number of degrees of freedom of the $i$-th particle. 

Adding the contribution from the scalars, the total energy momentum tensor is 
\begin{equation}
    T_{\mu\nu} = T_{\mu\nu}^{pl}+T_{\mu\nu}^{\phi}
\end{equation}
where 
\begin{equation}
T^{\phi}_{\mu \nu}= \sum_{j=1}^n \left[\partial_\mu \phi_j \partial_\nu \phi_j -g_{\mu \nu}\frac{(\partial \phi_j)^2}{2}  \right] + g_{\mu\nu} V_{_{T=0}} , 
\end{equation}
with $V_{_{T=0}}$ the zero-temperature effective potential and the index $j$ running over the $n$ scalars participating in the transition. It is thus apparent that we need three types of equations to fully describe the wall dynamics: (i) hydrodynamic equations for the space-dependent plasma profiles $T(x)$ and $v_p(x)$, (ii) equations for the scalar fields, (iii) equations allowing to determine the deviation from equilibrium $\delta f$. When working within the LTE approximation, equations (iii) are neglected, and one only solves the coupled plasma and scalar equations. The LTE dynamics has been thoroughly scrutinised in \cite{Branchina:2025jou}, while the present work focuses onto the corrections due to the OOE dynamics.   

\paragraph{Hydrodynamic equations.} Hydrodynamic equations are obtained from the conservation of $T_{\mu\nu}$. As customary, neglecting the expansion of the universe, and assuming the bubble radius to be much larger than its width, we can take a planar approximation for the wall and write the conservation equations in the domain wall reference frame as
\begin{equation}
\partial^z T_{z0}= \partial^z T_{zz} = 0.
\label{eq: EMT cons planar}
\end{equation}
The $z$-axis here is chosen to be aligned with the direction of propagation of the front. 

Integrating Eq.\,\eqref{eq: EMT cons planar} over $z$ we get
	\begin{align}
		\label{hydro eq 1}
		T_{30} \equiv w\, \gamma^2 v_p + T_{30}^{\rm OOE} = c_1, \\
		\label{hydro eq 2}
		T_{33} \equiv \sum_{j=1}^n\frac{(\partial_z \phi_i)^2}{2} - V(\phi_1,\dots,\phi_n, T) + w\, \gamma^2 v_p^2 + T_{33}^{\rm OOE} = c_2,
	\end{align}
where $V(\phi_1,\dots,\phi_n,T)\equiv V_{_{T=0}}-p$ is the finite temperature effective potential and $c_{1,2}$ are integration constants. The latter can be determined from the asymptotic values of the plasma temperature, $T_\pm$, and velocity, $v_\pm$, either in front or behind the wall (as usual in the literature, we use the $+$ subscript for quantities in front of the wall, and the $-$ subscript for quantities behind it), that in turn depend on the combustion regime of the transition and on the potential. For further details, we refer to our previous work \cite{Branchina:2025jou} and to \cite{Espinosa:2010hh,Gyulassy:1983rq}.

\paragraph{Scalar equations of motion. }

As a benchmark scenario to assess the impact of out-of-equilibrium effects on the wall dynamics, we consider the case of two-step phase transitions in the $\mathbb Z_2$-symmetric singlet extension of the Standard Model. This is one of the simplest BSM models where a first-order phase transition can be obtained, with only a slight modification in the particle content with respect to the SM. Namely, besides the Higgs field ($h$ below), the model contains a second scalar ($s$), that is CP-even, neutral and only couples to the Higgs through a quartic portal $h^2 s^2$. The tree-level potential $V_0$ is 
\begin{equation}
	\label{eq: V0}
	V_0(h,s) = \frac{\mu_h^2}{2}  h^2 + \frac{1}{4} \lambda_h h^4 + \frac{\mu^2_s}{2} s^2 + \frac{\lambda_s}{4} s^4 + \frac{\lambda_{hs}}{2} h^2 s^2.
\end{equation}
The free parameters are $\mu_s,\, \lambda_s$ and $\lambda_{hs}$, and we trade $\mu_s$ for the singlet mass in the electroweak vacuum ($h=v$, $s=0$), $m_s=\mu_s^2 +\lambda_{hs}  v^2$. For the determination of the effective potential (at one-loop order), we use an on-shell renormalisation scheme and adopt Parwani resummation. For further details, we refer again to our previous work\,\cite{Branchina:2025jou}.  

Under the assumption that the scale of variation of the scalars is sufficiently larger than the particle mean free path in the plasma, the equations of motion (EOMs) for $h$ and $s$ can be determined using a WKB approximation~\cite{Moore:1995ua,Moore:1995si}, 
\begin{align}
	\label{eq: scalar EOMs}
	E_h \equiv - \partial^2_z h + \frac{\partial V(h,s,T)}{\partial h} + F_h^{\rm OOE}(z) = 0, \\
	E_s \equiv - \partial^2_z s + \frac{\partial V (h,s,T)}{\partial s} +  F_s^{\rm OOE}(z) =0.
\end{align}
The term $F_{j}^{\rm OOE}$ (with $j=h,s$) stands for the OOE contribution due to the coupling to the wall, and is
\begin{equation}
\label{eq: F def}
	F^{\rm OOE}_j(z)=\sum_i \frac{n_i}{2}\frac{\partial m^2_i}{\partial \phi_j}\int\frac{d^3p}{(2\pi)^3 E_p} \,\delta f_i,
\end{equation}
where the sum is over the species and $n_i$ counts the number of degrees of freedom. It is readily seen, from the equation above, that the contribution from each field is proportional to its coupling to the wall, $\partial_{\phi_j}m^2$. As already mentioned, this means that only perturbations $\delta f_i$ from species that are strongly coupled to the scalars give a sizeable contribution to the wall dynamics. For simplicity, in this work we only consider perturbations from the top quark which has the largest coupling to the Higgs. Moreover, within this approximation the specifics of the model are such that $F_s^{\rm OOE}=0$. 

In the two step phase transition scenario, as the universe cools down there is a first transition from the trivial vacuum $(h,s)=(0,0)$ to an intermediate vacuum $(0,s)$, where the $\mathbb Z_2$ symmetry in the $s$-sector is broken. The EW vacuum is then reached from it after a first-order transition $(0,s) \to (h,0)$, which is the one we are interested in. Having this pattern in mind, we solve the equations of motion \eqref{eq: scalar EOMs} by taking a {\it tanh} ansatz for the fields   
\begin{align}
	h(z) = \frac{h_{-}}{2} \left( 1 + \tanh{\left( \frac{z}{L_h}\right)}\right),  \nonumber \\
	s(z) = \frac{s_{+}}{2} \left( 1 - \tanh{\left( \frac{z}{L_s} - \delta_s \right)}\right),
	\label{eq: ansatz scalars}
\end{align}
where $L_h$ and $L_s$ are the widths of the respective bubble walls, and $\delta_s$ defines the displacement of one wall compared to the other. 
The factors $h_{-}$ and $s_{+}$ represent the VEVs of the $h$ and $s$ fields in front and behind the wall, respectively
\begin{equation}
	\frac{\partial V(h_{-}, 0, T_-)}{\partial h}=0, \qquad \qquad \frac{\partial V(s_{+}, 0, T_+)}{\partial s}=0.
\end{equation}

The ansatz \eqref{eq: ansatz scalars} contains four parameters that need to be determined: $\delta_s, \,L_h,\, L_s$ and $T_-$, which is defined by the wall velocity $v_w$ through the matching equations, while $T_+$ is obtained from $T_n$.  We then trade the EOMs for four constraint equations by taking two moments for each field\footnote{More generally, the $\tanh$ ansatz shown here contains two parameters per field. The framework can be easily generalised to phase transitions with $n$ scalars, in which case $2n$ moments ($P_i, \, G_i$, $i=1,\dots,n$) are taken. Similarly, in the case of a more general ansatz with $l$ parameters per field, $l\times n$ moments should be retained.}, 
\begin{align}
	P_h &= \int dz E_h h' = 0, \qquad \qquad  G_h = \int dz E_h \left(  2\frac{h}{h_{-}} -1\right) h' = 0, \nonumber \\
	P_s &= \int dz E_s s' = 0, \qquad \qquad  G_s = \int dz E_s \left(  2\frac{s}{s_{+}} -1\right) s' = 0,
\label{eq: constraints}
\end{align}
where $'$ denotes derivation with respect to $z$.  The $P_j$ moments measure the pressure acting on the $\phi_j$ wall, while the $G_j$ moments are for the corresponding pressure gradients \cite{Moore:1995ua,Moore:1995si}.  
When solving the equations, we trade $P_h$ and $P_s$ for the combinations $P_{tot}\equiv P_h+P_s$ and $\Delta \equiv P_h - P_s$. This is convenient as the sum
\begin{equation}
\label{eq: Ptot complete}
P_{tot}=\Delta V-\int dz\,\partial_{_T}\,V\, T'+\int dz \,\widetilde F_h^{\rm OOE},
\end{equation}
with $\widetilde F_h^{\rm OOE}=\sum_i n_i/2\, (m^2_i)' \int d^3p/((2\pi^3)E_p)\, \delta f_i$,
determines the total pressure acting on the system, and its vanishing ensures balance between the outward pressure $\Delta V$ and the friction $\int \partial_{_T} V \,T'+\int dz \,\widetilde F_h^{out}$ is achieved. Numerically, we observe that, among the various parameters, $P_{tot}$ mainly depends on $v_w$. The difference $\Delta P$, on the other hand, is mainly related to the displacement $\delta s$: the distance between the centres of the fields adjusts to ensure vanishing of the pressure on each wall. Finally, the pressure gradient equations $G_j=0$ ensure that the solution has fixed widths $L_j$. 

\paragraph{Boltzmann equation. }

The OOE perturbation $\delta f_i$ of the $i$-th particle travelling with momentum $p$ and with space-dependent mass term $m_i$ is determined from the Boltzmann equation (here written in the wall reference frame)
\begin{equation}
\mathcal L[f_i]\equiv \left(\frac{p_z}{E_p}\partial_z-\frac{\left(m_i^2(z)\right)'}{2 E_p}\partial_{p_z}\right) f_i = - \mathcal C[f_l],
\end{equation}
where $\mathcal L$ is the Liouville operator and $f_l$ in the collision integral $\mathcal C$ generically indicates the distribution functions for all the species in the plasma. As before, the prime symbol\,$'$ indicates derivation with respect to $z$. Particles in the plasma experience a non-vanishing force from the gradient of their mass term $m_i(z)$, with the latter inheriting its space-dependence from the field profiles. Since we only consider OOE contributions from the top quark, in the following we drop the subscripts on the distribution functions and simply use $f$ for the top. \\
For $2\to 2$ scattering processes, that are the dominant ones we include here, the collision integral $\mathcal C[f]$ is 
\begin{equation}
\label{eq: collision integral}
\mathcal C[f]=\sum_i\frac{1}{4 N_p E_p} \int \frac{d^3 \vec k\, d^3\vec {p'}\,d^3\vec{k'}}{(2\pi)^5 2E_k \,2 E_{p'}\, 2 E_{k'}}\left|\mathcal M_i\right|^2 \delta^4(p+k-p'-k')\mathcal P[f]
\end{equation}
where $\mathcal M_i$ are the amplitudes for the processes under consideration ($t\bar t\to gg$, $tg\to tg$, $tq\to tq$, with $t$ the top, $g$ the gluons and $q$ light quarks), $N_p$ is the number of degrees of freedom of the incoming particle with momentum $p$, and $\mathcal P$ is (the $+$ sign is for bosons, the $-$ one for fermions)
\begin{equation}
\mathcal P[f]= f(p)f(k)\left(1\pm f(p')\right) \left(1\pm f(k')\right)-f(p')f(k')\left(1\pm f(p)\right)\left(1\pm f(k)\right).
\end{equation}
The Boltzmann equation is typically solved by linearisation. When evaluated on the LTE distributions the collision integral vanishes, and the equation becomes
\begin{equation}
	\label{eq: linearised Boltzmann}
\mathcal L\left[\delta f\right] = \frac{p_z}{E}\mathcal S - \overline{\mathcal C}[\delta f]
\end{equation}
where 
\begin{equation}
\frac{p_z}{E} \mathcal S\equiv \mathcal L\left[f_0\right]  
\end{equation}
acts as a source for $\delta f$ (effects beyond linear order are shown to be subdominant in \cite{DeCurtis:2024hvh}).  
Upon linearisation, the population factor $\mathcal P$ reduces to 
\begin{equation}
\label{eq: linearised population}
\overline{\mathcal P}=f_0(p)f_0(k)\left(1\pm f_0(p')\right)\left(1\pm f_0(k')\right)\sum_{l\in \left(p,\,k,\,p',\,k'\right)}\frac{\mp \delta f(l)}{\left(f_0(l)\right)'},
\end{equation} 
and so the linearised collision integral $\overline{\mathcal C}$ is explicitly found.

The linearised population factor  $\overline{\mathcal P}$ can be expressed as the sum of four different terms, one of which is proportional to $\delta f(p)$, with $p$ the momentum of the particle entering in the Boltzmann equation, and the others depending on $\delta f(q)$, with $q$ one of the momenta that are integrated over in \eqref{eq: collision integral}. Ref.\,\cite{DeCurtis:2022hlx} has shown that, splitting $\overline {\mathcal C}$ as $\overline{\mathcal C}= c_1 \,\delta f(p)+\langle \delta f\rangle$, with $\langle \delta f\rangle$ the bracket, an iterative strategy to numerically solve the Boltzmann equation directly, that is without resorting to any ad-hoc ansatz, can be devised. 

The bracket $\langle\delta f\rangle$, which is given by a nine-dimensional integration over the momenta $k$, $p'$ and $k'$,  represents the most challenging part of the calculation. A different approach to the one typically used in the literature was put forward in \cite{DeCurtis:2022hlx,DeCurtis:2023hil,DeCurtis:2024hvh}. We briefly sketch it below.

The bracket contains various contributions from both annihilation and scattering processes. Focusing for example on the former ones, we write
\begin{equation}
\langle \delta f\rangle \supset -\frac{f_0(p)}{4 N_p E_p} \int \frac{d^3 \vec k}{2E_k}\mathcal K_a\, f_0(k)\frac{\delta f(k)}{\left(f_0(k)\right)'},
\end{equation}
where we defined the kernel $\mathcal K_a$ 
\begin{equation}
\mathcal K_a=\int \frac{d^3\vec {p'}\,d^3\vec{k'}}{(2\pi)^5\,2 E_{p'}\, 2 E_{k'}}\left|\mathcal M_a\right|^2 \delta^4(P)  \left(1\pm f_0(p')\right)\left(1\pm f_0(k')\right).
\end{equation}
and $P\equiv p+k-p'-k'$. As explained in \cite{DeCurtis:2022hlx}, $\mathcal K_a$ is a scalar that depends only on the energy of the incoming particles $E_{p_*}$ and $E_{k_*}$ in the plasma reference frame (inside the integral particles are taken to be massless, so $E_q=q$), and on the angle between $\vec p_*$ and $\vec k_*$ (starred momenta are calculated in the plasma reference frame). Similar considerations can be made for the scattering channel, and, after some manipulations, the full bracket can be expressed in the simple form \cite{DeCurtis:2023hil,DeCurtis:2024hvh}
\begin{equation}
    \langle \delta f \rangle = \int\frac{d^3 \vec{k_*}}{2|\vec {k_*}|} \mathcal K(\beta|\vec{p_*}|, \beta|\vec{k_*}|,\cos\theta_{p_*k_*})\frac{\delta f(k_\bot, k_z,z)}{f'_0(\beta|\vec{k_*}|)}
\end{equation}
where $k_\bot$ is the momentum perpendicular to the propagation of the wall.

This expression for $\langle\delta f\rangle$ can be simplified to minimal terms by first exploiting its rotational invariance.
Thanks to that, we decompose $\delta f$ and $\mathcal K$ in spherical harmonics as
\begin{equation}
\label{eq: decomposition df harmonics}
	\delta f(k_\bot,k_z, z) = \sum_l\frac{2l + 1}{2}\psi_l(|\vec{ k_*}|,z)P_l(\cos\theta_{k_*})\,,
\end{equation}
\begin{equation}
\mathcal K(\beta|\vec{p_*}|, \beta|\vec{k_*}|,\cos\theta_{p_*k_*})=\sum_{l}\frac{2l+1}{2} \mathcal G_l(\beta|\vec{p_*}|, \beta|\vec{k_*}|)P_l(\cos\theta_{p_*k_*}),
\end{equation}
where $P_l$ are the Legendre polynomials. By integrating over the polar and azimuthal angles we get 
\begin{equation}
\label{eq: bracket final}
\langle\delta f\rangle = \pi \sum_{l=0}^\infty\frac{2l+1}{2} \int \mathcal D k_*\; \mathcal G_l(\beta |\vec{p_*}|,\beta|\vec{ k_*}|)\frac{\psi_l(|\vec{k_*}|,z)}{f'_0(\beta|\vec{k_*}|)}P_l(\cos\theta_{p_*})\,
\end{equation}
where the integration measure is $\mathcal D k_* =  f_0(|\vec{k_*}|)|\vec{k_*}| d|\vec{k_*}|$.
The great improvement first brought in \cite{DeCurtis:2023hil,DeCurtis:2024hvh}
stems from the observation that each multipole in $\langle \delta f \rangle$ in \eqref{eq: bracket final} can be interpreted as the result of the application of an Hermitian operator $\mathcal O_l$ on the perturbation. 
The action of the operator on a generic function $g$ is defined as  
\begin{equation}
{\cal O}_l[g] = \int\mathcal D k_*\; \mathcal G_l(\beta |\vec{ p_*}|,\beta|\vec{k_*}|)\,g(|\vec{k_*}|)\,,
\end{equation}
and, owing to its hermiticity, ${\cal O}_l$ can be diagonalised on the basis of its eigenfunctions $\zeta_{l,i}$ with real eigenvalues $\lambda_{l,i}$. 
The bracket is then expressed as 
\begin{equation}
\langle \delta f\rangle = -\pi\sum_l\sum_i \frac{2l+1}{2}\lambda_{l,i}\, \zeta_{l,i}(\beta|\vec{p_*}|)P_l(\cos\theta_{p_*})\,\phi_{l,i}(z)\,,
\end{equation}
where the function $\phi_{l,i}(z)$ corresponds to the projection of the $l$-th Legendre mode of $\delta f$ onto the the previously mentioned eigenbasis, 
\begin{equation}
    \phi_{l,i}(z) =  \int \mathcal D k_*\, \zeta_{l,i}(\beta |{\vec{k_*}}|)\frac{\psi_l (|{\vec{k_*}}|, z)}{f_0'(|{\vec{ k_*}}|)} \,.
\end{equation}

As reviewed above, besides allowing for a way to directly solve the Boltzmann equation, the approach pioneered in \cite{DeCurtis:2022hlx,DeCurtis:2023hil,DeCurtis:2024hvh} allows to significantly mitigate the complexity in the calculation of the collision integral, that, once the basis of eigenfunctions is found, is reduced to a one-dimensional integration. Needless to say, the set of eigenfunctions constitutes the natural basis for the decomposition of the collision integral and allows to control the accuracy of its evaluation by exploiting the hierarchy of the eigenavalues.
For instance, it was found in \cite{DeCurtis:2024hvh} that in the block $l = 0$ all the eigenvalues but the first four are suppressed by a factor of
$10^{-4}$ and the suppression is even stronger for higher modes. Consequently, 
to reconstruct the kernel with, for example, an accuracy of order 1\%, just a few eigenvectors belonging to the lowest modes are sufficient.
Moreover, the kernels only depend on the processes considered, so that, when performing a survey of a model, they only need to be computed once. Details on the numerical implementation of this calculation are given in \cite{DeCurtis:2024hvh}. Combining with the results in \cite{Branchina:2025jou}, this makes it possible to fully study the wall dynamics across the parameter space of BSM models. In the following, we present the results of our numerical investigation for one such model.

\section{Numerical analysis}
\label{sec: numerical results}

In this section, we present the results for the survey of the parameter space of the singlet extension of the Standard Model, that we use as a benchmark scenario to assess the impact of out-of-equilibrium contributions on the wall dynamics. To this end, we compare the results to those obtained in local thermal equilibrium, that were recently presented in \cite{Branchina:2025jou} using the same set-up. In particular, we mainly focus on the wall velocity $v_w$ and the $h$-wall width $L_h$, that are the parameters to which cosmological relics are more sensitive. As expected, we find that the effect of out-of-equilibrium contributions is to reduce the wall velocity and increase the wall widths. The displacement $\delta_s$ is the least affected of all parameters. 

\begin{figure}
\centering
\includegraphics[width=0.62\textwidth]{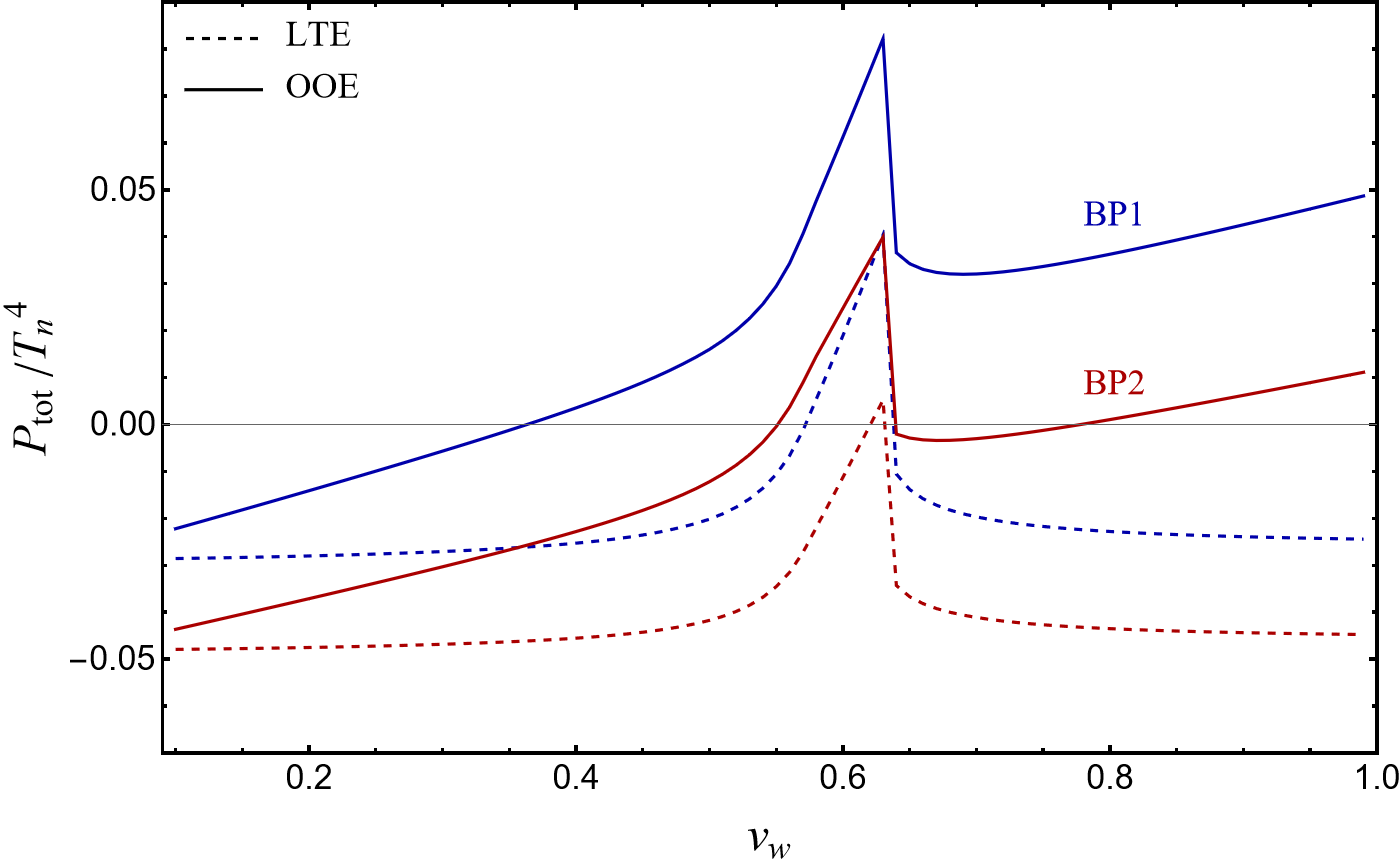}
\caption{Total pressure $P_{tot}$ in terms of the wall velocity $v_w$ for two benchmark points, BP1 with $m_s=80$ GeV and $\lambda_{hs}=0.36$ and BP2 with  $m_s=122$ GeV and $\lambda_{hs}=0.47$. Dashed lines are for the pressure evaluated in LTE, while solid lines are for the pressure evaluated including out-of-equilibrium contributions. The other three parameters $\delta_s$, $L_h$ and $L_s$ are fixed to the corresponding solution.}
\label{fig:Ptot}
\end{figure}

We find it useful to begin by discussing the $v_w$-dependence of the total pressure, that we display in Fig.\,\ref{fig:Ptot} for two benchmark points. When evaluating $P_{tot}(v_w)$, the other parameters are fixed to the value they take on the solution. 
In the figure, we compare the profiles of $P_{tot}(v_w)$ in the LTE approximation and including out-of-equilibrium contributions, to highlight how perturbations modify the shape. In local equilibrium, $P_{tot}(v_w)$ slowly grows with $v_w$ at low velocities. As the speed of sound is approached, the pressure starts to grow more rapidly, until a peak is reached for $v_w\to v_{_J}$ from below, where $v_{_J}$ is the Jouguet velocity. Beyond that point, the profile abruptly changes from a hybrid to a detonation, $P_{tot}$ suddenly drops and becomes negative\footnote{The fact that, within the LTE approximation, the pressure becomes negative beyond $v_{_J}$ was verified in our scan, and agrees with expectations built on analytic considerations for $P_{tot}$ as discussed in \cite{Branchina:2025jou}, where it was argued that detonations are unlikely to appear in LTE. However, there is no complete proof that $P_{tot}$ must necessarily drop to negative values for $v_w>v_{_J}$, and one should be open to the possibility that detonations might still appear in LTE with some tuning of the parameters.}. It then decreases as $v_w$ further increases, preventing the appearance of detonations. With this profile for $P_{tot}(v_w)$, a steady-state solution, corresponding to a deflagration/hybrid bubble, is found when the curve crosses the axis before reaching the peak. OOE perturbations modify this picture mainly in two ways: the slope for small $v_w$ gets larger, and beyond $v_{_J}$ the pressure is not monotonically decreasing. These features are such that (i) if a (deflagration/hybrid) solution exists in LTE, a solution with smaller wall velocity exists when OOE contributions are included; (ii) if no solution exists in LTE, a steady-state expansion can still be obtained within the full solution; (iii) in contrast to the LTE analysis,  detonations at intermediate wall velocities (as opposed to ultra-relativistic detonations) can be realised.  We stress that in this work we are only considering out-of-equilibrium contributions from the top quark; perturbations from other sufficiently strongly coupled species, especially $W$ bosons, can also have a significant impact and can further enlarge the region where a stationary expansion is realised \cite{DeCurtis:2024hvh}. 

\begin{figure}
	\centering
	\includegraphics[width=0.485\linewidth]{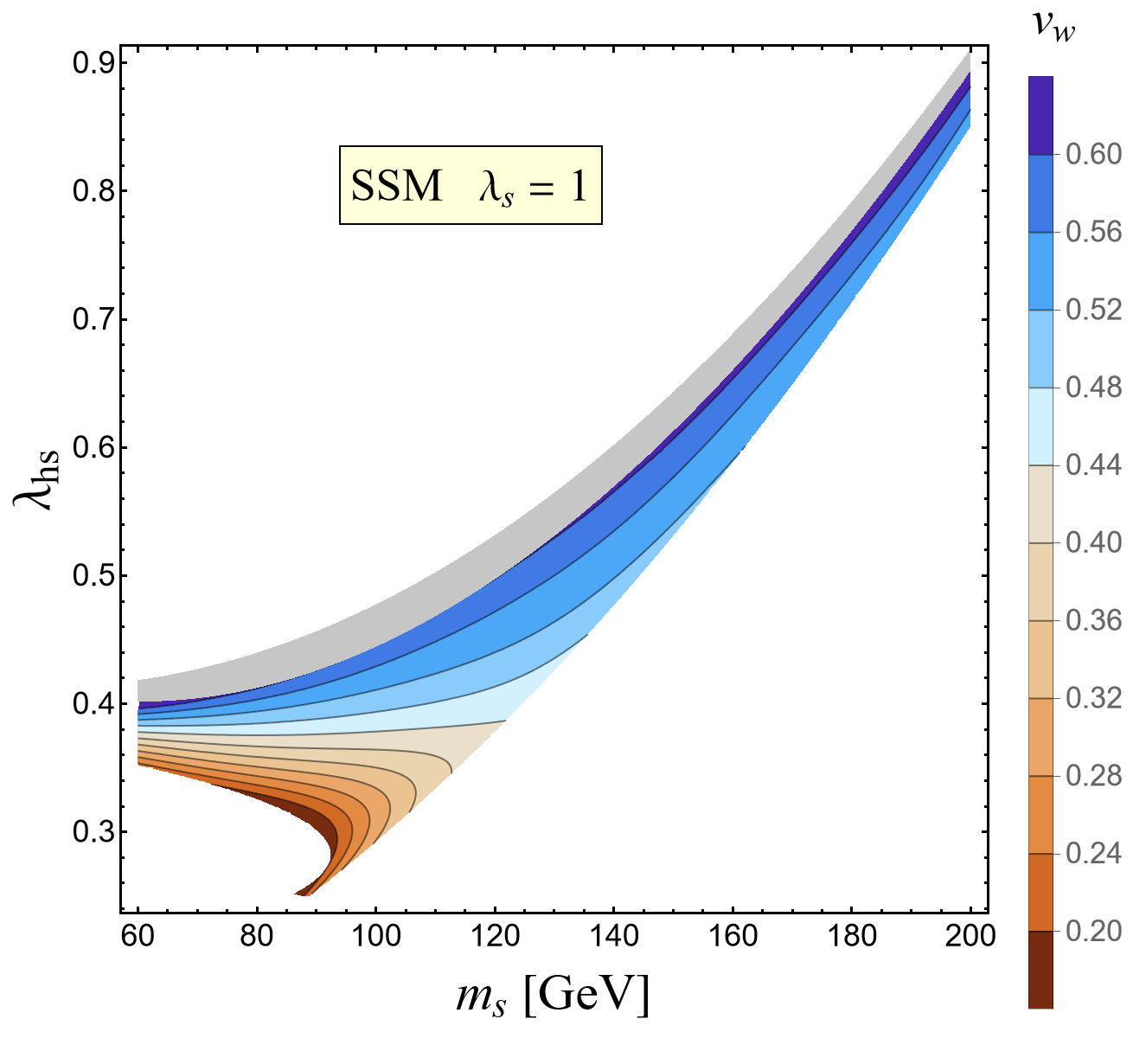}\,\,
    \includegraphics[width=0.495\linewidth]{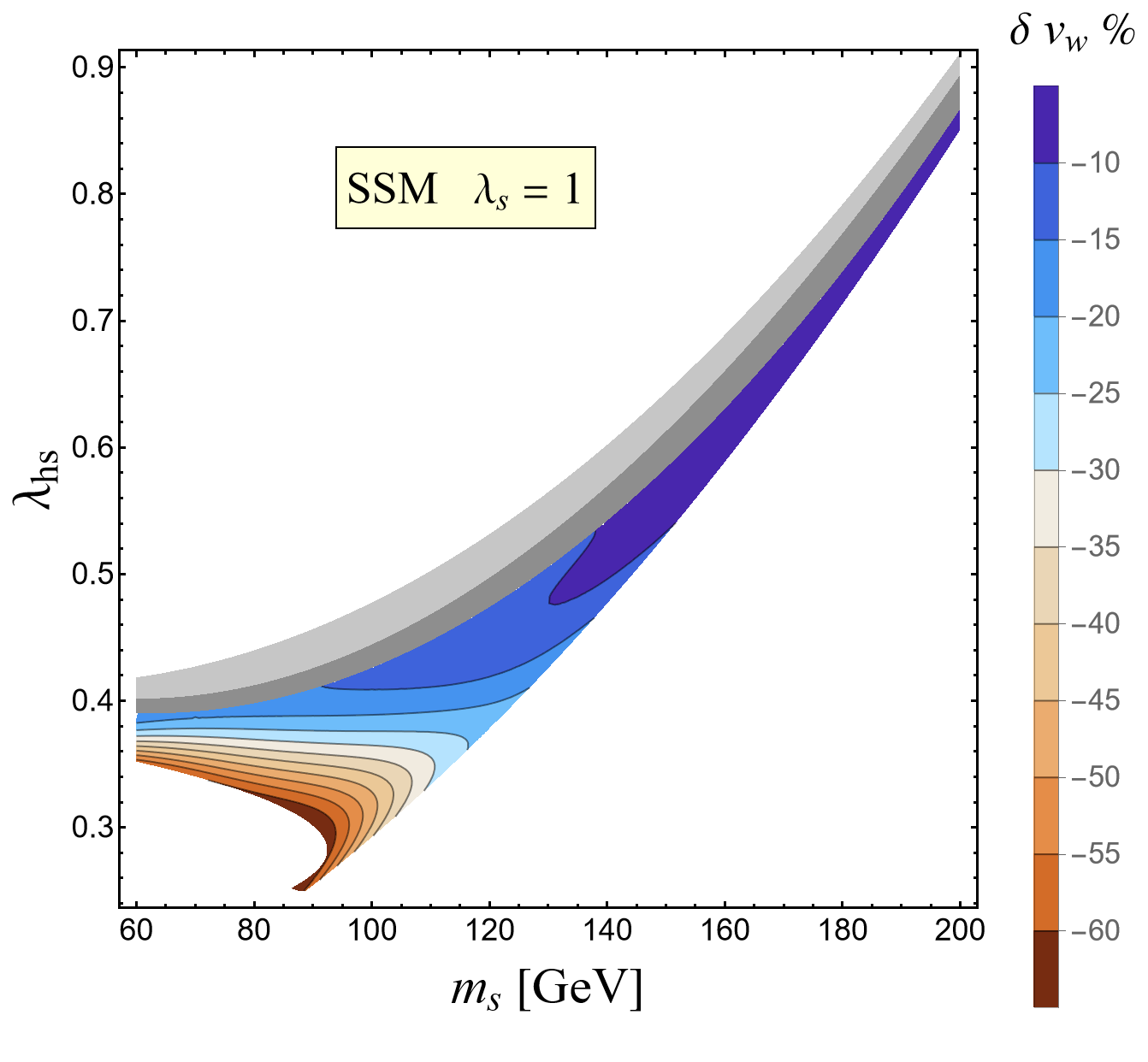}
	\caption{{\it Left panel}: Contour plot of the (out-of-equilibrium) wall velocity $v_w$ in the parameter space. The colour gradient shows the variation of $v_w$, with the colour bar serving as a legend and $v_w$ being constant on black lines. In the light grey region no deflagration/hybrid solution is found. {\it Right panel}: Relative correction $\delta v_w$ to the LTE wall velocity. The coloured region is restricted to the subspace where a solution exists in LTE. The dark grey band is where a deflagration/hybrid solution emerges when out-of-equilibrium contributions are included, while the light grey band is as in the left panel.}
	\label{fig:vw}
\end{figure}

These differences are due to the fact that in local equilibrium the only source of friction is given by the gradient of the temperature, and the latter is significantly overestimated on the LTE solution to balance the outward pressure. As we will see in a moment, the impact of this over-estimation on the wall dynamics depends on the values of the model parameters. To make this more transparent, it is convenient, within our ansatz for the particle distribution functions $f=f_0+\delta f$, to distinguish two different sources of friction. A first one, that we refer to as the back-reaction of the plasma, is the one that is already at work in LTE and is a hydrodynamic effect determined by $f_0$. As the bubble appears in a plasma with homogeneous temperature and velocity, it triggers a response from the plasma, that in turn develops non-trivial $T(z)$ and $v_p(z)$ profiles to ensure continuity conditions are respected. The role of the temperature profile is apparent in the expression of the total pressure $P_{tot}=\Delta V - \int dz \,\partial_{_T} V\, T'(z) \,- P_{\delta f}$, where $\Delta V$ is the difference in the potential inside and outside the bubble that drives the expansion and $P_{\delta f}>0$ stands for out-of-equilibrium contributions. In LTE a steady-state expansion is found when $\int dz \,\partial_{_T} V\, T'(z) = \Delta V$. The second source of friction, $P_{\delta f}$, is determined by the deviations from equilibrium $\delta f$, and is given by the term $F(z)$ defined above (see \eqref{eq: F def} and \eqref{eq: Ptot complete}) that enters in the scalar equations of motion. In the LTE approximation, this contribution is neglected, and the response of the plasma is fully encoded in the profile $T(z)$.

The results for the wall velocity $v_w$ are collected in Fig.\,\ref{fig:vw}. In the left panel, we present a contour plot of $v_w$ with out-of-equilibrium contributions in the parameter space. Throughout this work, we only show results for deflagration/hybrid walls. Points that have a sufficiently strong phase transition also develop a second (non ultra-relativistic) solution on the detonation branch\footnote{Points where the phase transition is not strong enough to generate a detonation solution within the set-up described in Section \ref{sec: set-up} will feature a ultra-relativistic detonation with $v_w\to1$. The latter is generated by $1\to2$ friction terms that strongly enhance the friction in the $v_w\to1$ limit \cite{Ai:2025bjw,Ramsey-Musolf:2025jyk} and stop the wall before it reaches $v_w=1$.}. This is expected to happen in a somehow narrow strip for sufficiently large values of $\lambda_{hs}$, and will be the object of future investigations. The light grey region in the plot is for the points where no solution was found. 

\begin{figure}
    \centering
    \includegraphics[width=0.58\linewidth]{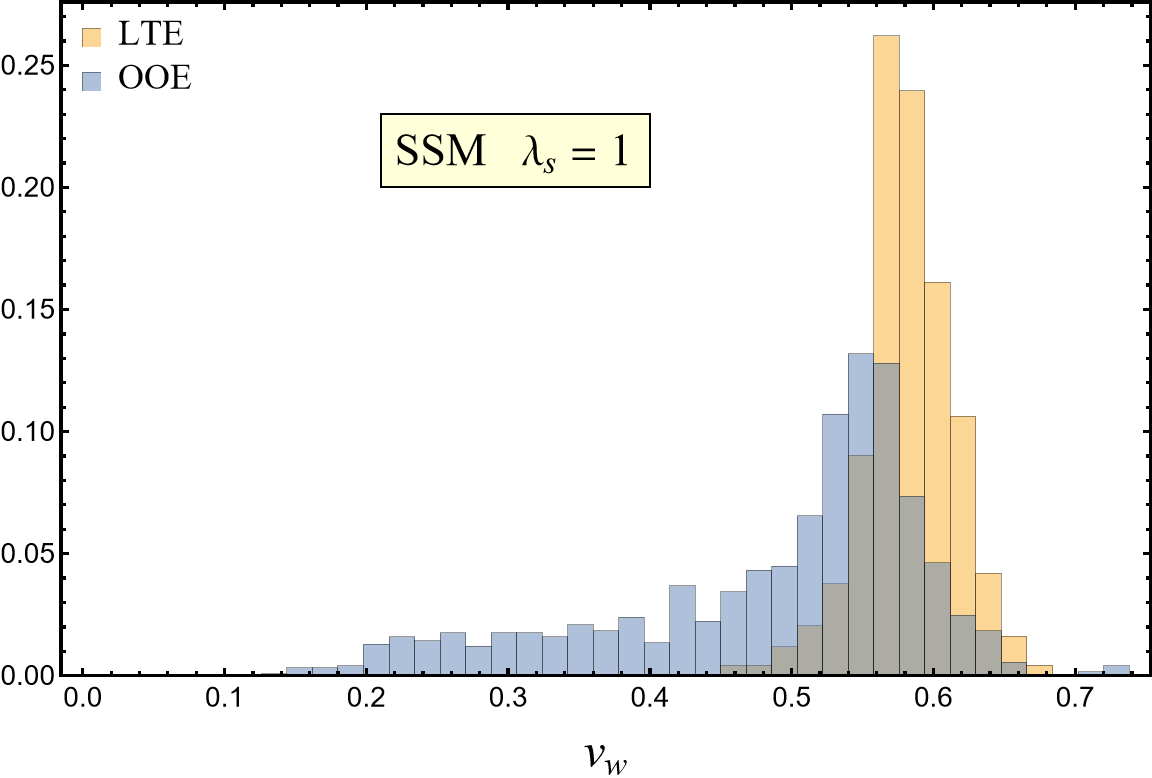}
    \caption{Histogram of the wall velocity $v_w$ in LTE  (yellow bins) and with out-of-equilibrium contributions included (blue bins).}
    \label{fig: histo distribution vw}
\end{figure}

In the right panel we show a contour plot of the relative deviation from LTE, {\small $\delta v_w \equiv \left(v_w^{({\rm OOE})}-v_w^{({\rm LTE})}\right)/v_w^{({\rm LTE})}$}, that determines the correction brought by out-of-equilibrium contributions to the LTE value of $v_w$. The plot is restricted to the subspace where a LTE solution exists\,\cite{Branchina:2025jou}. The dark grey band indicates the region where no steady-state expansion is found within the LTE approximation, but a solution exists when out-of-equilibrium contributions from the top quark are included 
(this region is then understood to be included in the coloured one in the left panel). This is due to the LTE plasma back-reaction not being sufficient to contrast the outward pressure, so that, within the LTE approximation, the expansion regime is erroneously predicted to be that of an ultra-relativistic detonation with $v_w\to1$. This can be easily understood from the behaviour of $P_{tot}(v_w)$ discussed above. As in the left panel, the light grey band is for the region where no solution is found even when perturbations from the top are considered. 

Besides the points that are erroneously classified as ultrarelativistic detonations in LTE, it is immediately apparent from Fig.\,\ref{fig:vw} that the largest corrections, up to $\sim -60\%$, are found in the lower and leftmost part of the parameter space, where the transition is weaker. For stronger transitions, the impact is smaller but seizable. It is also worth observing that the ``lines of constant correction" do not follow the isolines of $v_w^{({\rm LTE})}$ (see \cite{Branchina:2025jou}), so that the dependence of $v_w$ on the parameters $m_s$ and $\lambda_{hs}$ is significantly altered when perturbations are neglected.

Due to these large corrections, the interval where $v_w^{({\rm OOE})}$ varies is considerably wider than that of $v_w^{(\rm LTE)}$, and ranges from $v_w\sim 0.2$ to $v_w\sim 0.65$. This is further highlighted in Fig.\,\ref{fig: histo distribution vw}, where we show a histogram of the wall velocity for both $v_w^{(\rm LTE)}$ and $v_w^{(\rm OOE)}$. The LTE distribution of the wall velocity is quite sharply peaked around $v_w\sim 0.57$, with little spreading, such that most of the points lie in the range $v_w\sim 0.5-0.65$. This suggests a weak dependence of the LTE wall velocity on the potential and on the parameters of the transition. In this respect, it is worth mentioning that in \cite{Branchina:2025jou} we studied the dynamics in local equilibrium in three different models (the SSM, the real triplet extension, and the inert doublet model) and with two different choices of the additional scalar self-coupling, for each model, and found a substantial model-independence of the wall velocity. This agrees with the fact, observed in Fig.\,\ref{fig:Ptot}, that the LTE hydrodynamic obstruction is strongly enhanced in the region $v_w\sim (c_s,v_{_J})$, where $c_s$ is the speed of sound. Compared to the LTE one, the distribution of $v_w^{(\rm OOE)}$ is much wider, with a smaller peak around $v_w\sim 0.55$, and a persistent tail for small velocities down to $v_w\sim 0.2$. This suggests a much more pronounced dependence of $v_w$ on the potential and on the phase transition parameters, and that the model-independence observed in LTE ceases to be valid when out-of-equilibrium contributions are considered. 

\begin{figure}
    \centering
     \includegraphics[width=0.49\linewidth]{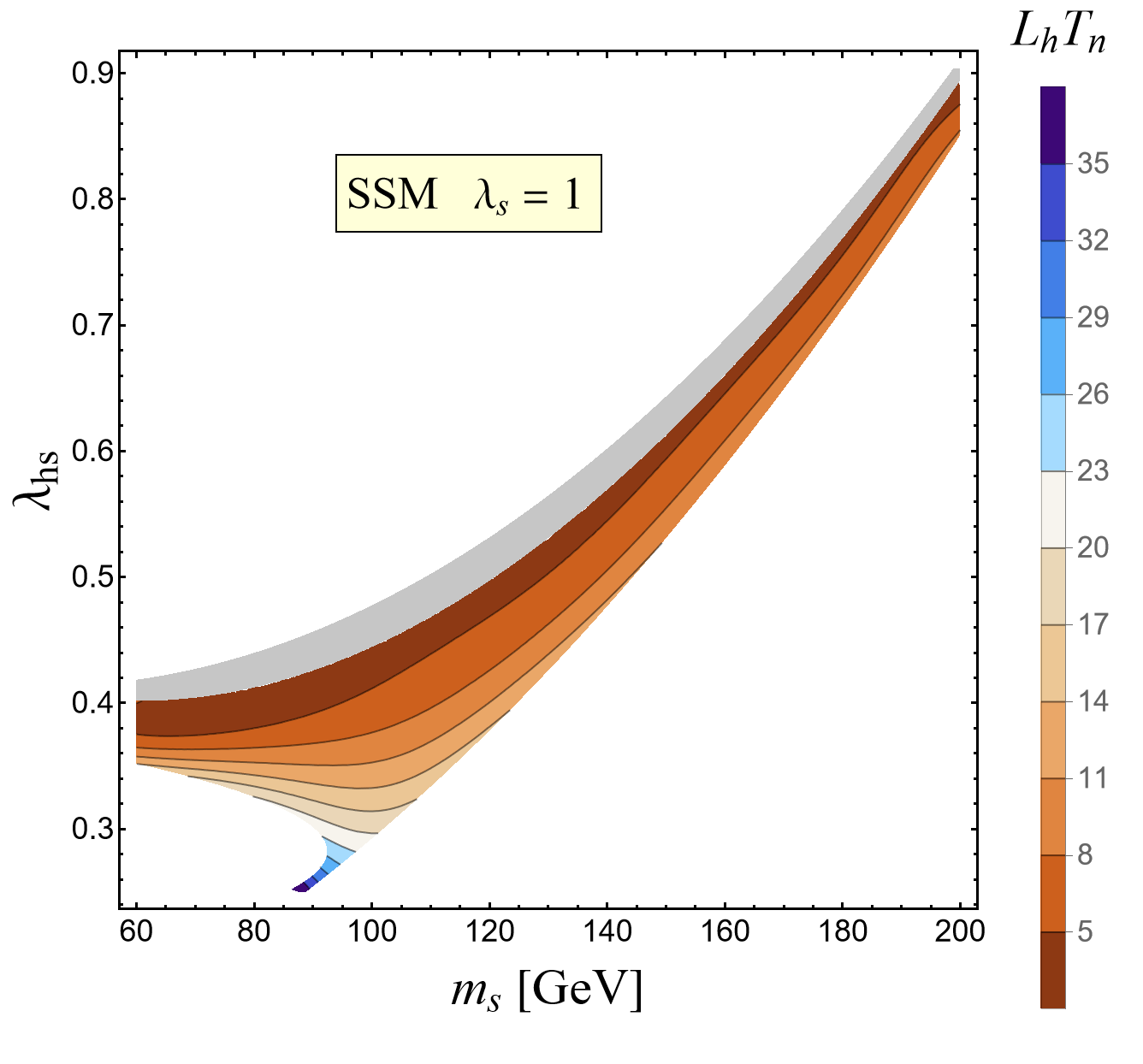}\,\,
    \includegraphics[width=0.49\linewidth]{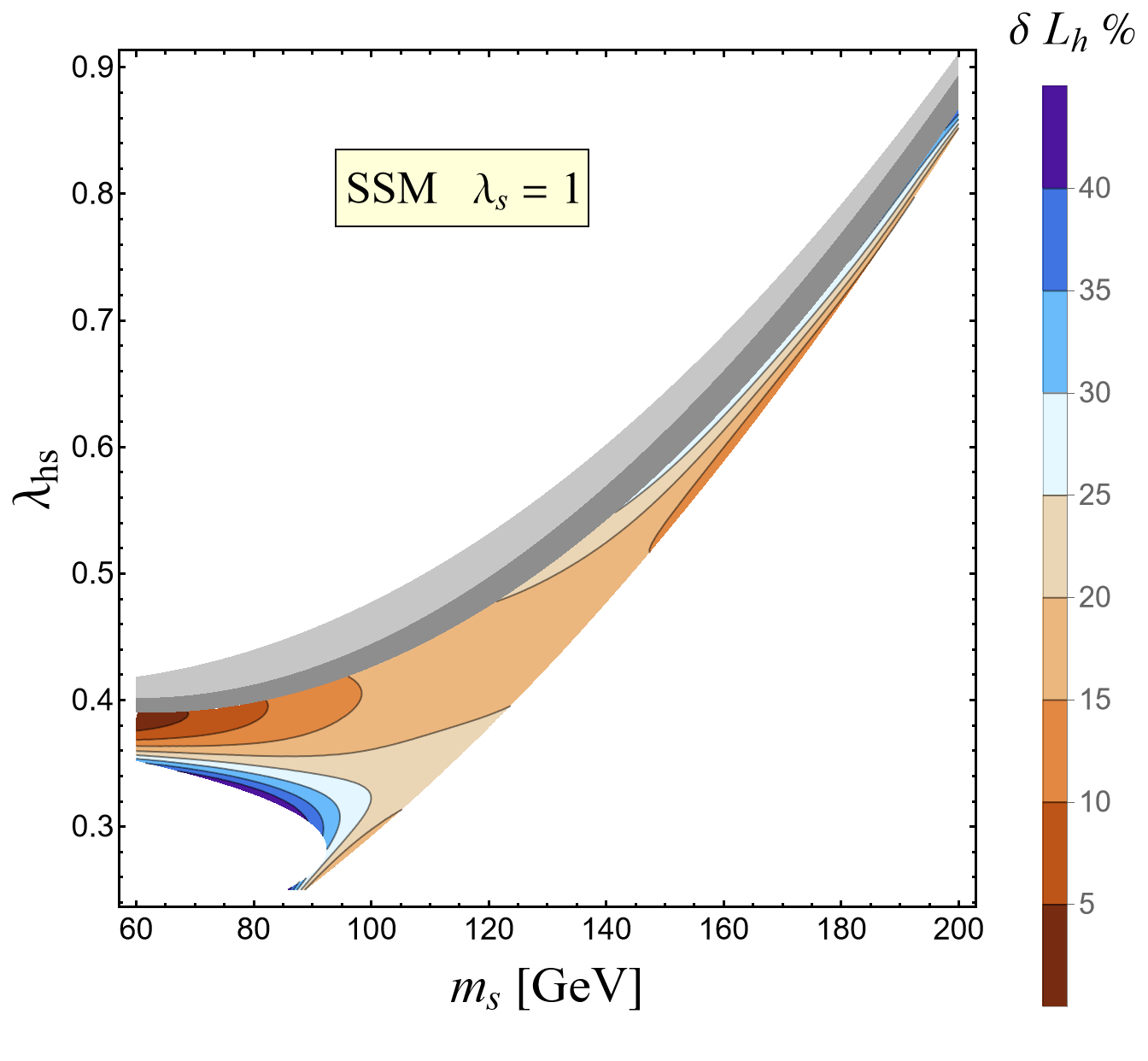}
    \caption{{\it Left panel}: Contour plot of the (out-of-equilibrium) wall width $L_h$ in the parameter space. In the light grey region no deflagration/hybrid solution is found. {\it Right panel}: Relative correction $\delta L_h$ to the LTE wall width. The coloured region is restricted to the subspace where a solution exists. The dark grey band is where a deflagration/hybrid solution emerges when out-of-equilibrium contributions are included, while the light grey band is as in the left panel.}
    \label{fig:Lh}
\end{figure}

Similar to Fig.\,\ref{fig:vw}, in Fig.\,\ref{fig:Lh} we present the results for the width $L_h$. The values found for $L_h$ are in the left panel, while the relative correction  {\small $ \delta L_h\equiv \left(L_h^{({\rm OOE})}-L_h^{({\rm LTE})}\right)/L_h^{({\rm LTE})}$} determined by OOE perturbations is in the right panel. Out-of-equilibrium contributions lead to an increase in the width, that in most of the parameter space is $\sim 20\%$. Smaller corrections are found for the fastest LTE walls, while larger corrections affect the slowest ones. These findings nicely agree with the common lore expectation that slower walls allow for a greater diffusion of field profiles in the plasma.

\begin{figure}
\centering
\includegraphics[width=0.485\linewidth]{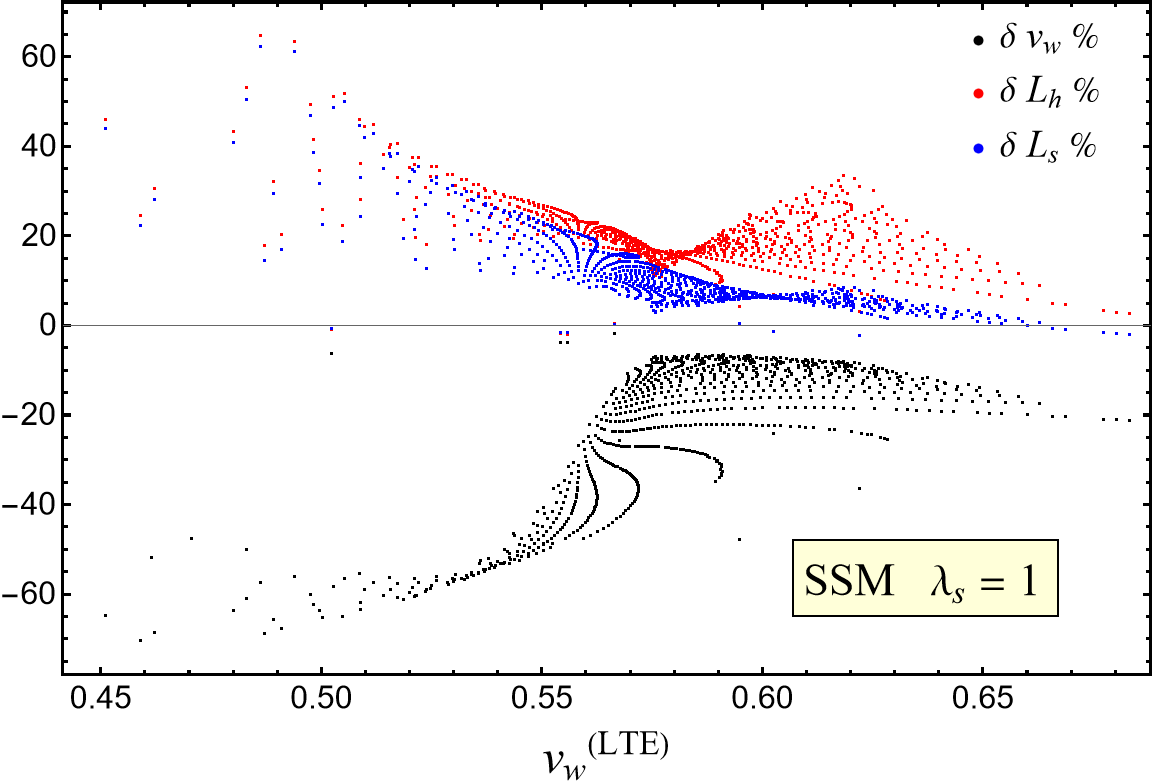}\,\,
\includegraphics[width=0.495\linewidth]{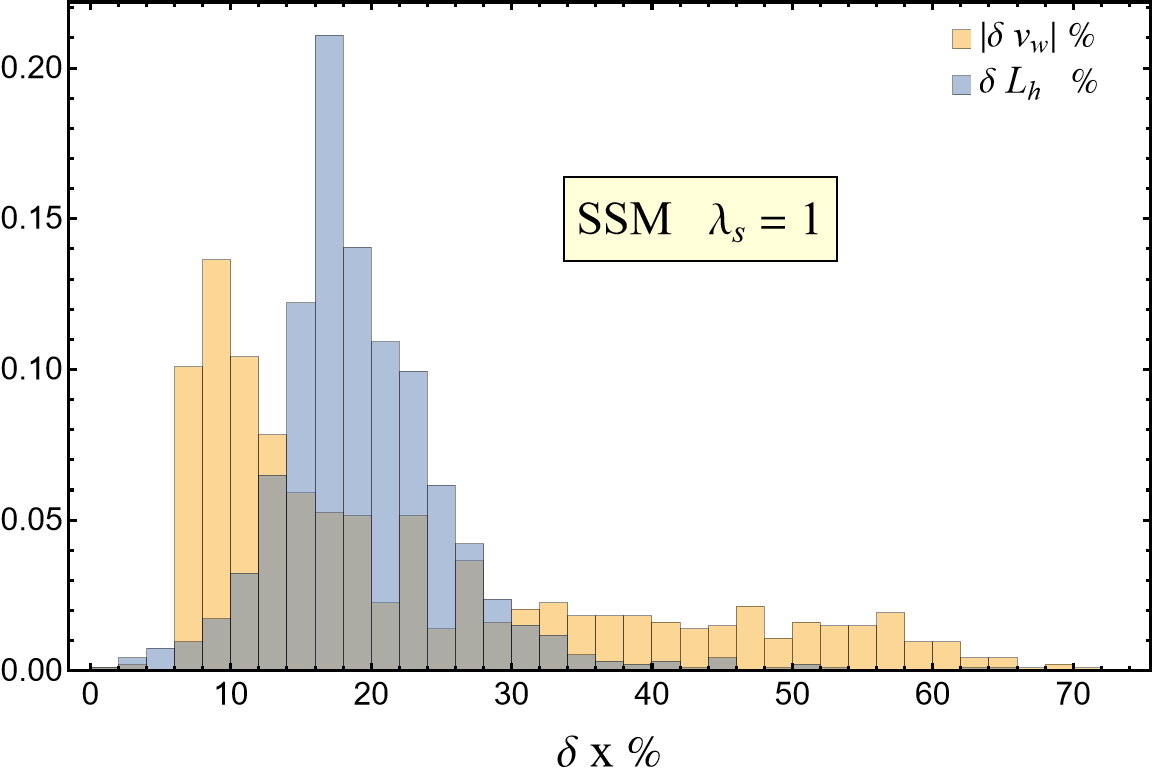}
 \caption{{\it Left panel}. Scatter plot of the relative corrections $\delta v_w\%$, $\delta L_h\%$ and$\delta L_s\%$ in terms of the LTE wall velocity $v_w^{(\rm LTE)}$. {\it Right panel}. Histogram of the relative corrections $\delta v_w\%$ and $|\delta L_h|\%$ normalised to one. The inside figure is a histogram of the value we find for $v_w$ in LTE and including out-of-equilibrium contributions.}
    \label{fig:scatter&histo}
\end{figure}

In Fig.\,\ref{fig:scatter&histo} we show a scatter plot (left panel) of the $\delta v_w$ and $\delta L_h$ corrections in terms of the LTE wall velocity, and the corresponding histogram (right panel). As a showcase of the corrections to $L_s$, we also include $\delta L_s$.  The patterns that appear in the scatter plot for the various parameters result from the fact that we use a square grid to explore the model parameter space.  
The figure shows that the correction $|\delta v_w|$ tends to be larger for slower (within the LTE characterisation) walls, with $|\delta v_w|\gtrsim 50\%$ for $v_w^{({\rm LTE})}\lesssim 0.55$, though faster walls ($v_w^{({\rm LTE})}\gtrsim 0.59$) can still have sizeable corrections, somewhere in the range $\sim 7-20\%$. The width corrections are more evenly distributed around $\sim 10\%$ for $L_s$, and $20\%$ for $L_h$. This last feature is also clearly highlighted by the histogram on the right panel of Fig.\,\ref{fig:scatter&histo}. There, one can also see that the histogram of the $|\delta v_w|$ corrections is approximately flat between  $\sim 30\%$ and $\sim 60\%$, indicating once more that a non-negligible number of points with large corrections to the wall velocity was found in our scan.

\begin{figure}
\centering
\includegraphics[width=0.485\textwidth]{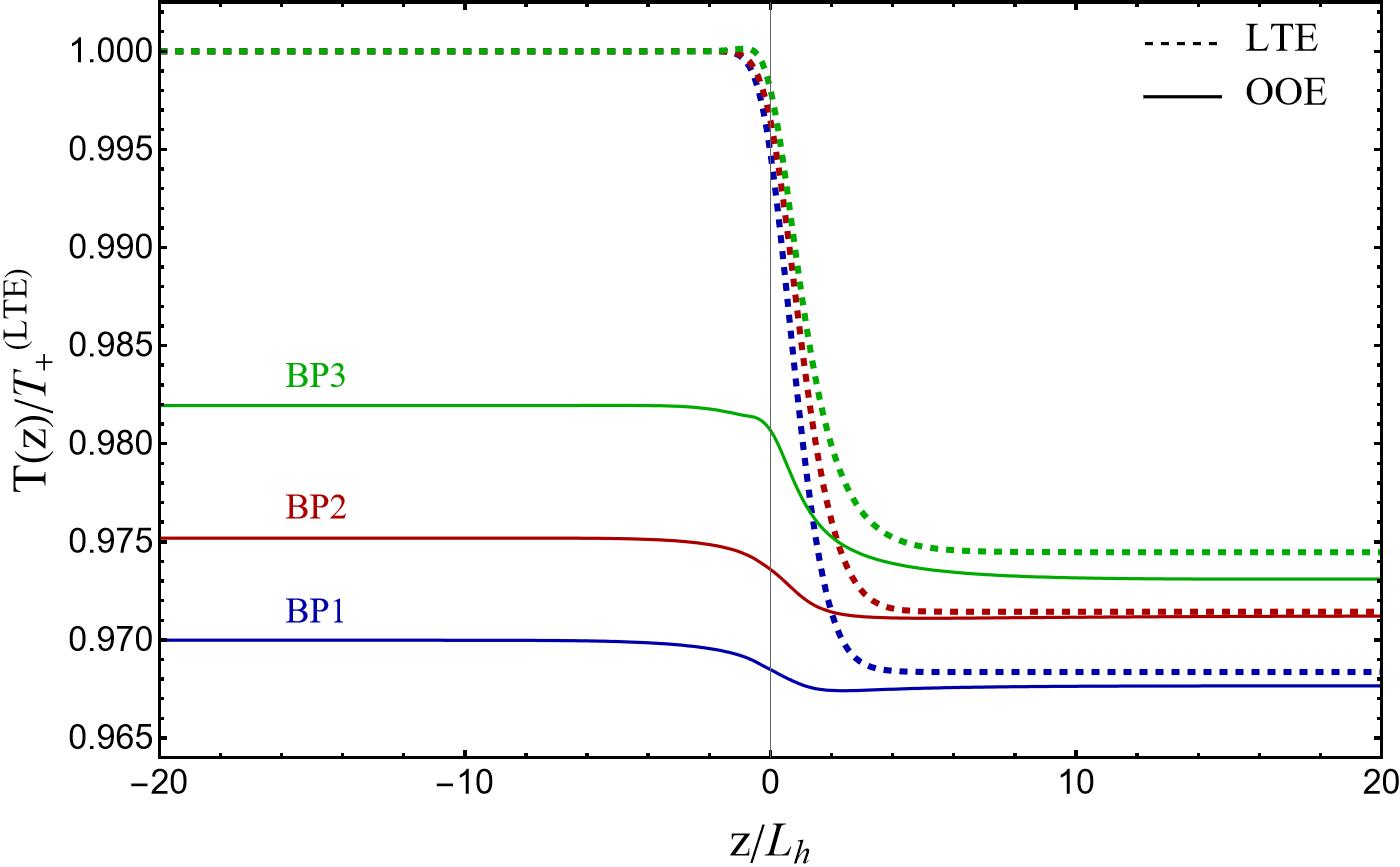}\,\,
\includegraphics[width=0.495\textwidth]{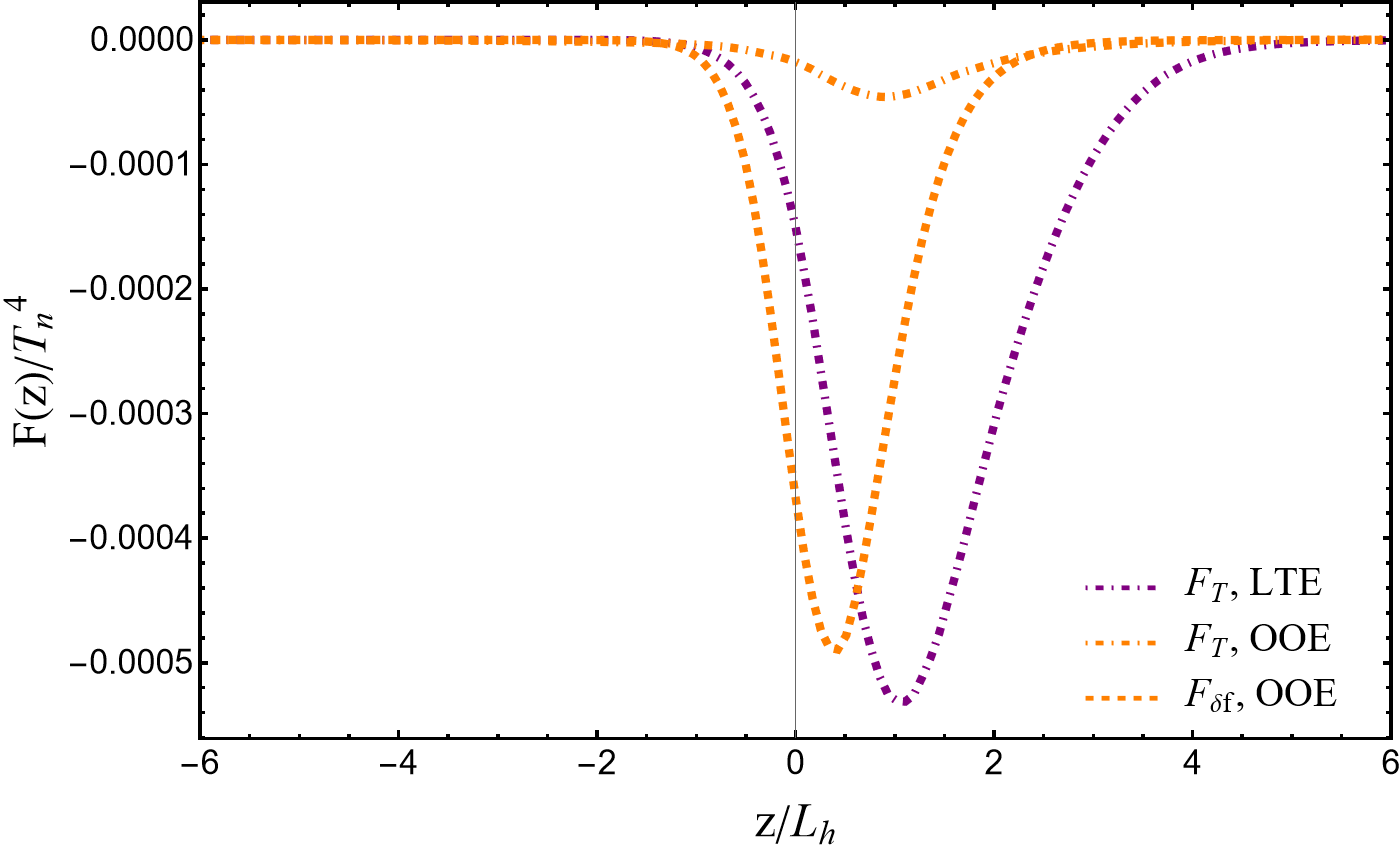}
    \caption{{\it Left panel}. Plot of the temperature profile $T(z)$ in the LTE approximation (dashed lines) and with out-of-equilibrium contributions (full lines) for three benchmark points: BP1 ($m_s=83$ GeV, $\lambda_{hs}=0.35$), BP2 ($m_s=112$ GeV, $\lambda_{hs}=0.39$), and BP3 ($m_s=176$ GeV, $\lambda_{hs}=0.69$). {\it Right panel}. Plot of the friction terms contributing to $P_{tot}$ for the second benchmark point BP2. The purple dot-dashed line is for the LTE friction $F_{_T}$ on the LTE solution, the orange dot-dashed line is for the LTE friction on the full solution, and the orange dashed line is for the out-of-equilibrium friction $F_{\delta f}$ on the full solution.}
	\label{fig:profiles}
\end{figure}

The results presented in this section show that the friction generated by purely out-of-equilibrium effects is not negligible, and provides a sizeable contribution to the wall dynamics. In turn, as already discussed, this also suggests that temperature gradients for the plasma are overestimated in LTE, where no mechanism other than the hydrodynamic obstruction can induce a friction on the expanding wall. As an example of this, we show the LTE and out-of-equilibrium temperature profiles for some benchmark points in the left panel of Fig.\,\ref{fig:profiles}. The three benchmarks are chosen to represent scenarios where the LTE parameters receive corrections of different size. In particular, we have (in percentage) $\delta v_w\sim-40\%$ for the blue curve, $\delta v_w\sim-20\%$ for the red one, and $\delta v_w\sim-10\%$ for the green one. It is immediately apparent, from the figure, that the LTE determination provides a fairly good approximation to the temperature inside the wall, but fails to reproduce the temperature outside of it. The overestimation of the temperature gradients is more significant for weaker phase transitions, but remains sizeable even when the OOE corrections are milder.

This is further displayed in the right panel of Fig.\,\ref{fig:profiles}, where we present the various contributions to the friction for BP2. In particular, we show the friction arising on the LTE solution, denoted ``$F_{_T}$, LTE", the friction arising from the temperature gradient on the full solution, denoted ``$F_{_T}$, OOE", and the purely out-of-equilibrium friction, ``$F_{\delta f}$, OOE". It is readily seen that, on the actual OOE solution, the purely out-of-equilibrium friction is dominant with respect to the one provided by the temperature gradient, that only amounts for a small contribution. A comparison between the purple and orange dot-dashed curves reveals how much the $F_{_T}$ contribution is overestimated by the LTE solution. By-products of this overestimation are the fact that the area below the purple line is slightly larger than the sum of the area below the two orange ones, and the fact that the profile of $F_{_T}$ in LTE is shifted to the right compared to that of $F_{\delta f}+F_{_T}$ on the full solution. The first feature is due to the fact that a larger variation of the temperature also induces a (slightly) larger variation in the potential difference between the asymptotic states; the second feature has to do with the fact that $\partial_{_T} V$ is larger for larger values of the Higgs field, and finding a solution in LTE is then facilitated if the temperature gradient is larger inside the wall.

\section{Gravitational waves and electroweak baryogenesis}
\label{sec- applications}

In this section we apply our results on the bubble wall dynamics to determine some cosmological relics of the phase transition as described in our reference model (actually, for the second application, we will consider a slight modification). In particular, we present results for the emitted spectrum of gravitational waves, and for the generation of the matter asymmetry within the framework of EWBG.   

\subsection{Gravitational waves}

\begin{figure}
    \centering
\includegraphics[width=0.49\linewidth]{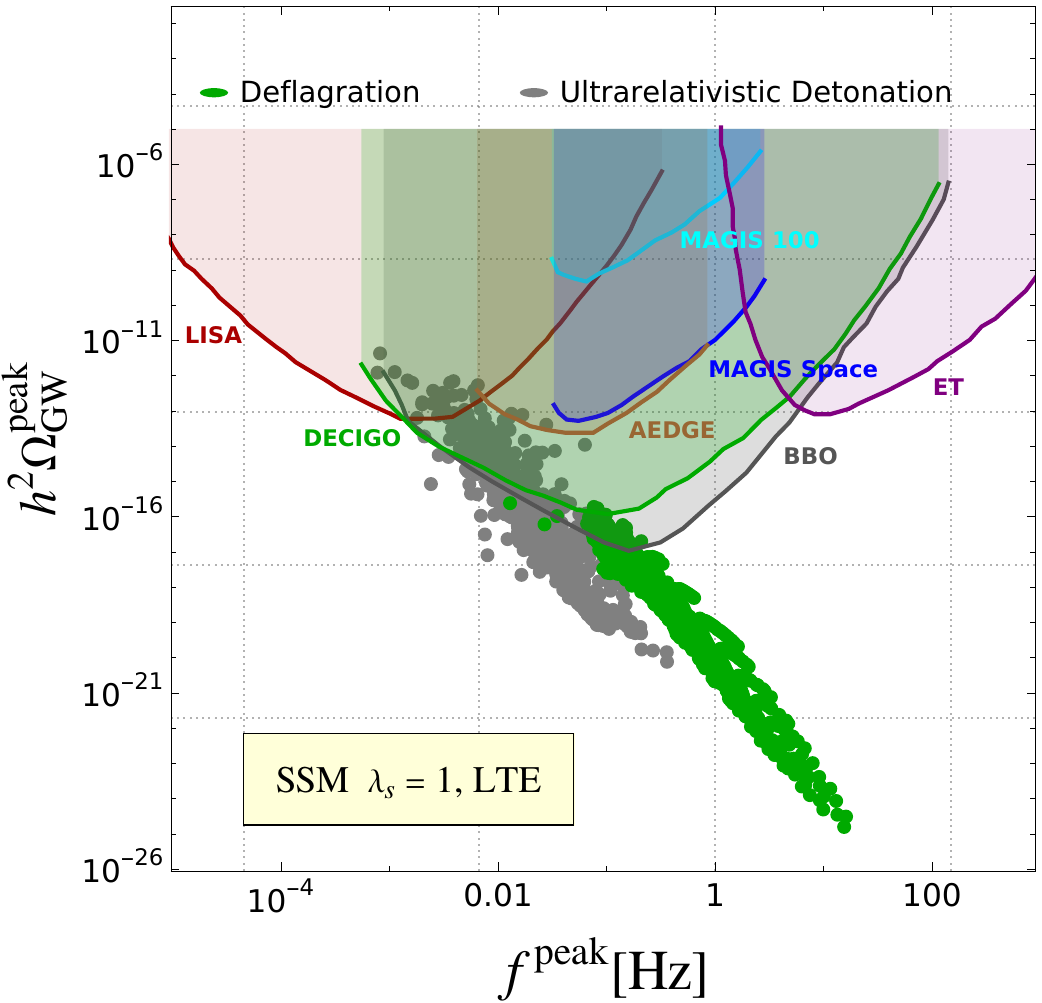}\,\,
\includegraphics[width=0.49\linewidth]{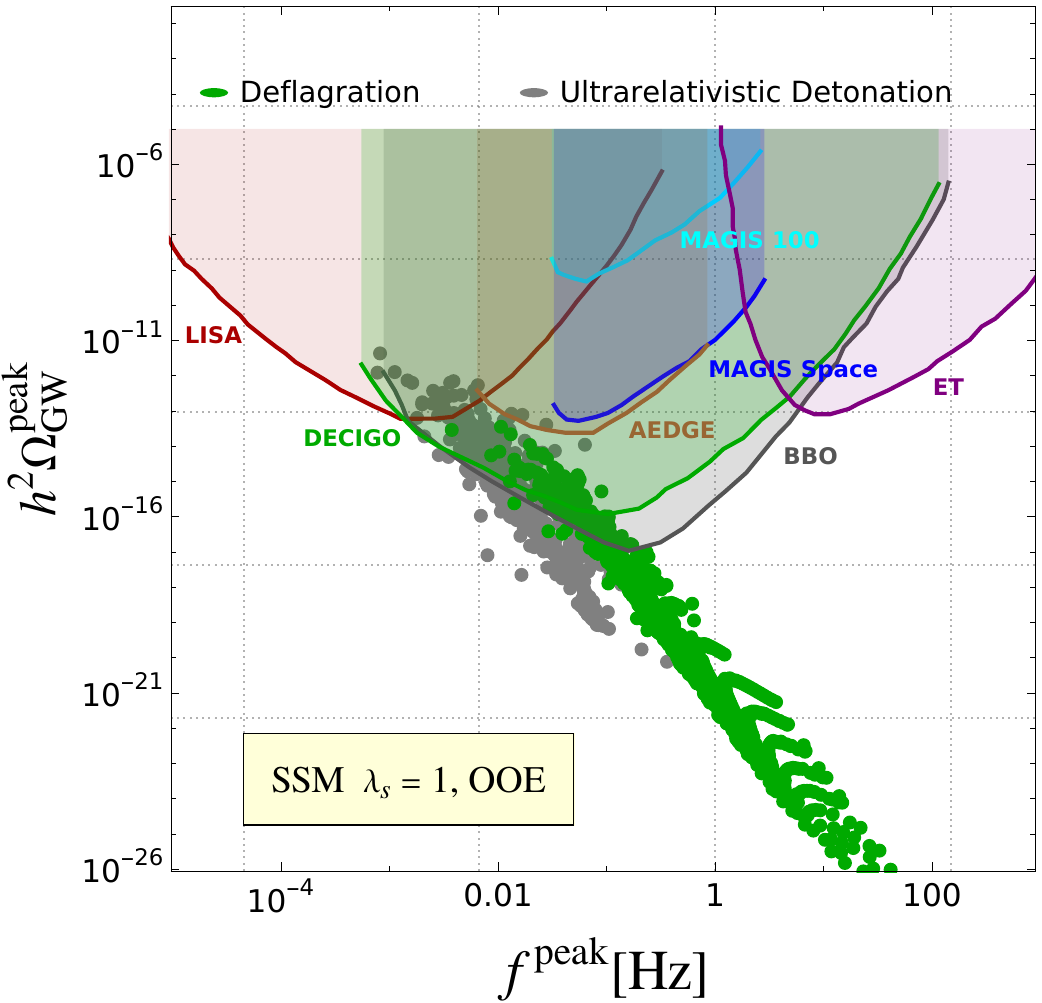}
\label{fig:GWspectrum}
\caption{Scatter plot of the GW peak frequency $f^{\rm peak}$ versus the peak energy density $h^2\Omega_{\rm GW}^{\rm peak}$ for the SSM with $\lambda_s=1$ within the LTE LTE approximation (\textit{left panel}) and including OOE contributions (\textit{right panel}). The coloured curves represent the sensitivity of future detectors.}
\label{fig: GWspectrum}
\end{figure}

A stochastic background of gravitational waves (GW) is produced during a first-order phase transition, when vacuum bubbles appear and expand. Their dynamics leaves an imprint on the GW energy spectrum $h^2 \Omega_{\rm GW}$, and the final GW signal results from the combination of three main channels
\begin{equation}
    h^2 \Omega_{\text{GW}} \simeq h^2 \Omega_{\text{sw}} + h^2 \Omega_{\text{col}} + h^2 \Omega_{\text{turb}}.
\end{equation}
The three terms above stand for contributions arising from acoustic waves travelling through the plasma after collisions between bubbles, those directly generated by such collisions, and those generated by turbulent fluid motion, respectively. The resulting GW signal is dominated by sound wave contributions \cite{Hindmarsh:2013xza,Hindmarsh:2015qta}, and the spectrum at the peak is given by
\begin{equation}
    h^2 \Omega_{\text{sw}}= 2.65 \times 10^{-6} \left( \frac{k_{sw} \alpha_n}{1+ \alpha_n}\right)^2 \left( \frac{H}{\beta}\right) \left( \frac{100}{3}\right)^{1/3} (H \tau_{sw}) \ v_w,
\end{equation}
where $\alpha_n$ and $\beta$ are the transition strength and the inverse transition duration, $H$ is the Hubble parameter, and $\tau_{sw}$ is the sound wave formation timescale, with $H \tau_{sw}$ 
\begin{equation}
    H \tau_{sw}= \text{Min}\left( 1, (8 \pi)^{1/3} \left( \frac{\text{Max}(v_w, c_s)}{\beta/H}\right) \left( \frac{4}{3}\frac{1+ \alpha_n}{k_{sw} \alpha_n}\right)^{1/2}\right).
\end{equation}
For the definition of the fraction of released energy $k_{sw}$, we refer to \cite{Espinosa:2010hh}. The expression for the wave peak frequency $f_{sw}$ is
\begin{equation}
    f_{sw}=1.9 \times 10^{-5} \ \text{Hz}\left( \frac{1}{v_w}\right) \left( \frac{\beta}{H} \right) \left( \frac{g_*}{100} \right)^{1/6} \left( \frac{T_n}{100 \ \text{GeV}}\right), 
\end{equation}
with $g_*$ the effective number of relativistic degrees of freedom. 
In Fig.\,\ref{fig: GWspectrum} we show the GW signals obtained for the sampled points in the SSM parameter space. In particular, for each sampled point, we display the peak value of the GW power spectrum $h^2 \Omega^{\text{peak}}_{\rm GW}$ versus the peak frequency $f^{\text{peak}}$. We additionally present the power-law integrated curves corresponding to future GW observatories: LISA \cite{LISA:2017pwj,Baker:2019nia}, DECIGO \cite{Yagi:2011wg,Kawamura:2020pcg}, AEDGE \cite{AEDGE:2019nxb,Badurina:2021rgt}, MAGIS 100, MAGIS Space \cite{Graham:2016plp,Graham:2017pmn}, BBO \cite{Crowder:2005nr,Corbin:2005ny}, and ET \cite{Punturo:2010zz,Hild:2010id}. In the figure, green points correspond to solutions in the deflagration regime, while grey points are for ultrarelativistic detonations. On the left, the calculation of the bubble velocity is limited to the LTE regime. In contrast, in the right-hand panel, the calculation includes the OOE contributions. When OOE effects are taken into account, deflagration points tend to be shifted to the right and downward, indicating an increase of $f^\textrm{peak}$ and a reduction in the predicted GW amplitude. This reduction stems from the decrease of the bubble wall velocity induced by non-equilibrium dynamics.

To assess the detectability of the predicted stochastic GW background, we computed the signal to noise ratio (SNR) for the LISA and BBO interferometers, which is defined as 
\begin{equation}
    \text{SNR} = \sqrt{t_{\rm obs} \int_{f_{\text{min}}}^{f_{\text{max}}} df \left[ \frac{h^2 \Omega_{\rm GW}(f)}{h^2 \Omega_{\text{Sens}}(f)} \right]^2},
\end{equation}
where $h^2 \Omega_{\text{GW}}(f)$ is the predicted GW spectrum, $h^2 \Omega_{\text{Sens}}(f)$ denotes the nominal sensitivity curves of the detectors, and $t_{\rm obs}$ is the effective observation time (3 years for LISA, 1 year for BBO). The SNR provides a measure of how the GW signal is compared to the instrumental noise, with a higher value indicating a greater probability of detection. 
The frequency range over which the integration is performed spans from $f_{\text{min}}= 10^{-6} \ \text{Hz}$ to $f_{\text{max}}= 10 \ \text{Hz}$ for LISA, and  $f_{\text{min}}= 10^{-3} \ \text{Hz}$ to $f_{\text{max}}= 100 \ \text{Hz}$ for BBO. 
The sensitive curve is related to the spectral density $S_h(f)$ via
\begin{equation}
    \Omega_{\text{Sens}}(f)= \frac{2 \pi^2}{3 H_0^2}f^3 S_h (f), 
\end{equation}
with the present-day Hubble parameter taken as $H_0 = 2.19 \times 10^{-18} \ \text{s}^{-1}$. For each detector, $S_h(f)$ can be found in \cite{Babak:2021mhe} for LISA and \cite{Yagi:2011wg} for BBO. 

\begin{figure}
    \centering
\includegraphics[width=0.48\linewidth]{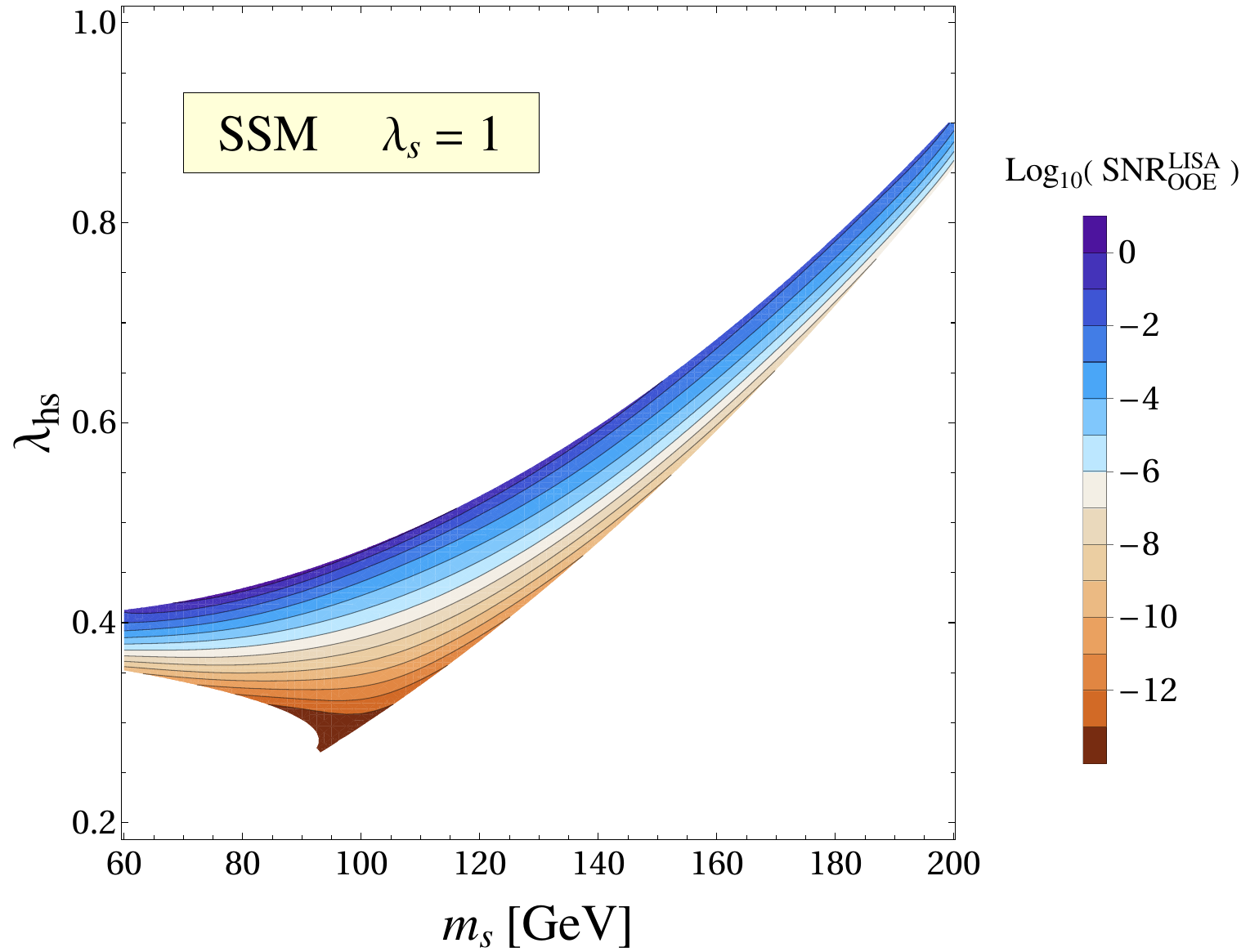}\,\,
\includegraphics[width=0.48\linewidth]{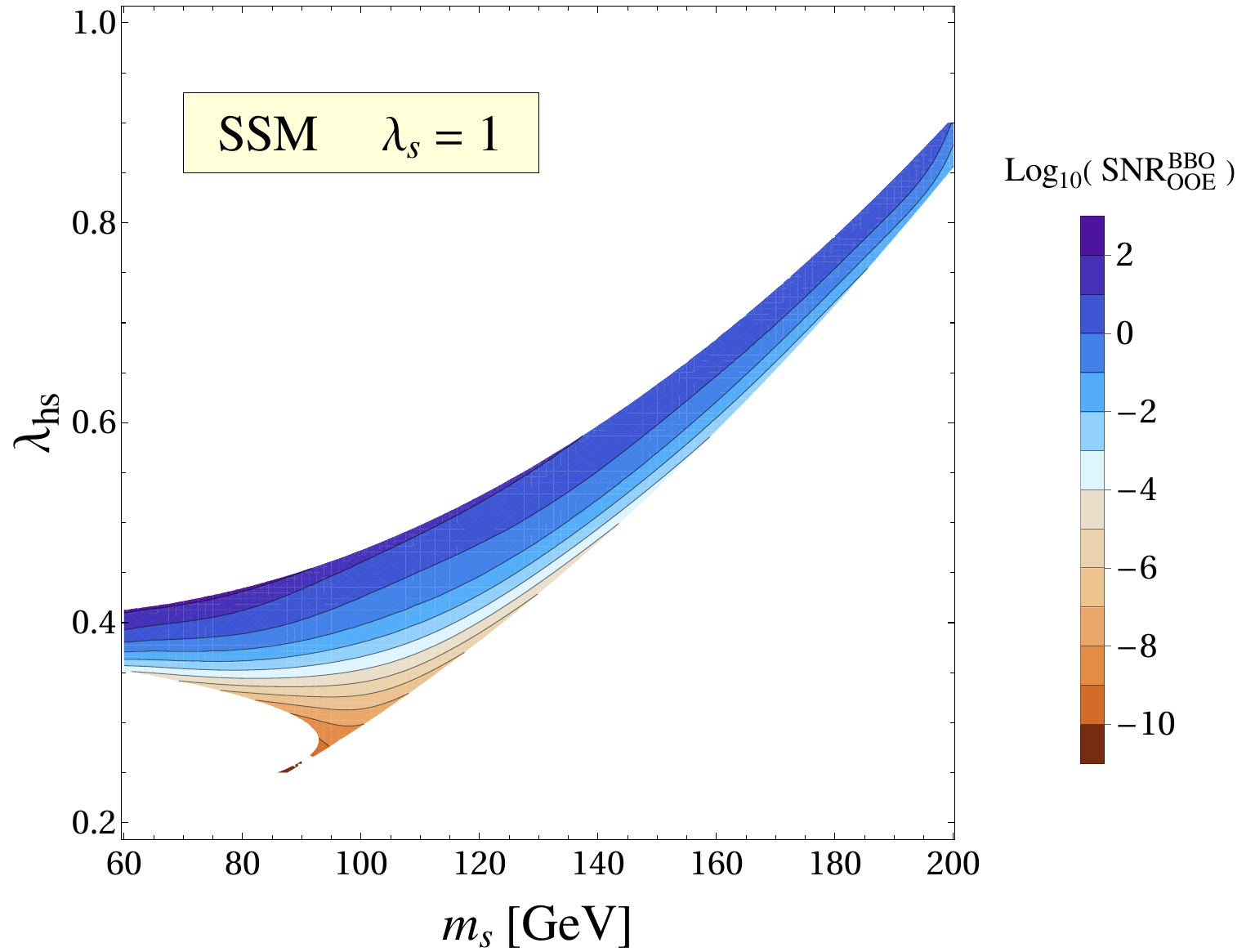}\,\,
\caption{Contour plots of the logarithm of the SNR for the SSM with $\lambda_s =1$. The plots show the dependence of the SNR on the singlet scalar mass $m_s$ and the portal coupling $\lambda_{hs}$. The left panel corresponds to the LISA experiment, while the right panel corresponds to the BBO experiment.}
\label{fig: SNRcontour}
\end{figure}

\begin{figure}
    \centering
\includegraphics[width=0.4\linewidth]{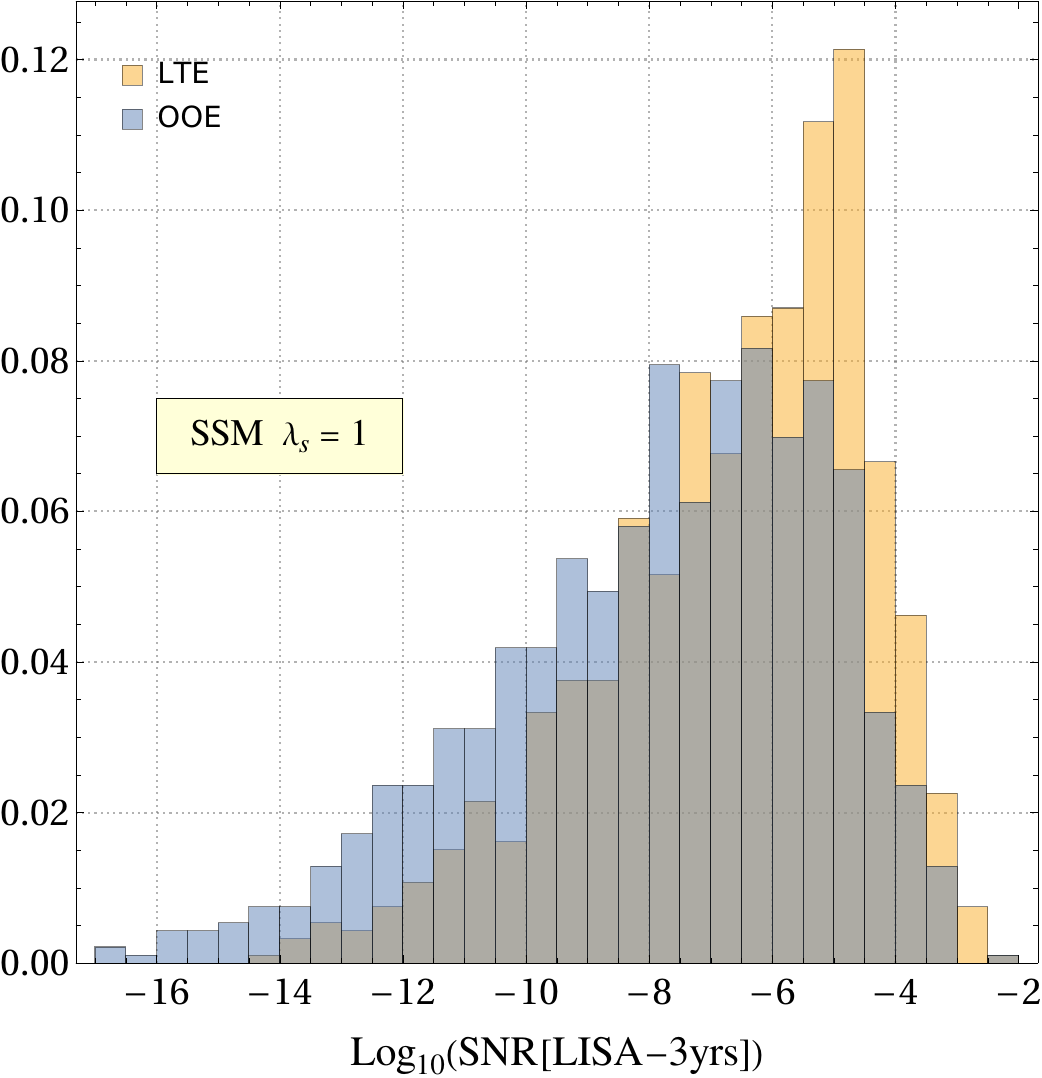}\,\,\,\,\,\,\,\,\,\,\,
\includegraphics[width=0.4\linewidth]{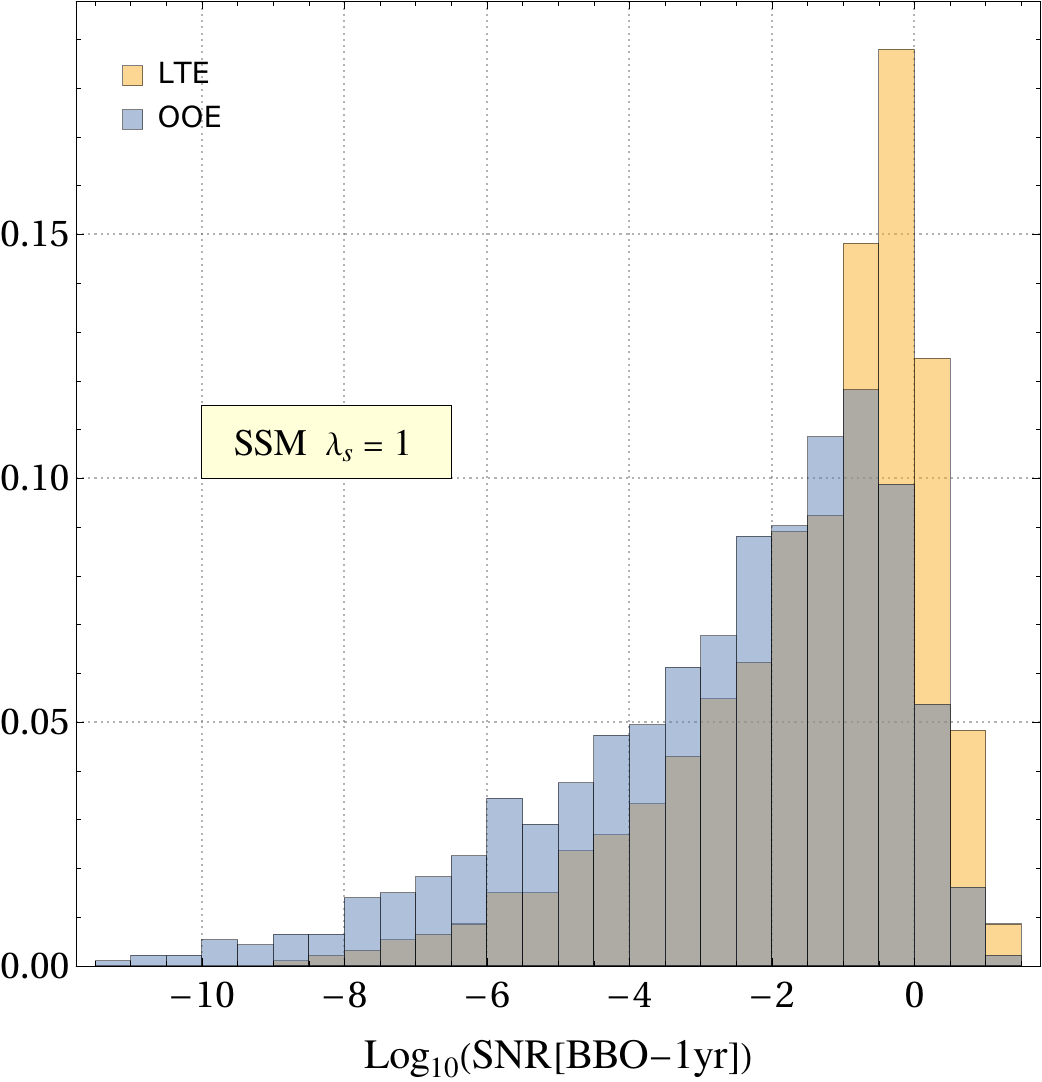}

\caption{Probability distribution of $\log_{10}\text{SNR}$ for the SSM with $\lambda_s = 1$, comparing the LTE and OOE cases for points with deflagration solutions in LTE. The left panel is for the LISA detector, assuming 3 years of observation, while the right panel shows the corresponding results for BBO with a one-year observation time.}
\label{fig: SNR_histo}
\end{figure}

In Fig.\,\ref{fig: SNRcontour} we show the results for the SNR relative to LISA (left panel) and to BBO (right panel) in the model parameter space.
For the LISA configuration, the predicted signal remains very weak in most of the parameter space except for a tiny strip. For BBO (right panel), the sensitivity is improved. In the same parameter range, the predicted SNR increases by approximately two orders of magnitude, up to $\text{SNR} \sim 10-100$.

The histograms in 
Fig.~\ref{fig: SNR_histo} show the distribution of $\log_{10} \text{SNR}$, restricted to the points having a deflagration solution in LTE, for LISA (left panel) and BBO (right panel), with a comparison between the LTE and OOE estimates. For the sake of this comparison, we are not showing in the histograms ultrarelativistic detonations and those OOE deflagrations that originate from detonations in LTE, as discussed in Section \ref{sec: numerical results}.
For LISA both the LTE and OOE scenarios predict low SNR, with LTE systematically yielding slightly higher values than OOE. This reflects the impact of non-equilibrium effects, which tend to reduce the efficiency of GW production during the phase transition. 
Compared to LISA, the $\log_{10} \text{SNR}$ distribution for BBO is shifted to higher values. Again, LTE predictions are larger than the OOE one. Comparing the results obtained with the two detectors, the SNR for BBO is a few orders of magnitude larger than that for LISA across the parameter space.

\subsection{Electroweak baryogenesis}
\label{sec: baryo}

A first-order electroweak phase transition offers a unique possibility to generate the baryon asymmetry of the universe (BAU) with very few ingredients. The mechanism of electroweak baryogenesis requires CP-violating interactions between matter and the wall. These are typically activated by the non-trivial profiles of the scalars, and the CP-asymmetries are then converted into a matter asymmetry by sphaleron processes.  

In the simplest SSM scenario\footnote{As shown in \cite{McDonald:1995hp,Espinosa:2011eu}, a small explicit breaking of $Z_2$ is also needed to bias the population of one of the two $Z_2$-broken minima which arise in the two-step process after the first step. This ensures that the net baryon asymmetry is not even out across different patches of the Universe. The necessary explicit breaking can be safely taken sufficiently small to avoid EDM bounds, see for instance \cite{DeCurtis:2019rxl}.}, these CP-violating interactions can be described by an effective dimension-five operator $s \,\bar Q_L H t_R$ with a complex coefficient, where $s$ is a (BSM) scalar, $H$ the Higgs doublet, $Q_L$ the left-handed top quark doublet and $t_R$ the right-handed top quark field. For instance, such an operator was shown to emerge in models with composite Higgs from their non-linear dynamics\,\cite{DeCurtis:2019rxl,Espinosa:2011eu}. The dimension-five operator generates the quadratic term
\begin{equation}
\mathcal L \supset \frac{y_t}{\sqrt 2} h \bar t_L\left(1+ic_5\frac{s}{\Lambda}\right)t_R + \, \text{h.c.}, 
\label{eq: dim5 operator}
\end{equation}
for the top, where $c_5$ is the Wilson coefficient and $\Lambda$ a scale. The field-dependent mass term for the top encoded in the equation above can be trivially  rewritten in terms of the Dirac fields $t$ and $\bar t$ as $\bar t \left(\widetilde m_t e^{i\theta\gamma_5}\right)t$, with  \begin{equation}
   \widetilde m_t = \frac{y_t h}{\sqrt 2} \sqrt{1+c_5^2\frac{s^2}{\Lambda^2}}\,,  \qquad \theta = \arctan \left(c_5\frac{s}{\Lambda}\right). 
   \label{eq: mass EWBG}
\end{equation}
CP-violation is activated when $s$ has a non-trivial spatial profile, that in turn induces a space-dependent phase $\theta(z)$ that cannot be reabsorbed by a field redefinition of the top field. 

We have verified that the addition of the dimension-five operator to the SSM has a negligible impact on the wall dynamics. In the following we will then safely use the results of the previous section to describe EWBG within the ``augmented SSM" (that is the SSM $+$ the dimension-five operator). For the calculation of the BAU, we closely follow the method of \cite{Cline:2000nw,Fromme:2006wx,Cline:2020jre}, that we briefly summarise below, and report more at length in Appendix \ref{sec: appendix baryo}.

Solving the equations of motion by means of a WKB ansatz, it is readily seen that the presence of a space-dependent mass term of the form $\widetilde m(z) e^{i\theta(z)\gamma_5}$ modifies the usual dispersion relation, with the canonical momentum $p_c$ not coinciding with the physical momentum $p$. This introduces a series of subtleties in the determination of the group velocity $v_g$ and the force $F$ that enter in the Boltzmann equation
\begin{equation}
\left(v_g \partial_z +F \partial_{p_z}\right)f=-\mathcal C[f]. 
\label{eq: Boltzmann for baryo text}
\end{equation}
At first order in a gradient expansion, these are eventually found to be 
\begin{align}
    v_g&=\frac{p_z}{\omega} \\
    F&=-\frac{(\widetilde m^2)'}{2 \omega}+\bar s s_{cp}\frac{(\widetilde m^2 \theta ')'}{2 \omega \omega_{z}},
    \label{eq: CP-violating force text}
\end{align}
where the $'$ denotes derivation with respect to $z$ and $\omega$ is the conserved wall-frame energy
\begin{equation}
\omega = E_0\left(1-\frac{\bar s s_{cp}}{2}\frac{\widetilde m^2\theta'}{E_0^2 E_{0,z}}\right),
\end{equation}
with $E_0=\sqrt{p_{\parallel}^2+p_z^2+\widetilde m^2}$, $E_{0,z}=\sqrt{p_z^2+\widetilde m^2}$, and $\omega_z=\sqrt{\omega^2-p_{\parallel}^2}$. The symbol $\bar s$ is the spin, while $s_{cp}$ is introduced to account for the difference in the equations of motion of particles and antiparticles, with $s_{cp}=1$ for particles and $s_{cp}=-1$ for antiparticles.

As $\omega$ is the conserved energy, in \eqref{eq: CP-violating force text} the first term is the analogous of the usual CP-conserving force that arises from the $z$-dependence of $\widetilde m$. Focusing only on this term, an increase in $\widetilde m$ leads to a decrease of the momentum ($\dot p=F$): as a particle gets heavier, it decelerates. The second term is CP-breaking and is non-vanishing when the gradient of the phase is so. Its sign depends on the particle/antiparticle nature of the state, so that if a particle is accelerated, the corresponding antiparticle is decelerated. In a nutshell, this is the seed for the generation of the asymmetry.

Expressing $v_g$ and $F$ in terms of physical momentum variables, we have (at first order in the gradient expansion)
\begin{align}
    v_g&=\frac{p_z}{E_0}\left(1+\frac{\bar s s_{cp}}{2}\frac{\widetilde m^2\theta'}{E_0^2 E_{0,z}}\right) \\
    F&=-\frac{(\widetilde m^2)'}{2 E_0} +\frac{\bar s s_{cp}}{2}\frac{\left(\widetilde m^2\theta'\right)'}{E_0 E_{0,z}}-\frac{\bar s s_{cp}}{4}\frac{\widetilde m^2 (\widetilde m^2)'\theta'}{E_0^3E_{0,z}}.
\end{align}
Inserting these expressions in \eqref{eq: Boltzmann for baryo text}, we extract from it a CP-even and a CP-odd equation. The first one corresponds to the one presented in Section \ref{sec: set-up}, that we solve through the iterative method described there (in practice, we take the results of Section \ref{sec: numerical results}). For the CP-odd equation, we limit ourselves to solve it through the moment expansion method by taking a two-moment truncation (see \cite{Cline:2000nw,Fromme:2006wx,Cline:2020jre} and Appendix \ref{sec: appendix baryo} for details), as typically done in the literature. In fact, our main goal in this work is to assess the impact of out-of-equilibrium contributions to the wall dynamics on the BAU. 

\begin{figure}
    \centering
\includegraphics[width=0.49\linewidth]{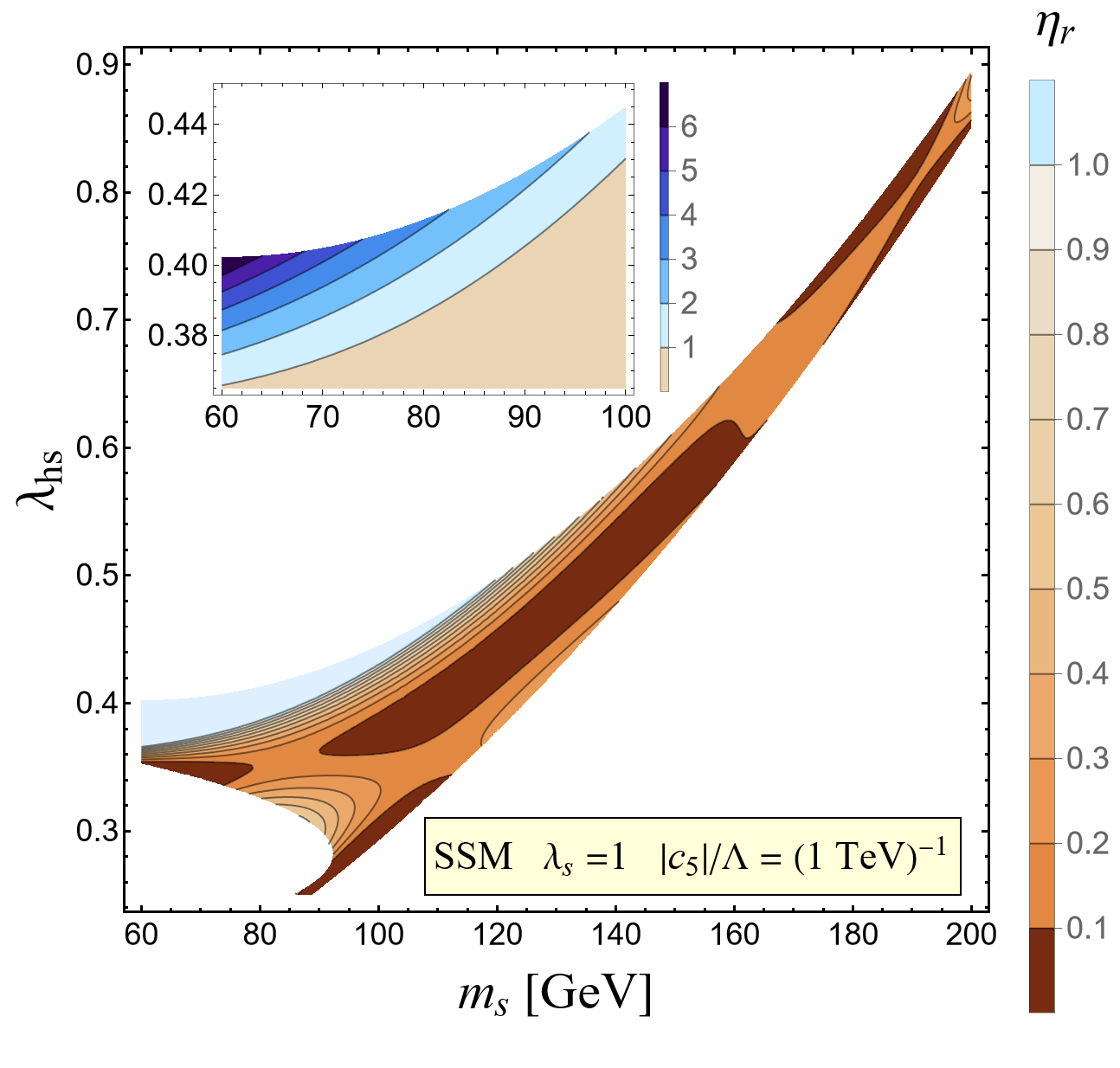}\,\,
\includegraphics[width=0.49\linewidth]{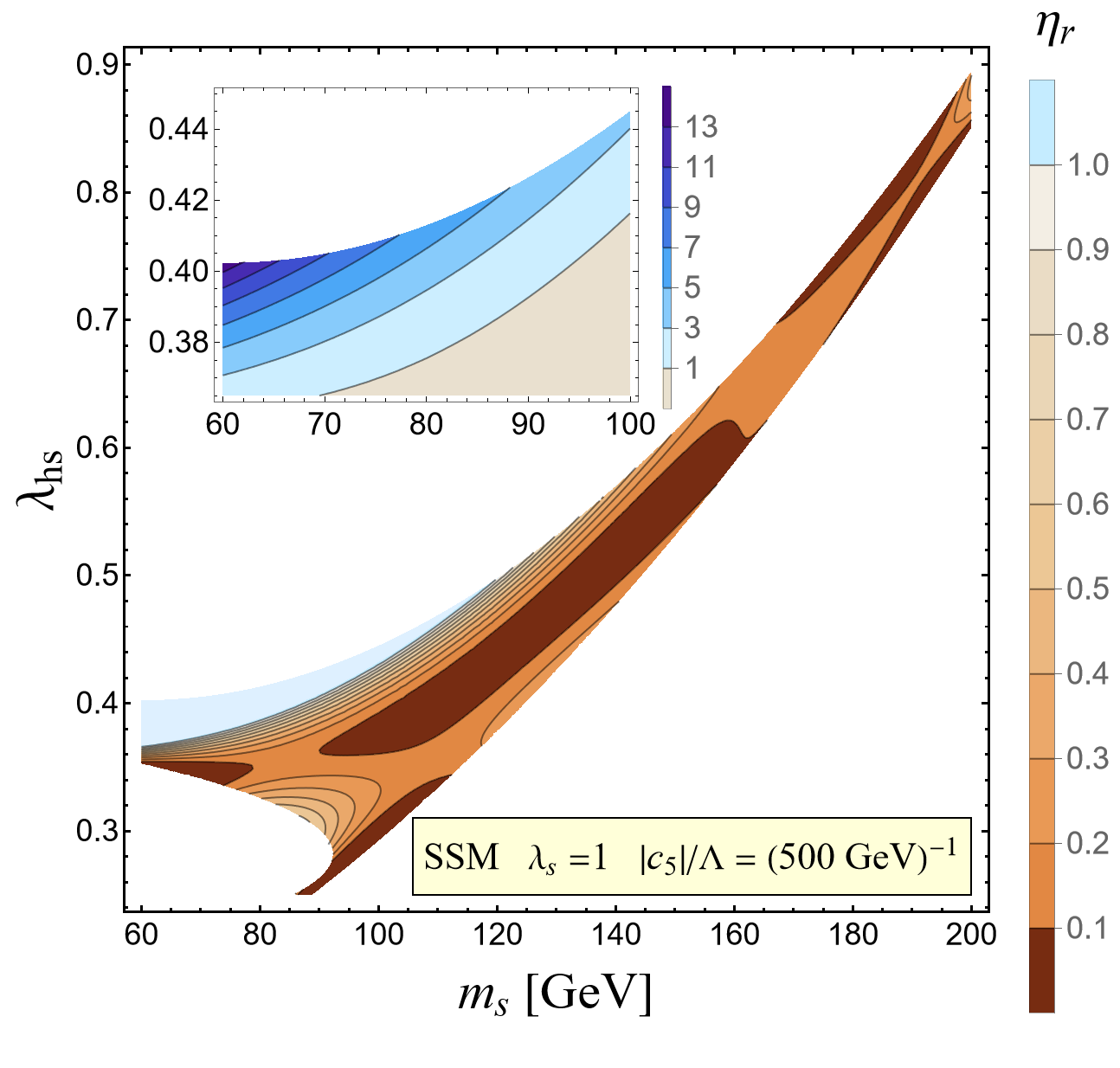}
\caption{Contour plot of the normalised BAU $\eta_r$ in the parameter space for $|c_5|/\Lambda=(1$ TeV$)^{-1}$ (\textit{left panel}) and $|c_5|/\Lambda=(500$ GeV$)^{-1}$ (\textit{right panel}) with out-of-equilibrium contributions from the top quark. In the main figure, the light blue region corresponds to $\eta_r> 1$. In both panels, the inside plot provides a zoom inside the light blue region of the corresponding figure, where $\eta_r$ varies rapidly.}
\label{fig: baryo param space}
\end{figure}

The results for the generated baryon asymmetry $\eta_B$ (see Appendix \ref{sec: appendix baryo}) normalised to the observed baryon asymmetry $\eta_{\rm obs} \sim 8.7\times 10^{-11}$ are shown in Fig.\,\ref{fig: baryo param space} for $|c_5|/\Lambda = (1\,{\rm TeV})^{-1}$ (left panel) and $|c_5|/\Lambda = (500\,{\rm GeV})^{-1}$ (right panel). Focusing for definiteness on $|c_5|/\Lambda = (1\,{\rm TeV})^{-1}$ first, we see that in a large part of the parameter space the desired value $\eta_r\equiv \eta_B/\eta_{\rm obs}=1$ is hardly reproduced, and most of the points have $\eta_r\lesssim 0.2$. However, the asymmetry rapidly grows in the upper left corner of the region we analysed, and values even larger than $1$ are obtained therein. We insert a zoom of this region inside the plot. The same qualitative features are found for $|c_5|/\Lambda = (500\,{\rm GeV})^{-1}$, with higher values of $\eta_r$ across the parameter space. This is in agreement with our results in \cite{Branchina:2025jou}, where we found the BAU to be decreasing with $|c_5|/\Lambda$ within the LTE approximation. We will further comment on the dependence of $\eta_r$ on $|c_5|/\Lambda$ below. 

\begin{figure}
    \centering
\includegraphics[width=0.49\linewidth]{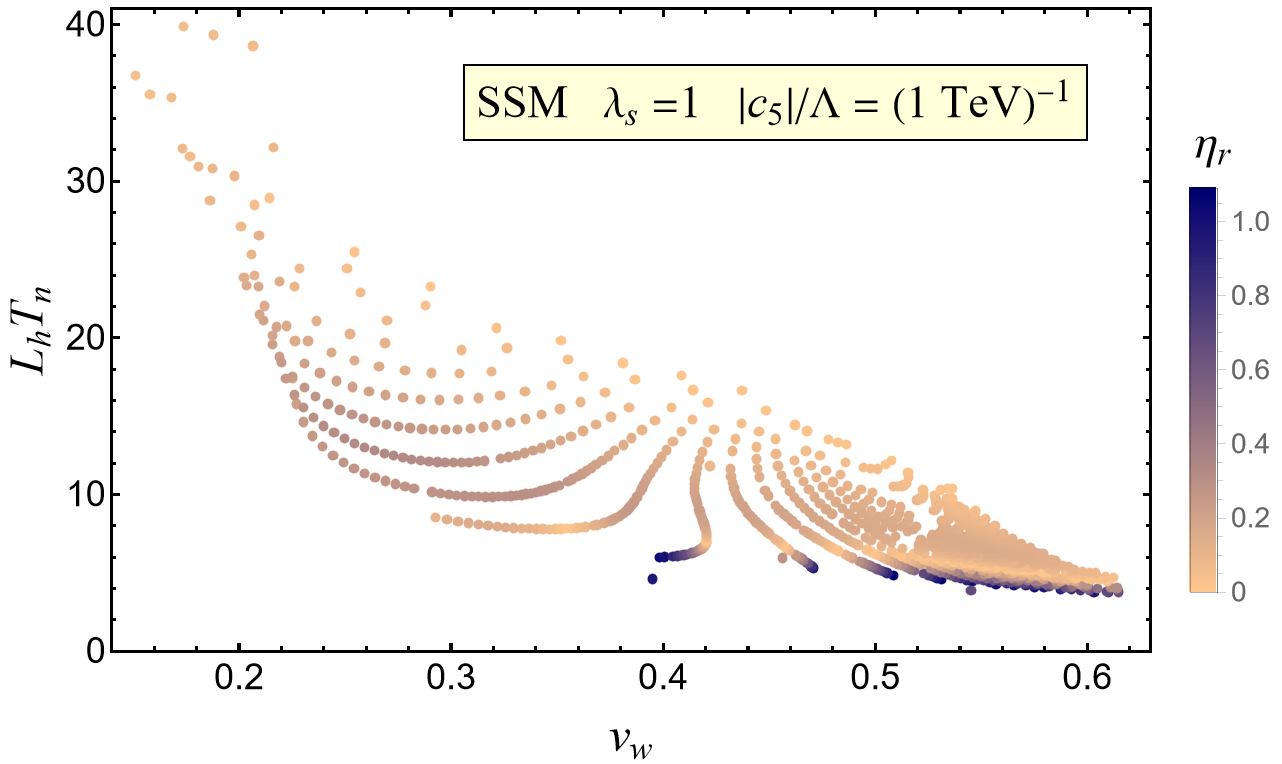}\,\,
\includegraphics[width=0.49\linewidth]{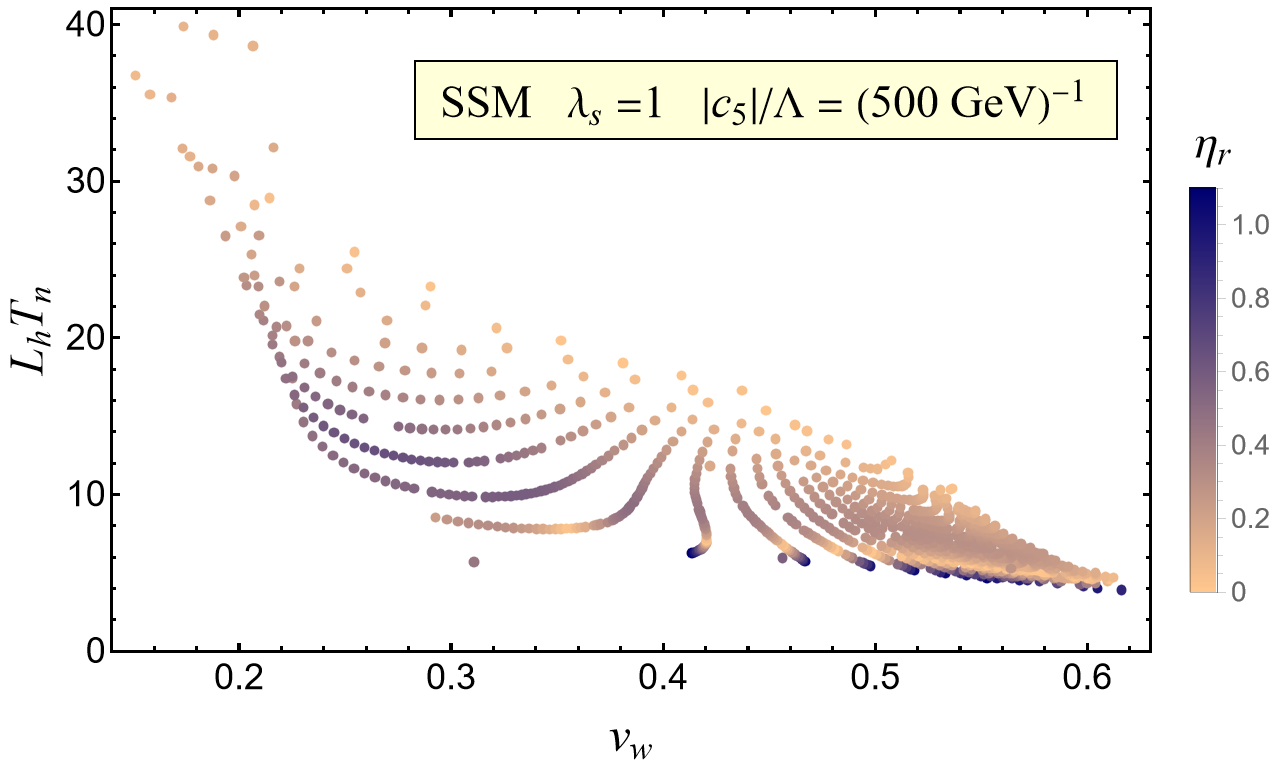}
\includegraphics[width=0.49\linewidth]{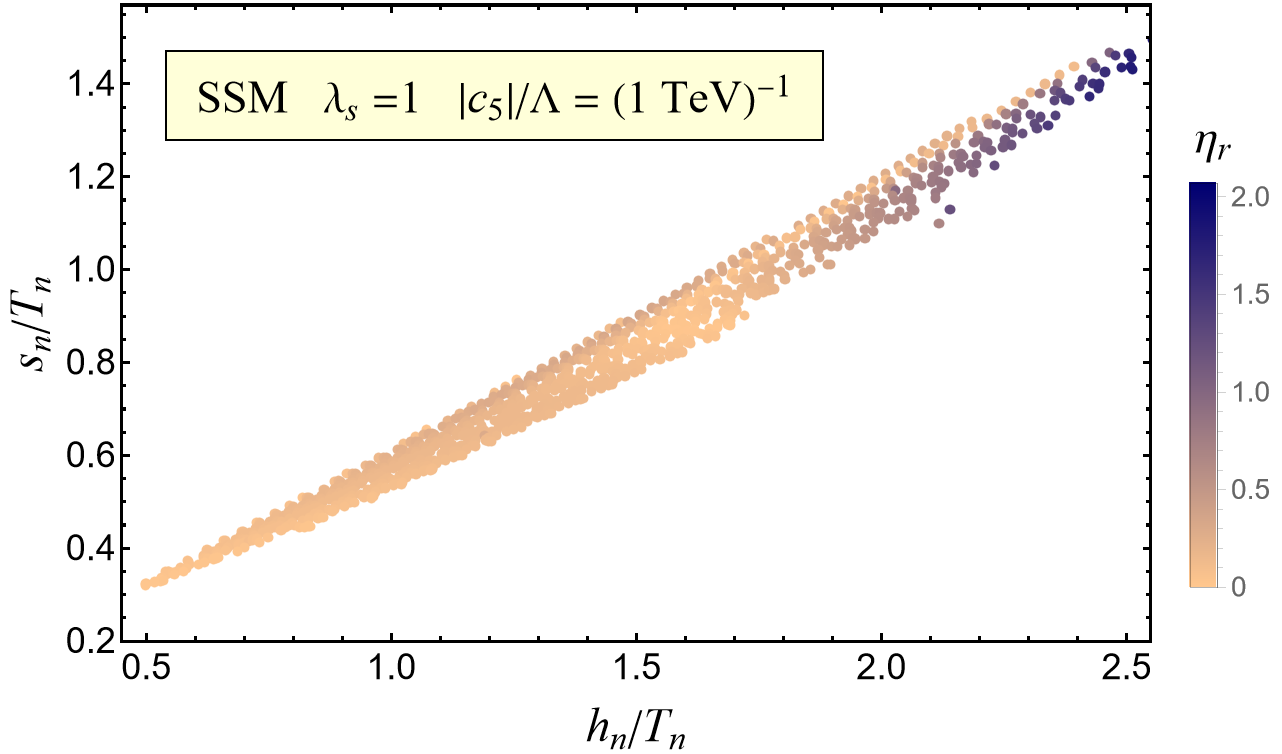}\,\,
\includegraphics[width=0.49\linewidth]{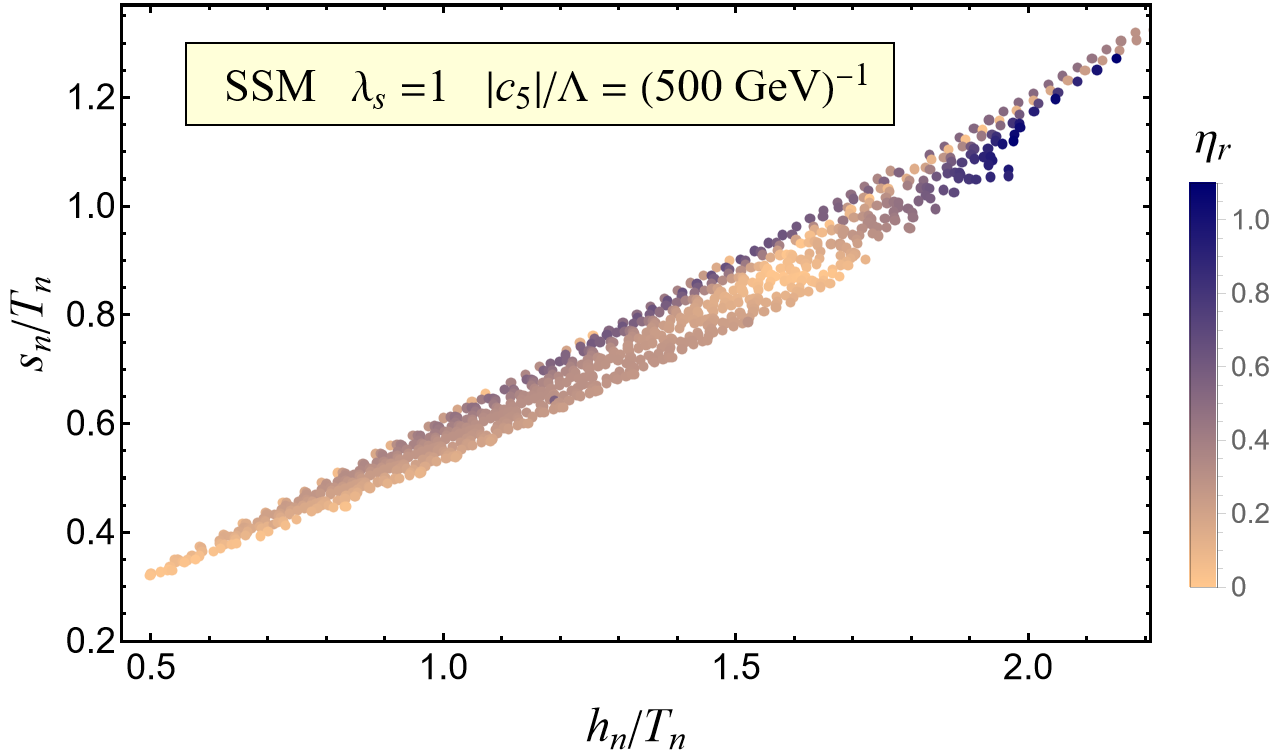}
    \caption{\textit{Upper row}. Scatter plot of the normalised BAU $\eta_r$ in terms of $v_w$ and $L_hT_n$ for $|c_5|/\Lambda = (1\,{\rm TeV})^{-1}$ (\textit{left panel}) and $|c_5|/\Lambda = (500\,{\rm GeV})^{-1}$ (\textit{right panel}) up to $\eta_r=1$. The colour gradient shows the variation of $\eta_r$. 
    \textit{Lower row}. Scatter plot of the normalised BAU $\eta_r$ in terms of $h_n/T_n$ and $s_n/T_n$ for $|c_5|/\Lambda = (1\,{\rm TeV})^{-1}$ (\textit{left panel}) and $|c_5|/\Lambda = (500\,{\rm GeV})^{-1}$ (\textit{right panel}) up to $\eta_r=1$.} 
\label{fig: BAU_vw_Lh}
\end{figure}

A comparison between Fig.\,\ref{fig: baryo param space} and Fig.\,\ref{fig:vw} shows that the region with the highest obtainable $\eta_r$ corresponds to the region where the wall velocity reaches its largest values. In the literature it is typically observed that the BAU decreases with $v_w$ for $v_w\sim \mathcal O\left(10^{-1}\right)$ (see for instance Fig.\,3 of \cite{Cline:2020jre}), but this result is obtained by fixing all the other parameters relevant to baryogenesis. Our result is a manifestation of the fact that $\eta_r$ has a complicate dependence on the wall velocity, the profiles (especially the widths), as well as the equilibrium parameters of the transition, and the overall variation of these is such that the generated asymmetry is maximised in that specific corner of the parameter space. 

This feature can be better appreciated in Fig.\,\ref{fig: BAU_vw_Lh}, where we show the results for $\eta_r$ (up to $\eta_r=1$ for clarity) in terms of $v_w$ and $L_hT_n$ in the upper row, and in terms of $h_n/T_n$ and $s_n/T_n$ in the lower one, for both $|c_5|/\Lambda = (1\,{\rm TeV})^{-1}$ (left panel) and $|c_5|/\Lambda = (500\,{\rm GeV})^{-1}$ (right panel). The peculiar pattern that emerges for the colour gradient, i.\,e.\,for the value of $\eta_r$, reflects the non-trivial dependence of $\eta_r$ discussed above. 
In agreement with \cite{Cline:2020jre} we see that, given a point with $\eta_r\simeq 1$, if one moves to either larger wall velocities at fixed $L_h T_n$, or larger widths at fixed $v_w$, the baryon asymmetry decreases. Configurations with intermediate values of $\eta_r$ (approximately $\eta_r\simeq 0.3$ for $|c_5|/\Lambda=(1 $ TeV$)^{-1}$ and $\eta_r\simeq 0.6$ for $|c_5|/\Lambda=(500 $ GeV$)^{-1}$) are found for the lowest velocities $0.2\lesssim v_w\lesssim 0.3$ when the width is not too large. Overall, these results suggest that the dependence of  $\eta_r$ on $L_h\,T_n$ is as relevant as that on $v_w$, and having a low enough ratio between the wall width and the average particle mean free path ($\sim T_n^{-1}$) is crucial to achieve successful baryogenesis. Concerning the dependence on the nucleation VEV-to-temperature ratios, the figure shows that increasing $h_n/T_n$ and $s_n/T_n$ enhances the baryon asymmetry. This is due to the fact that larger values of the fields inside or outside the wall produce steeper gradients for $\widetilde m^2$ and $\theta$, and thus a stronger source for the CP-odd perturbations (see Eq.\,\ref{eq: source_CPodd}).

\begin{figure}   
\centering
\includegraphics[width=0.49\linewidth]{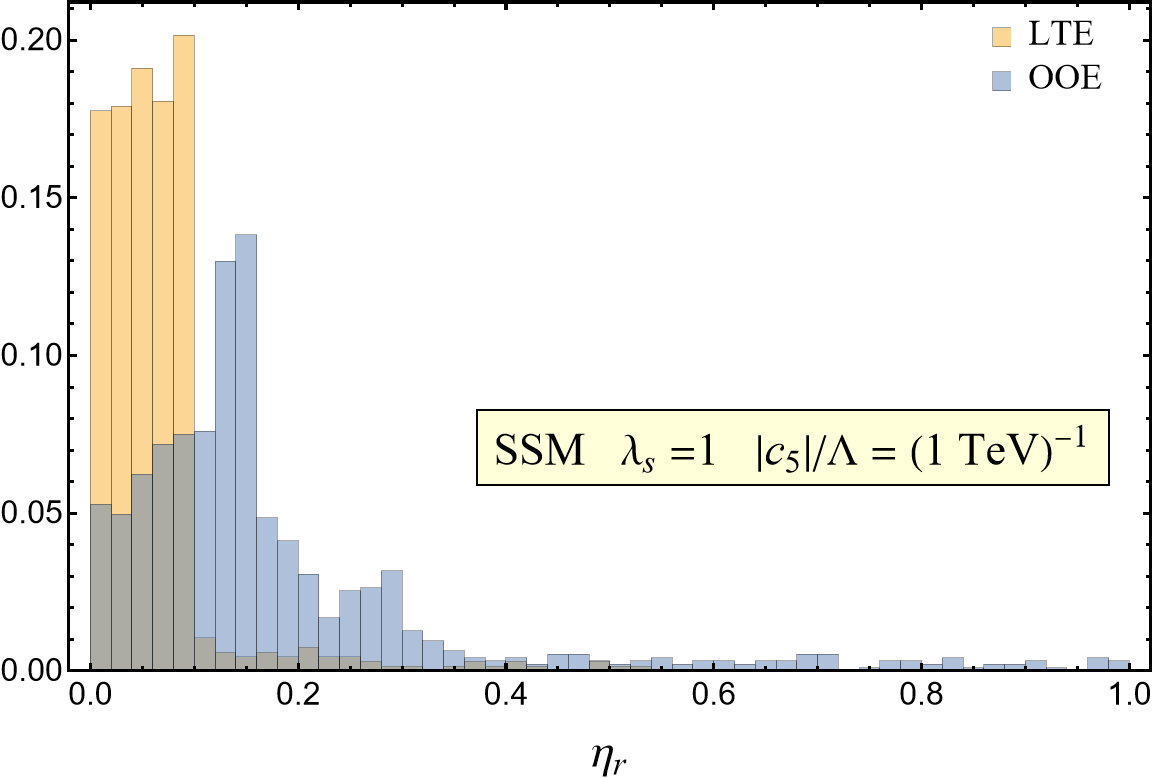}\,\,
\includegraphics[width=0.49\linewidth]{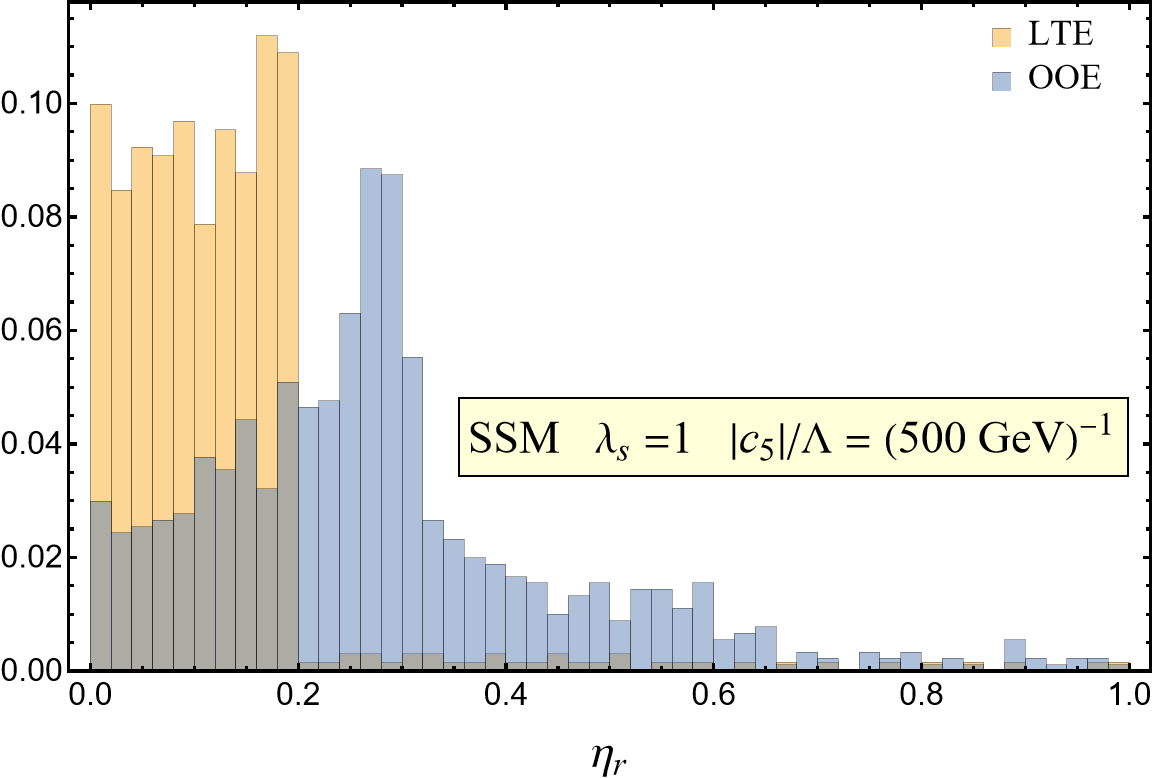}
    \caption{{\it Left panel}. Histogram of the normalised BAU $\eta_r$ calculated with the parameters obtained within the LTE approximation (yellow bins) and with the inclusion of the out-of-equilibrium effects (blue bins) for $|c_5|/\Lambda=(1$ TeV)$^{-1}$. {\it Right panel}. Same histogram as in the left panel, with the choice $|c_5|/\Lambda=(500$ GeV)$^{-1}$.}
    \label{fig: baryo histo 1}
\end{figure}

\begin{figure}
    \centering
\includegraphics[width=0.48\linewidth]{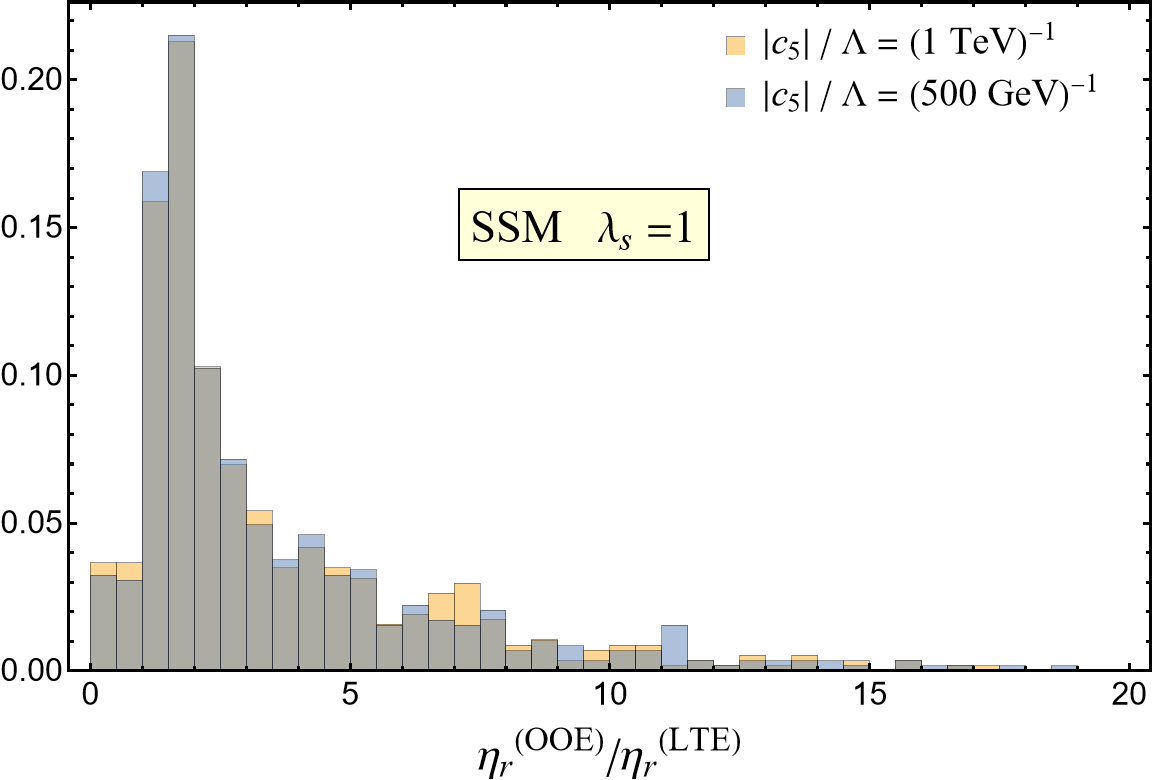}\,\,
\includegraphics[width=0.499\linewidth]{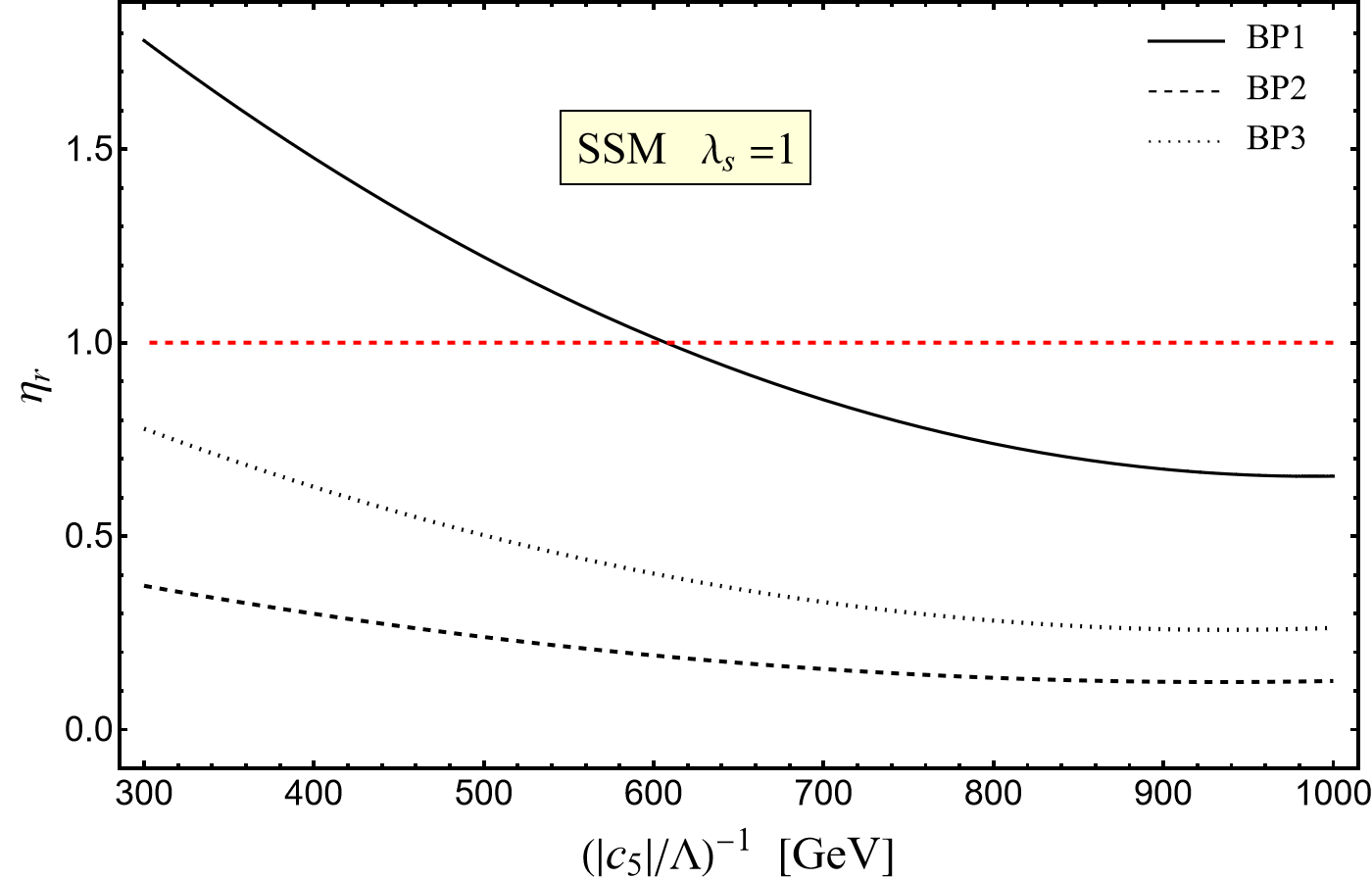}
    \caption{{\it Left panel}. Histogram of the ratio between the normalised BAU $\eta_r$ generated within the LTE approximation and the one obtained including out-of-equilibrium contributions with $|c_5|/\Lambda=(1$ TeV$^{-1})$ (yellow bins) and $|c_5|/\Lambda=(500$ GeV$^{-1})$ (blue bins).  {\it Right panel}. Plot of the relative BAU $\eta_r$ versus $|c_5|/\Lambda$. The full, dashed and dotted lines refer to three different benchmark points, BP1 with $m_s= 74$ GeV and $\lambda_{hs}=0.37$, BP2 with $m_s= 106$ GeV and $\lambda_{hs}=0.42$, and BP3 with $m_s= 131$ GeV and $\lambda_{hs}=0.52$. The red dashed line is for successful baryogenesis, $\eta_r=1$.}
 \label{fig: baryo histo 2}
\end{figure}

In Fig.\,\ref{fig: baryo histo 1} and in the left panel of Fig.\,\ref{fig: baryo histo 2} we present histograms where we compare the results for the BAU obtained within the LTE approximation and when OOE perturbations are included. The first of the two figures clearly shows that, across the parameter space, the inclusion of out-of-equilibrium effects tends to make the BAU larger. This is true for both the choices $|c_5|/\Lambda = (1$ TeV$)^{-1}$ (left panel) and $|c_5|/\Lambda = (500$ GeV$)^{-1}$ (right panel). As already observed above, a comparison between the two also reveals that the asymmetry increases with $|c_5|/\Lambda$, in agreement with the LTE results presented in \cite{Branchina:2025jou}. This is further displayed in the right panel of Fig.\,\ref{fig: baryo histo 2}, where we plot the evolution of the BAU with the effective scale of the dimension-five operator. From the right panel of Fig.\,\ref{fig: baryo histo 1}, one can see that values of $\eta_r\gtrsim 0.5$ are not so rare within the augmented SSM for $|c_5|/\Lambda = (500$ GeV)$^{-1}$ when out-of-equilibrium contributions from the top are taken into account. Ref.\,\cite{DeCurtis:2024hvh} has also shown that gauge bosons have an impact on the wall dynamics that is comparable to that of top quark. Combined with the results we have found, this suggests that the achievement of $\eta_r\simeq1$ within the augmented SSM is much more feasible than the LTE analysis indicates. 

The impact of out-of-equilibrium contributions can be appreciated even better in the left panel of Fig.\,\ref{fig: baryo histo 2}, where we show a histogram of the ratio  between the BAU with $\delta f$ from the top included ($\eta_r^{(\rm OOE)}$) and the BAU in LTE ($\eta_r^{(\rm LTE)}$). Needless to say, the plot is restricted to the points that were found to have a steady-state solution in LTE. The results are fairly similar for both the choices of $|c_5|/\Lambda$ considered above. 
It is readily seen that the out-of-equilibrium asymmetry is $4-10$ times larger than the LTE one for a non-negligible number of sampled points. More extreme scenarios  are also found in some rare cases. 

\begin{figure}
    \centering
\includegraphics[width=0.49\linewidth]{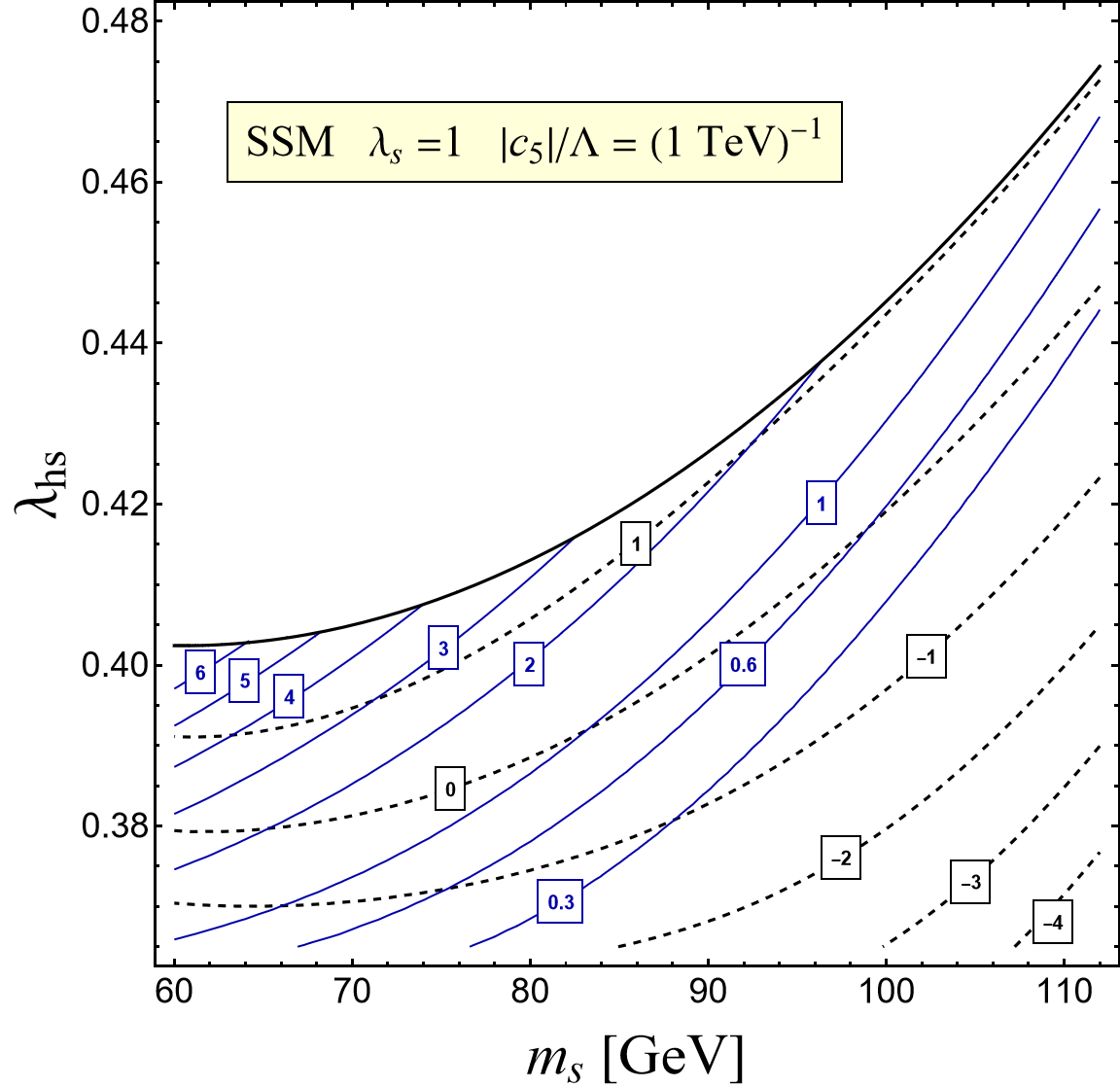}\,\,
\includegraphics[width=0.49\linewidth]{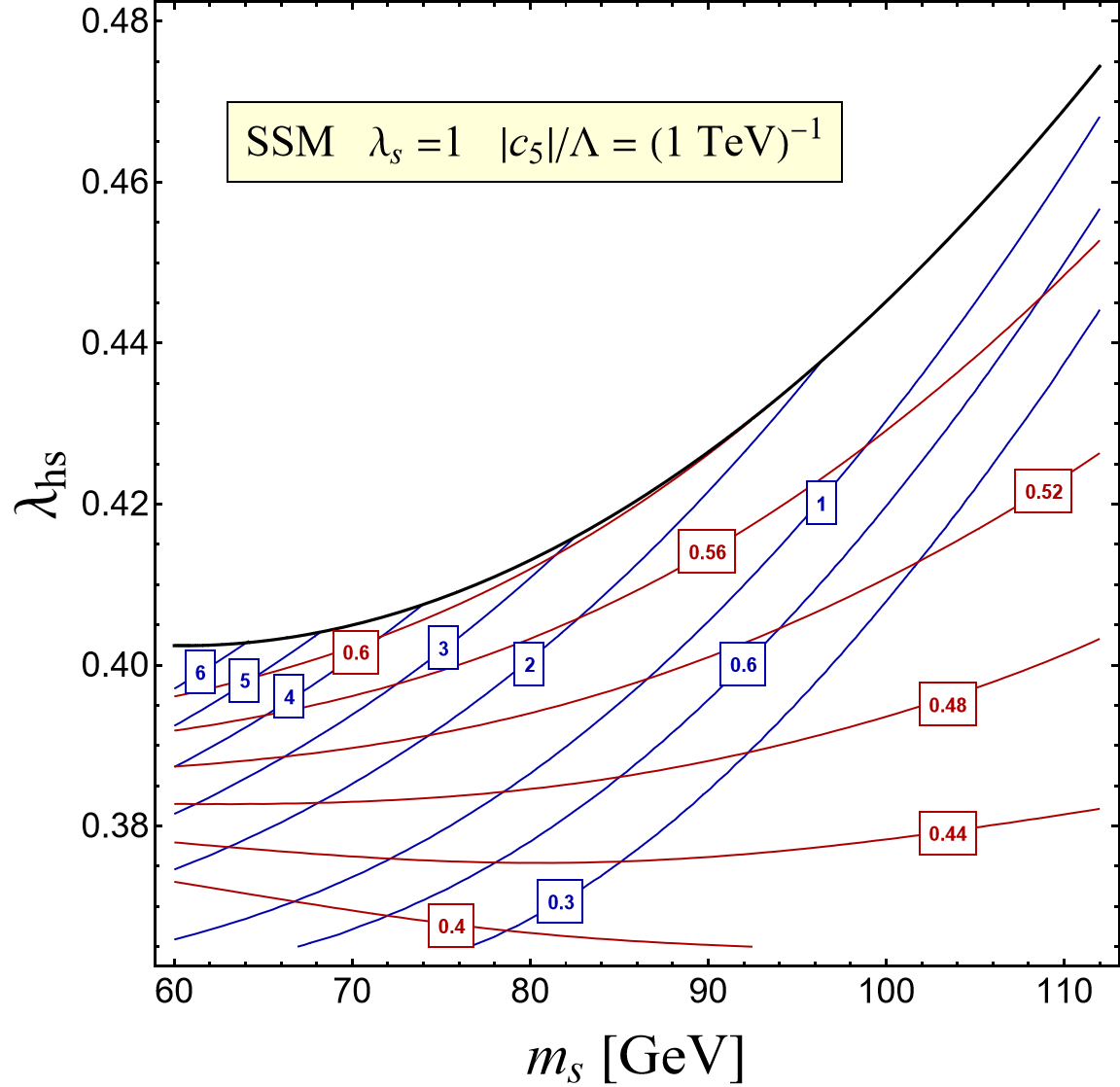}
    \caption{\textit {Left panel}. Contour plot of the relative BAU $\eta_r$ (blue solid lines) and of the BBO $\log_{10} \textrm{SNR}$ (black dashed lines) in the region of parameter space around $\eta_r=1$ for $|c_5|/\Lambda =(1$ TeV$)^{-1}$. \textit {Right panel}. Contour plot of the relative BAU $\eta_r$ (blue lines) and of the wall velocity $v_w$  (red lines) in the same region of parameter space for $|c_5|/\Lambda =(1$ TeV$)^{-1}$.}
    \label{fig: BAU and SNR}
\end{figure}

Combining the BAU results from this section with the GW signal-to-noise ratio obtained in the previous one, Fig.\,\ref{fig: BAU and SNR} (left panel) presents a contour plot of $\eta_r$ (blue solid lines) and of the BBO $\log_{10}\textrm{SNR}$ (black dashed lines) in the region of parameter space around $\eta_r = 1$ for the conservative choice $|c_5|/\Lambda =(1$ TeV$)^{-1}$. The figure shows that both the baryon asymmetry and the gravitational wave signal reach their maximum values in the same corner of the $m_s$–$\lambda_{hs}$ plane. Our results therefore indicate that successful baryogenesis and a potentially observable GW spectrum can be achieved simultaneously in a region where the phase transition is strong and the bubble wall velocity is larger than $v_w \sim 0.4$, as shown in Fig.\,\ref{fig: BAU and SNR} (right panel)
where the red curves represent the isolines of the velocity.

This outcome appears somewhat at odds with the expectation commonly expressed in the literature that an inherent tension between observable gravitational waves and successful baryogenesis exists. As discussed above, our results do confirm the common expectation that GW signals grow with $v_w$, and that, for fixed values of the other parameters, the BAU decreases with increasing $v_w$. However, owing to the complicate dependence of $\eta_r$ on all such parameters, we find that $\eta_r$ reaches its largest values for fast, thin walls. This behaviour allows to open up the possibility to simultaneously produce the baryon asymmetry of the Universe and an observable GW spectrum. As a general trend, our analysis suggests that scenarios featuring a small singlet mass and a large portal coupling are favoured in regard to the generation of cosmological relics of the phase transition.

\section{Conclusions}
\label{sec- conclusions}

In this work, we presented a detailed analysis of the dynamics of bubble walls during a first-order electroweak phase transition. We took as a benchmark the singlet extension of the Standard Model, that allows to embed the EWPhT in a two-step scenario. We performed a numerical investigation across the parameter space of the model, adopting the optimised algorithm previously introduced in \cite{DeCurtis:2022hlx,DeCurtis:2023hil,DeCurtis:2024hvh} to iteratively solve the Boltzmann equation for the  distribution functions of the species together with the scalar and hydrodynamic equations. This allows us to extract key properties of the phase transition dynamics, such as the wall velocity $v_w$, the plasma profiles $T(z)$ and $v_p(z)$, as well as the scalar field profiles, that we parametrise through three parameters: the widths $L_h$, $L_s$ and the displacement $\delta_s$. For each sampled point in the model parameter space, the results found within the local thermal equilibrium approximation in \cite{Branchina:2025jou} were used as starting point of the iterative procedure. Points for which a LTE deflagration solution was not found (i.e. ultrarelativistic detonations in LTE) were processed using the results obtained on one of their nearest neighbours as initial input for the iterative procedure. 

By means of a comparison between the results found in LTE and those obtained including OOE contributions, our analysis has shown that out-of-equilibrium effects have an important impact on the wall dynamics across the parameter space at both qualitative (enlargement of the region where a steady state expansion is realised) and quantitative level.  In turn, this determines also a large impact on the theoretical estimate of the corresponding gravitational wave signal and of the generated baryon asymmetry. As the out-of-equilibrium perturbations enter into the determination of the dynamics through their coupling to the scalar fields, in this work we only included OOE contributions from the most strongly coupled species, i.e.\,the top quark. 

Within this framework, the correction to the LTE determination of the parameters was found to be up to $\sim -65\%$ for $v_w$ and to $\sim 40\%$ for $L_h$, the two parameters to which cosmological relics are the most sensitive. The distribution of the wall velocity was shown to be considerably wider in comparison to the one determined in LTE. 
Large differences ($\gtrsim 100\%$) in the predicted baryon asymmetry were found, with the distribution of the ratio between the OOE and LTE results, $\eta_r^{\rm (OOE)}/\eta_r^{\rm (LTE)}$, peaked around $\sim 2$, with a persistent tail up to $\sim 10$. The determination of the signal-to-noise ratio of gravitational waves has shown that OOE contributions reduce the probability of observation with respect to the LTE estimate. In particular, our analysis finds it unlikely that any GW signal arising from deflagration solutions with the SSM as underlying particle physics model can be detected by the LISA interferometer. On the other hand, the BBO detector was found to have a sufficiently large signal-to-noise ratio in a region of parameter space with strong phase transitions.
In such region, we found that successful baryogenesis can also be achieved with bubble wall velocity larger than $v_w \sim 0.4$. This can be ascribed to the fact that the BAU not solely depends on $v_w$, but instead has a non-trivial dependence on the other parameters too.

As shown in \cite{DeCurtis:2024hvh}, out-of-equilibrium contributions from other species, in particular from the $W$ gauge bosons, can provide additional corrections to the phase transition parameters that could be comparable in size to those induced by the top quark. The inclusion of such perturbations leads to a further decrease in $v_w$. As such, the values found for $v_w$ provide upper bounds to the full wall velocity that are considerably closer to the complete $v_w$ than those obtained in LTE.
In turn, we expect this to determine a further enhancement of the generated baryon asymmetry, while simultaneously leading to a larger reduction of the GW signal strength. Combining this observation with the results found in this work for $\eta_r$ and for the ratio $\eta_r^{\rm (OOE)}/\eta_r^{\rm (LTE)}$, it becomes readily apparent that successful baryogenesis within the augmented SSM is considerably easier to achieve than the LTE results alone would suggest. In this respect, it seems reasonable to expect that when OOE perturbations are fully included, viable baryogenesis is attained in a non-negligible region of the model parameter space. Concerning the GW signals, whether those produced by strong phase transitions would remain observable at the BBO interferometer is a question that should be addressed when all relevant OOE contributions to the dynamics are included. 

This work represents a further step toward a quantitative understanding of bubble dynamics, enabling a more realistic phenomenological analysis of the impact of first-order phase transitions on cosmological relics such as GW signals and the baryon asymmetry.
It is also particularly encouraging that, even within the simplest extension of the SM, there are regions of parameter space that can simultaneously yield potentially detectable GW signatures and a viable mechanism for generating the matter–antimatter asymmetry. While accounting for the latter is necessary to address one of the open problems of the SM, the former offers an intriguing observational opportunity for future GW interferometers.

 \acknowledgments

We thank Giulio Barni and Peach Conaci for useful discussions. The work has been funded by the European Union – Next Generation EU
through the research grant number P2022Z4P4B “SOPHYA - Sustainable Optimised PHYsics
Algorithms: fundamental physics to build an advanced society” under the program PRIN 2022
PNRR of the Italian Ministero dell’Universit\`{a} e Ricerca (MUR) and by the research grant number 20227S3M3B “Bubble Dynamics in Cosmological Phase Transitions” under the program PRIN 2022 of the Italian Ministero dell’Universit\`{a} e Ricerca (MUR). The work has been also partially supported by
ICSC – Centro Nazionale di Ricerca in High Performance Computing, Big Data and Quantum
Computing.

\appendix
\section{Transport equations for electroweak baryogenesis}
\label{sec: appendix baryo}

In this appendix, we go through the derivation of the equations  used in Section \ref{sec: baryo} to calculate the BAU. We closely follow references \cite{Cline:2000nw,Fromme:2006wx,Cline:2020jre}.

\subsection{Motion of a fermion with complex mass}

To calculate the baryon asymmetry generated by the phase transition, we shall begin by describing the motion of a fermion with complex mass. In particular, we need to determine suitable expressions for the speed $v_g$ and the force $F$ felt by a particle that enter in the Boltzmann equation \eqref{eq: Boltzmann for baryo text}, that we report here for completeness,  
\begin{equation}
    \left(v_g \partial_z +F \partial_{p_z}\right)f=-\mathcal C[f]. 
    \label{eq: Boltzmann baryo}
\end{equation}
The $z$-dependent chemical potentials for the various species will be extracted from their distribution functions $f$, and the particle asymmetries calculated from them. 

The equation of motion for a Dirac field $\psi$ with mass $m(z)=\widetilde m(z)e^{i\theta\gamma_5}$ can be written as
\begin{equation}
\left(i\slashed\partial -\overline m \,P_R-\overline m^*\,P_L\right)\psi=0, 
\end{equation}
with $\overline m=\widetilde m(z)e^{i\theta}$. Assuming a planar solution that only depends on $z$, and working in the frame where the momentum parallel to the wall vanishes $p_{\parallel}=0$, we find an approximate solution to the equation for particles with momentum $p\gg L_w^{-1}$ ($w=h,s$) using a WKB ansatz 
\begin{equation}
    \psi=e^{-i\omega t} \left(\begin{array}{c} \mathcal L_s  \\ \mathcal R_s \end{array} \right)\otimes \chi_s,
\end{equation}
where $\chi_s$ is the spin eigenstate $\sigma_3\chi_s=\bar s \chi_s$, with $\bar s$ the spin. The equations for $\mathcal L_s$ and $\mathcal R_s$ read
\begin{align}
    \left(\omega +i\bar s\partial_z\right)\frac{1}{\overline m}\left(\omega-i\bar s\partial_z\right)\mathcal L_s&=\overline m^*\mathcal L_s \\
    \left(\omega -i\bar s\partial_z\right)\frac{1}{\overline m^*}\left(\omega+i\bar s\partial_z\right)\mathcal R_s&=\overline m\, \mathcal R_s. 
\end{align}
We solve these equations in a gradient expansion using the ansatz (same for $\mathcal R_s$)
\begin{equation}
\mathcal L_s=w_s(z) e^{i\int^z dz'p_c(z')}\,.
\label{eq: Ls ansatz}
\end{equation}
Two equations are obtained for the real and imaginary part, 
\begin{align}
\label{eq: first eq Dirac}
    \omega^2-\widetilde m^2-p_c^2+(\bar s\omega+p_c)\theta'-\frac{\widetilde m'}{\widetilde m}\frac{w_s'}{w_s}+\frac{w_s''}{w_s}&=0 \\
    \label{eq: second eq Dirac}
    2p_c w_s'+w_s p_c'-\frac{\widetilde m'}{\widetilde m}(p_c+\bar s\omega)w-w'\theta'&=0.
\end{align}
Eq.\,\eqref{eq: second eq Dirac} can be used to determine $w_s$, but we will not consider it further. From \eqref{eq: first eq Dirac} one recovers, at zeroth order in gradients, the usual dispersion relation $\omega^2=p_c^2+\widetilde m^2$, while at first order 
\begin{equation}
    p_c^2+\widetilde m^2-\omega^2-\theta'(p_c+\bar s\omega)=0.
\end{equation}
For the antiparticle solution, $\theta\to-\theta$, and, to first order approximation, we can write in compact form for both particles and antiparticles
\begin{equation}
    p_c=p_0+\frac{1}{2}s_{cp}\frac{p_0+\bar s\omega}{p_0}\theta'+\alpha',
\end{equation}
with $s_{cp}=1$ for particles and $s_{cp}=-1$ for antiparticles, $p_0=\sqrt{\omega^2-\widetilde m^2}$ and $\alpha$ accounting for the gauge ambiguity in the definition of $p_c\, (\psi\to e^{i\alpha}\psi)$.  

Applying the same procedure to the equation for $\mathcal R_s$, one gets, at first order in gradients,
\begin{equation}
p_c^2+\widetilde m^2-\omega^2-\theta'(p_c-\bar s\omega)=0, 
\end{equation}
from which
$p_c= p_0+\frac{1}{2}s_{cp}\frac{-p_0+\bar s\omega}{p_0}\theta'+\alpha'$. The equations derived above can also be used to express $\omega$ in terms of the other parameters,
\begin{align}
\label{eq: omega Ls}
    \omega&=\sqrt{\left(p_c-\alpha'-s_{cp}\frac{\theta'}{2}\right)+\widetilde m^2}-\bar s s_{cp}\frac{\theta'}{2} \qquad\qquad\qquad  (\mathcal L_s) \\
\label{eq: omega Rs}
    \omega&=\sqrt{\left(p_c-\alpha'+s_{cp}\frac{\theta'}{2}\right)+\widetilde m^2}-\bar s s_{cp}\frac{\theta'}{2} \qquad\qquad\qquad  (\mathcal R_s)
\end{align}

Canonical equations of motion can be derived from the explicit expression of $\omega$ as $v_g=\partial_{p_c}\omega$ and $\dot p_c=-\partial_z \omega$ ($v_g$ is a group velocity, the time derivative of a ``collective coordinate" $q$, $v_g\equiv \dot q$).  At first order in derivatives for $\mathcal L_s$, one gets
\begin{equation}
    v_g=\frac{p_c-\alpha'-s_{cp}\theta'/2}{\omega +\bar s s_{cp}\theta'/2}\sim\frac{p_0}{\omega}\left(1+\frac{\bar s s_{cp}}{2}\frac{\widetilde m^2}{\omega p_0^2}\theta '\right),
\end{equation}
where the final expression is obtained writing $p_c$ in terms of $\omega$ through Eq.\,\eqref{eq: omega Ls}. Note that $v_g$ is independent of $\alpha$. The second Hamilton equation is easily obtained at leading order as 
\begin{equation}
    \dot p_c= v_g \left(\alpha'+s_{cp}\theta'\right)'-\frac12\,\frac{(\widetilde m^2)'}{\omega+\frac{\bar s s_{cp}}{2}\theta'}+\frac{\bar s s_{cp}}{2}\theta''.
\end{equation}
The calculation can be repeated for $\mathcal R_s$, and one finds that $v_g$ and $\dot p_c$ have a similar formal expression in terms of the corresponding $\omega$,
\begin{equation}
    v_g=\frac{p_c-\alpha'+s_{cp}\theta'/2}{\omega +\bar s s_{cp}\theta'/2}\sim\frac{p_0}{\omega}\left(1+\frac{\bar s s_{cp}}{2}\frac{\widetilde m^2}{\omega p_0^2}\theta '\right)
\end{equation}
and 
\begin{equation}
    \dot p_c= v_g \left(\alpha'-s_{cp}\theta'\right)'-\frac12\,\frac{(\widetilde m^2)'}{\omega+\frac{\bar s s_{cp}}{2}\theta'}+\frac{\bar s s_{cp}}{2}\theta''.
\end{equation}
One can use a compact notation for both $\mathcal L_s$ and $\mathcal R_s$ by defining the parameter $\alpha_{cp}\equiv \alpha'\pm \frac{s_{cp}}{2}\theta'$, with the $+$ sign referring to $\mathcal L_s$ and the minus sign to $\mathcal R_s$. 

For a particle with the usual dispersion relation, $E_0=\sqrt{p^2+m^2}$, $v_g=\partial_p E_0=p/E_0$, with  $p$  the physical kinetic momentum of the particle. In the same way, we define here the physical momentum of the particle of mass $\widetilde m e^{i\theta\gamma_5}$ as $p\equiv \omega v_g$. The force acting on the particle is thus individuated as $\dot p = \omega \dot v_g$ ($\omega$ is conserved). The expression of $\dot v_g$ is easily found using the chain rule $\dot v_g=v_g\partial_z v_g+\partial_{p_c}v_g\,\dot p_c$, with 
\begin{align}
\partial_z v_g &= -\frac12\frac{v_g}{\left(\omega+\frac{\bar  s s_{cp}}{2}\theta'\right)^2}(\widetilde m^2)' - \frac{\widetilde m^2}{\left(\omega+\frac{\bar s s_{cp}}{2}\theta'\right)^3}\alpha_{cp}'\\
\partial_{p_c}v_g&=\frac{\widetilde m^2}{\left(\omega+\frac{\bar s s_{cp}}{2}\theta'\right)^3},
\end{align}
and one gets $\dot p=-\omega (\widetilde m^2)'/2(\omega +\bar s s_{cp}\theta'/2)^2+\bar s s_{cp}\,\omega\,\widetilde m^2\theta''/2(\omega +\bar s s_{cp}\theta'/2)^3$. Finally, expanding to linear order in gradients, the expression
\begin{equation}
    \dot p=-\frac{(\widetilde m^2)'}{2 \omega }+\frac{\bar s s_{cp}}{2\omega^2}\left(\widetilde m^2\theta'\right)'
\end{equation}
is found. 

As a final step, we should now boost the results to a more general frame, where $p_x$ and $p_y$ are not necessarily vanishing. As shown in \cite{Fromme:2006wx}, the effect of the boost is to transform the dispersion relations \eqref{eq: omega Ls} and \eqref{eq: omega Rs} to 
\begin{equation}
    \omega = \sqrt{(p_{c,z}-\alpha_{cp})^2+p_\parallel^2+\widetilde m^2} - \bar s s_{cp}\frac{\theta'}{2}\frac{\sqrt{(p_{c,z}-\alpha_{cp})^2+\widetilde m^2}}{\sqrt{(p_{c,z}-\alpha_{cp})^2+p_\parallel^2+\widetilde m^2}},
\end{equation}
where $p_\parallel^2=p_x^2+p_y^2$ and $p_x=p_{c,x}$, $p_y=p_{c,y}$. Ref.\,\cite{Cline:2017qpe,Cline:2020jre} also pointed out that in the wall frame $\bar s$ must be replaced with $h s_p$, where $h$ is the helicity and $s_p\equiv p_z \,\omega /|\vec p|\omega_z $.
The $z$-component of the group velocity and of the physical kinetic momentum defined above are derived from the boosted dispersion relation as $v_{g,z}=\partial_{p_{c,z}}\omega$ and $p_z=\omega v_{g,z}$. Repeating calculations similar to those above, we find 
\begin{align}
   v_{g,z}&=\frac{p_{c,z}-\alpha_{cp}}{\omega_0}\left(1-\bar s s_{cp}\frac{\theta'}{2}\frac{\omega_0^2-\omega_{0,z}^2}{\omega_0^2\omega_{0,z}^2}\right),\\
   p_z&=\left(p_{c,z}-\alpha_{cp}\right)\left(1-\bar s s_{cp}\frac{\theta'}{2\omega_{0,z}}\right), 
\end{align}
where  $\bar s$ is understood to be boosted and we defined $\omega_0\equiv \sqrt{(p_{c,z}-\alpha_{cp})^2+p_\parallel^2+\widetilde m^2}$, and $\omega_{0,z}=\sqrt{(p_{c,z}-\alpha_{cp})^2+\widetilde m^2}$. 

At first order in gradients, the energy and the force $F=\dot p_z$ are then
\begin{align}
 \omega&=E_0\left(1-\bar s s_{cp}\frac{\widetilde m^2\theta'}{2E_0^2E_{0,z}}\right)\\
    F&=-\frac{(\widetilde m^2)'}{2 \omega}+ \bar s s_{cp}\frac{(\widetilde m^2 \theta ')'}{2 \omega \omega_{z}},
    \label{eq: CP-violating force}
\end{align}
where $E_0\equiv \sqrt{p_z^2+p_\parallel^2+\widetilde m^2}$, $E_{0,z}\equiv \sqrt{p_z^2+\widetilde m^2}$ and $\omega_z\equiv \omega^2-p_\parallel^2$. Equation \eqref{eq: CP-violating force} corresponds to \eqref{eq: CP-violating force text}.

Finally, expressing $v_g$ and $F$ in terms of physical momentum variables, we have
\begin{align}
\label{eq: vg final}
    v_g&=\frac{p_z}{E_0}\left(1+\frac{ \bar s s_{cp}}{2}\frac{\widetilde m^2\theta'}{E_0^2 E_{0,z}}\right) \\
    \label{eq: F final}
    F&=-\frac{(\widetilde m^2)'}{2 E_0} +\frac{\bar s s_{cp}}{2}\frac{\left(\widetilde m^2\theta'\right)'}{E_0 E_{0,z}}-\frac{\bar s s_{cp}}{4}\frac{\widetilde m^2 (\widetilde m^2)'\theta'}{E_0^3E_{0,z}}.
\end{align}

\subsection{Transport equations}

To calculate the baryon asymmetry generated by the transition, we now insert the expressions \eqref{eq: vg final} and \eqref{eq: F final} into the Boltzmann equation \eqref{eq: Boltzmann baryo} to get, in the wall rest frame,
{\small \begin{equation}
\label{eq: Boltzmann full}
    \left\{\left(\frac{p_z}{E_0}\partial_z -\frac{(\widetilde m^2)'}{2 E_0}\partial_{p_z}\right)+ \frac{h s_p s_{cp}}{2 E_0 E_{0,z}}\left(\frac{p_z}{E_0^2}\widetilde m^2 \theta'\partial_z+\left[\left(\widetilde m^2\theta'\right)'-\frac12\frac{\widetilde m^2(\widetilde m^2)'\theta'}{E_0^2}\right]\partial_{p_z}\right)\right\}f=-\mathcal C[f]. 
\end{equation}
}The first round bracket contains CP-even terms, while the second one, together with its prefactor, is for CP-odd terms. Given the hierarchy in $\theta$ gradients between the two, we can extract a CP-even and a CP-odd equation from \eqref{eq: Boltzmann full} and solve them separately.

Writing the collision integral as $\mathcal C=\mathcal C_e+s_{cp} \,\mathcal C_o$, where $\mathcal C_e$ and $\mathcal C_o$ stand for CP-even and CP-odd components, the CP-even equation is then obtained, to leading order, as ($f_e$ indicates the even part of $f$)
\begin{equation}
\left(\frac{p_z}{E_0}\partial_z -\frac{(\widetilde m^2)'}{2 E_0}\partial_{p_z}\right)f_e =-\mathcal C_e. 
\end{equation}The latter coincides with the Boltzmann equation presented in Section \ref{sec: set-up}, that we solve using the iterative method described there to determine the wall dynamics. As stressed in the main text, we have verified that for the augmented SSM the additional term in $m$ (with respect to the SSM) has a negligible impact on the wall dynamics so that, in practice, we take the results presented in Section \ref{sec: numerical results} for the CP-even Boltzmann equation. Note that this is different from what is typically done in the literature,
where the ansatz $f=f_0+\delta f_e+s_{cp}\,\delta f_o$, with
\begin{equation}
    f_0=\frac{1}{e^{\beta\left[\gamma_w(\omega-v_wp_z)-\mu(z)\right]}\pm 1}
\end{equation}
is taken. The Boltzmann equation is then extracted expanding for small $\delta f$, small perturbations $\delta T(z)$ around a fixed temperature $T=\beta^{-1}$, and around  $\mu=0$, $\omega-E_0=0$ ($\delta T$ is usually omitted for CP-odd perturbations as it does not play any significant role \cite{Laurent:2020gpg}). It is then used to define a set of weighted moment equations. For the CP-even sector, this method has been recently shown to have severe drawbacks, such as the appearance of unphysical features for sonic bubbles (singularities or peaks). The iterative method that we use was developed in \cite{DeCurtis:2022hlx,DeCurtis:2023hil,DeCurtis:2024hvh} to address these issues and provide for the first time a way to directly solve the Boltzmann equation without resorting to any ansatz. 

As for the CP-odd equation, our primary goal is to check the impact of out-of-equilibrium contributions to the wall dynamics on the BAU. For this purpose, it is sufficient to use the parametrisation introduced above and the moment expansion method to compute the BAU. 
To extract the CP-odd equation, we expand the distribution function as 
\begin{equation}
    f\sim \widetilde f_{0}+\left[\delta f_e - \mu_e \widetilde f'_{0}\right]+s_{cp}\left[\delta f_o+\left(h s_p\gamma_w\Delta E-\mu_o\right)\widetilde f'_{0}-h s_p\gamma_w\Delta E \widetilde f''_{0}\mu_e\right],
\end{equation}
where $\Delta E=\omega - E_0$, $\widetilde f_{0}\equiv f_0|_{_{\mu=0, \omega=E_0}}$ and the $'$ denotes derivative with respect to $\gamma_w E_0$ (to avoid confusion, we stress here that the symbol $'$ denotes derivatives with respect to $z$ for all functions but the distribution function; this notation agrees with the one typically adopted in the literature). CP-even terms are included in the first square bracket, while CP-odd terms are in the second one. The equation for $\delta f_o$ and $\mu_o$ is then finally obtained as\,\cite{Cline:2020jre}
\begin{equation}
\label{eq: Boltzmann eq CP-odd}
\left(\frac{p_z}{E_0}\partial_z -\frac{(\widetilde m^2)'}{2 E_0}\partial_{p_z}\right)\delta f_o +\left(-\frac{p_z}{E_0}\widetilde f'_{0}\partial_z+v_w\gamma_w\frac{(\widetilde m^2)'}{2 E_0}\widetilde f''_{0}\right)\mu_o=\mathcal S_o-\mathcal C_o, 
\end{equation}
where the source term $\mathcal S_o$ is 
\begin{equation}
    \mathcal S_o =-v_w\gamma_w h s_p \left(\frac{(\widetilde m^2\theta ')'}{2 E_0 E_{0,z}}\widetilde f'_{0} -\frac{\widetilde m^2(\widetilde m^2)'\theta'}{4E_0^2E_{0,z}}\left(\frac{\widetilde f'_{0}}{E_0}-\gamma_w\widetilde f''_{0}\right)\right).
    \label{eq: source_CPodd}
\end{equation}

We reduce this equation to a set of moment equations by integrating it over the three-momentum $p$ with a set of weights given by powers of $p_z/E_0$. The ratio $p_z/E_0$ is chosen ad-hoc as it corresponds to the group velocity obtained with the usual dispersion relation, that is when CP-odd perturbations vanish. The $l$-th moment $u_l\equiv\langle (p_z/E_0)^l\delta f_o\rangle$ (see Eq.\,\ref{eq: Liouville moments} below for the definition of the bracket) is then referred to as the $l$-th velocity perturbation. We divide by the normalisation factor
\begin{equation}
    N_1\equiv\int d^3p \,\widetilde f'_{0,0}=\gamma_w\int d^3p \,(\widetilde f'_{0,0})^{FF}=-\gamma_w \frac{2\pi^2}{3}T^2,
\end{equation}
where $\widetilde f_{0,0}$ is the equilibrium distribution function for a massless fermion\footnote{$N_1$ is used as normalisation factor with the fermion massless distribution function also for bosons.} and the superscript $FF$ indicates that $\widetilde f$ is evaluated in the fluid frame. Indicating with $L$ the left-hand side of \eqref{eq: Boltzmann eq CP-odd}, as typically done in the literature, the moment equations for the CP-odd sector take the form 
\begin{equation}
\label{eq: Liouville moments}
\left\langle{\left(\frac{p_z}{E_0}\right)^l L}\right\rangle = \left\langle\left(\frac{p_z}{E_0}\right)^l(\mathcal S_o-\mathcal C_o)\right\rangle, 
\end{equation}
where the brakets are defined as 
\begin{equation}
    \langle A\rangle=\frac{1}{N_1}\int d^3p \,A.
\end{equation}

In our analysis, we consider a two-moment truncation, as was done in \cite{Cline:2020jre}, that we largely follow for the presentation of the moment equations. More recent works have shown the importance of higher moments in the determination of the baryon asymmetry \cite{Kainulainen:2024qpm}, that tend to be overestimated when only two moments are included. Our results should thus be understood as upper bounds for the BAU in the scenarios we  describe.  

For $l=0$ and $l=1$, the left-hand-side of \eqref{eq: Liouville moments} reads
\begin{align}
\label{eq: L moment 0}
    \left\langle L\right\rangle &= -D_1\,\mu_o'+u_1'+v_w\gamma_w (\widetilde m^2)'Q_1\,\mu_o\\
\label{eq: L moment 1}
    \left\langle \frac{p_z}{E_0} L\right\rangle &=-D_2\,\mu_o'+u_2'+v_w\gamma_w (\widetilde m^2)'Q_2\,\mu_o +(\widetilde m^2)'\left\langle\frac{\delta f}{2 E_0^2}\right\rangle
\end{align}
where we defined the symbols $D$ and $Q$ for 
\begin{equation}
    D_l\equiv \left\langle\left(\frac{p_z}{E_0}\right)^l \widetilde f'_0\right\rangle, \qquad Q_l\equiv \left\langle\frac{p_z^{l-1}}{E_0^l} \widetilde f''_0\right\rangle. 
\end{equation}
Equations \eqref{eq: L moment 0} and \eqref{eq: L moment 1} have two clear issues that need to be dealt with. The first problem is that last term in \eqref{eq: L moment 1} does not take the form of a velocity perturbation. We express it in terms of $u_1$ using the factorisation ansatz first proposed in \cite{Fromme:2006wx}. For a generic $A$, we define $\langle A\,\delta f\rangle $ by making the replacement
\begin{equation}
    \left\langle A \,\delta  f\right\rangle\to \left[A\,\frac{E_0}{p_z}\right]\left\langle\frac{p_z}{E_0}\delta f\right\rangle,\qquad [X]\equiv \frac{1}{N_0}\int d^3p\, X\,\widetilde f_0,
\end{equation}
with $N_0$ a normalisation factor $N_0\equiv \int d^3p\, \widetilde f_0 = \gamma_w \int d^3p \,(\widetilde f_0)^{FF}$, and $f_0$ the massive distribution function of the species under consideration in the wall frame. The term $\langle\delta f/2 E_0^2\rangle$ in \eqref{eq: L moment 1} is then expressed as
\begin{equation}
    \left\langle\frac{\delta f}{2 E_0^2}\right\rangle \to \left[\frac{1}{2 p_z E_0}\right]u_1,  
\end{equation}
and, using the principal value for the angular integration,  
\begin{equation}
\bar R\equiv \left[\frac{1}{2 p_z E_0}\right]=\frac{\pi}{\gamma_w}\int_0^\infty dp\,p\frac{f_0}{E_0} \log \left|\frac{p-v_w E_0}{p+v_w E_0}\right|. 
\end{equation}
The second issue is given by the fact that, in the moment equation of order $l$, the velocity perturbation of order $l+1$ appears. When choosing a $l$-order truncation, one should then also define how to express $u_{_{l+1}}$ in terms of $u_n$, $n\le l$. For the case at hand, we need to express $u_2$ in terms of $u_1$. We can do this by using the factorisation ansatz and write 
\begin{equation}
    u_2=\left\langle\frac{p_z^2}{E_0^2}\delta f\right\rangle\to \left[\frac{p_z}{E_0}\right]\left\langle\frac{p_z}{E_0}\delta f\right\rangle \equiv R \,u_1. 
\end{equation}
As first discussed in \,\cite{Cline:2020jre}, $R$ is the expectation value of the plasma velocity in the wall frame, and we have $R=-v_w$.

The right-hand side of \eqref{eq: Liouville moments} remains to be discussed. Adopting the notation of the literature, the source term is
\begin{equation}
\mathcal S_{o,l} \equiv   \left \langle \left(\frac{p_z}{E_0}\right)^{l-1} \mathcal S_o  \right\rangle =-v_w\gamma_w h  \left((\widetilde m^2\theta ')'Q_l^{8o} -\widetilde m^2(\widetilde m^2)'\theta'Q_l^{9o}\right), 
\end{equation}
with 
\begin{equation}
    Q_l^{8o}\equiv \left\langle\frac{s_p\, p_z^{l-1}}{2 E_0^lE_{0,z}}\widetilde f'_0\right\rangle, \qquad Q_l^{9o}\equiv \left\langle\frac{s_p p_z^{l-1}}{4 E_0^{l+1}E_{0,z}}\left(\frac{\widetilde f'_0}{E_0}-\gamma_w \widetilde f''_0\right)\right\rangle. 
\end{equation}
The collision integrals as derived in \cite{Cline:2000nw} are 
\begin{equation}
  \mathcal C_{o,1}\equiv  \left\langle\mathcal C_o\right\rangle = -K_0 \sum_m \Gamma_i\sum_n \frac{s_{mn } \,\mu_n}{T}, \qquad \mathcal C_{o,2} \equiv \left\langle\frac{p_z}{E_0}\mathcal C_o\right\rangle =\Gamma_{tot} \,u_1+v_w \,\mathcal C_{o,1},
\end{equation}
where $K_0$ is a normalisation factor $K_0\equiv-\langle\widetilde f_0\rangle =-N_0/N_1$, $\Gamma_i$ are the interaction rates and $s_{mn}=\pm 1$ depending on whether the particle $n$ is in the initial ($+1$) or final ($-1$) state in the $m$-th interaction. Finally, $\Gamma_{tot}$ stands for the total interaction rate.

\subsection{Application to the model}

For the augmented SSM, we consider the transport equation for four species: left and right-handed top, left-handed bottom and Higgs field. The top-quark is the only one that gets a complex mass, and thus have a non-vanishing source $\mathcal S_o$. The bottom and the Higgs  appear in the collision integral and play a crucial role in the equilibration processes. We write the moment equations for the four species in compact form adopting a vector notation $w_i\equiv\left(\mu_{o,i},u_{1,i}\right)^T$, $\mathcal S_i=\left(\mathcal S_{o,1}^{(i)},\mathcal S_{o,2}^{(i)}\right)^T$ and $\mathcal C_{o,i}\equiv \left(\mathcal C_{o,1}^{(i)},\mathcal C_{o,2}^{(i)}\right)^T$ for each species, $i=t_L,\,b_L,\,t_R,\,h$,  
\begin{align}
\label{eq: moments eqs network}
A_{t} w_{t_L}' + (\widetilde m_t^2)'B_t\, w_{t_L}  &=   \mathcal S_t -  \mathcal C_{o,t_L} ,  \\
A_{b} w_{b_L}' + (m_b^2)'B_b\, w_{b_L}  &= -  \mathcal C_{o,b_L}, \nonumber \\
A_{t} w_{t_R}' + (\widetilde m_t^2)'B_t\, w_{t_R} &=  - \mathcal S_t  -  \mathcal C_{o,t_R} , \nonumber \\
A_{h} w_{h}' + (m_h^2)'B_t\, w_{t_L}  &=  -  \mathcal C_{o,h} \nonumber 
\end{align}
where $A$ and $B$ are two matrices defined as
\begin{equation}
    A_i\equiv \left(\begin{array}{c c} -D_1^{(i)} & 1 \\ \mathcal -D_2^{(i)} & R^{(i)}\end{array} \right) \qquad B_i\equiv\left(\begin{array}{c c} -v_w\gamma_w Q_1^{(i)} & 0 \\ -v_w\gamma_w Q_2^{(i)} & \bar R^{(i)}\end{array} \right). 
\end{equation}
\noindent
The collision integrals are $\left(\overline {\mathcal C}_{o,1}^{(i)}\equiv \mathcal C_{o,1}^{(i)}/K_0^i \right)$
\begin{align}
    \overline {\mathcal C}_{o,1}^{(t_L)}&=-\Gamma_y\left(\mu_{o,t_L}-\mu_{o,t_R}+\mu_{o,h}\right)-\Gamma_m\left(\mu_{o,t_L}-\mu_{o,t_R}\right)-\Gamma_W\left(\mu_{o,t_L}-\mu_{o,b_L}\right)-\widetilde \Gamma_{SS}[\mu_{o,i}] \nonumber\\
     \overline {\mathcal C}_{o,1}^{(b_L)}&=-\Gamma_y\left(\mu_{o, b_L}-\mu_{o,t_R}+\mu_{o,h}\right)-\Gamma_W\left(\mu_{o,b_L}-\mu_{o,t_L}\right)-\widetilde \Gamma_{SS}[\mu_{o,i}] \nonumber \\
       \overline {\mathcal C}_{o,1}^{(t_R)}&=\Gamma_y\left(\mu_{o,t_L}+\mu_{o,b_L}-2\mu_{o,t_R}+2\mu_{o,h}\right)-\Gamma_m\left(\mu_{o,t_R}-\mu_{o,t_L}\right)+\widetilde \Gamma_{SS}[\mu_{o,i}] \nonumber \\
       \overline {\mathcal C}_{o,1}^{(h)}&=-\frac32\Gamma_y\left(\mu_{o,t_L}+\mu_{o,b_L}-2\mu_{o,t_R}+2\mu_{o,h}\right)-\Gamma_h \mu_{o,h}
\end{align}
and $\mathcal C_{o,2}^{(i)}=-\Gamma_{tot}^{(i)}\, u_{1,i}-v_w \mathcal C_{o,1}^{(i)}$ (no sum over the index $i$ is intended here and in the following), where we take $\Gamma_y=4.2\cdot 10^{-3}T$, $\Gamma_m=m_t^2/63 \,T$, $\Gamma_h=m_W^2/50\,T$ and $\Gamma_W=-\Gamma_{tot}^{(h)}$ for the interaction rate, with $m_W$ the $W$ boson field-dependent mass $m_W^2=g^2h^2/4$, and $\Gamma_{tot}^{(i)}=D_2^{(i)}/(d_i \,D_0^{(i)})$ for the total interaction rates, with $d_i$ the diffusion constant ($d_i=6/T$ for quarks and $d_i=20/T$ for the Higgs) and $D_0^{(i)}=\langle\widetilde f_0'^{(i)}\rangle$ \cite{Fromme:2006wx,Cline:2020jre}. The strong sphaleron rate is $\widetilde \Gamma_{SS}[\mu_{o,i}]=\Gamma_{SS}\sum_q\left(\mu_{o,q_L}-\mu_{o,q_R}\right)$, where $\Gamma_{SS}=4.9\cdot 10^{-4}T$ and the sum extends to all the quarks. The contribution from the light quarks that are not included in the moment equations network \eqref{eq: moments eqs network} is determined analytically to a good approximation combining baryon conservation $B=\sum_q n_q-\bar n_q=0$ with the observation that when Yukawa mixing is neglected, $\mu_{o,q_L}=-\mu_{o,q_R}$. One is then able to express the chemical potential of all quark species, and in turn $\widetilde \Gamma_{SS}$, in terms of only $\mu_{o,t_L}, \,\mu_{o,b_L}$ and $\mu_{o,t_R}$ as   \cite{Fromme:2006wx,Cline:2020jre,vandeVis:2025efm} 
\begin{equation}
\label{eq: chem pot light quarks}
    \mu_{o,q_L}=-\mu_{o,q_R}=D_0^{(t)}\mu_{o,t_L}+D_0^{(b)}\mu_{o,b_L}+D_0^{(t)}\mu_{o,t_R}
\end{equation}
and 
\begin{equation}
    \widetilde \Gamma_{SS}=\left(\left(1+9 D_0^{(t)}\right)\mu_{o,t_L}+\left(1+9 D_0^{(b)}\right)\mu_{o,b_L}-\left(1-9 D_0^{(t)}\right)\mu_{o,t_R}\right)\Gamma_{SS}.
\end{equation}

Grouping the quantities for all the species into single vectors and matrices, $\mathcal U\equiv \left(w_{t_L},w_{b_L},w_{t_R}, w_h\right)^T$, $\mathcal S\equiv \left(\mathcal S_t, 0,\mathcal S_t,0\right)^T$, $A={\rm Block}\left(A_t,A_b,A_t,A_h\right)$ and combining the action of $(m^2_i)'B_i$ and of the collision terms $\mathcal C_{o,i}$ into a matrix $\Gamma_o$ acting on $\mathcal U$,  we can write the moment equation network as a single matrix equation 
\begin{equation}
    A\,\mathcal U'-\Gamma_o\, \mathcal U =\mathcal S.
\end{equation}
The matrix $A$ is tridiagonal, independently of the number of species considered. Far from the wall, the field and plasma profiles asymptote to homogeneous functions, causing the source term to vanish. The equation simplifies to $  \mathcal U'-A^{-1} \Gamma_o\,\mathcal U=0$, so that growing and decaying modes at $\pm\infty$ can be determined as the eigenmodes of the matrix $A^{-1}\Gamma_o$. We use this feature, in our numerical code, to solve the moment equations and determine the chemical potential of the species.

\begin{figure}
\centering
\includegraphics[width=0.485\linewidth]{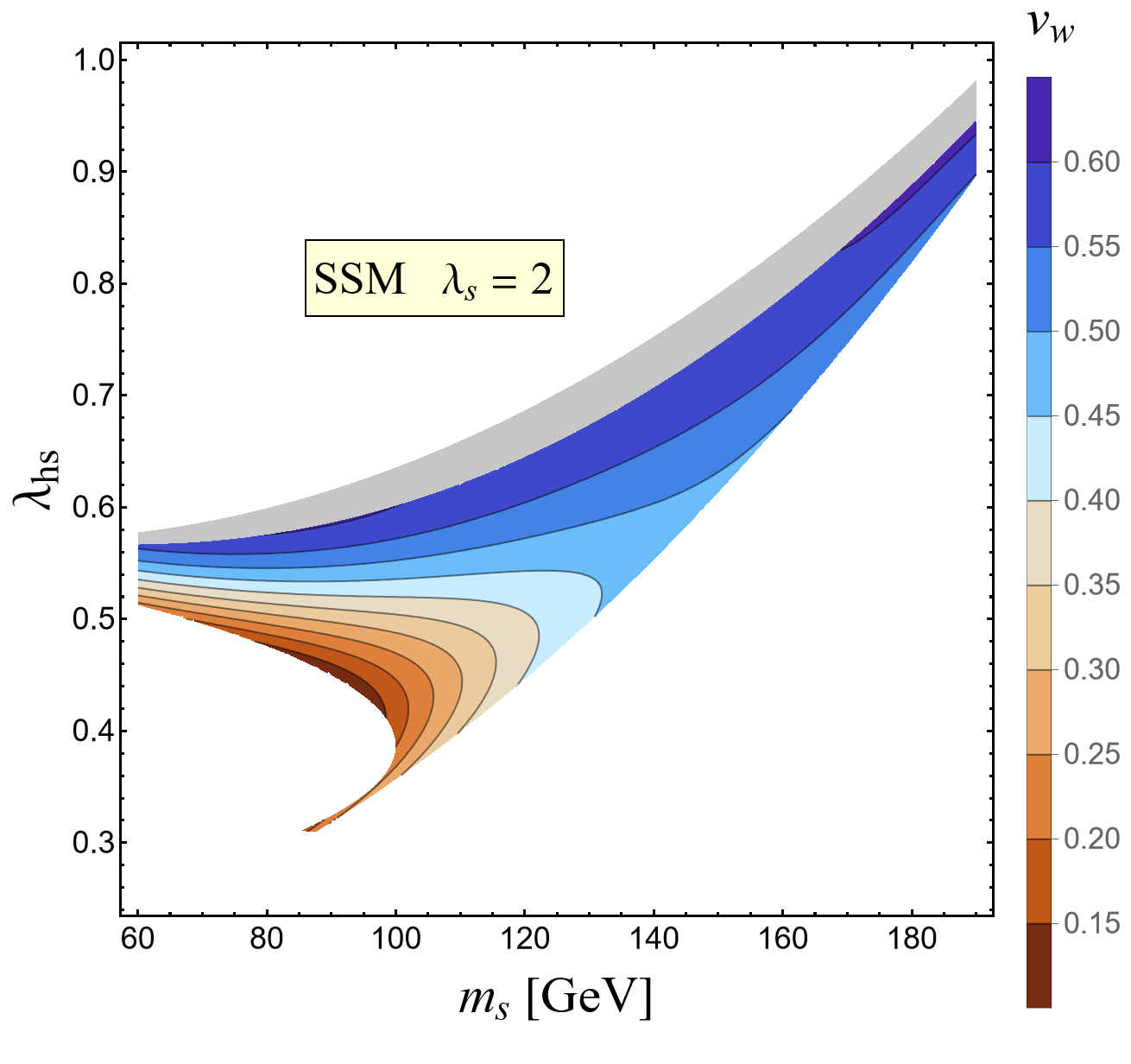}\,\,
\includegraphics[width=0.495\linewidth]{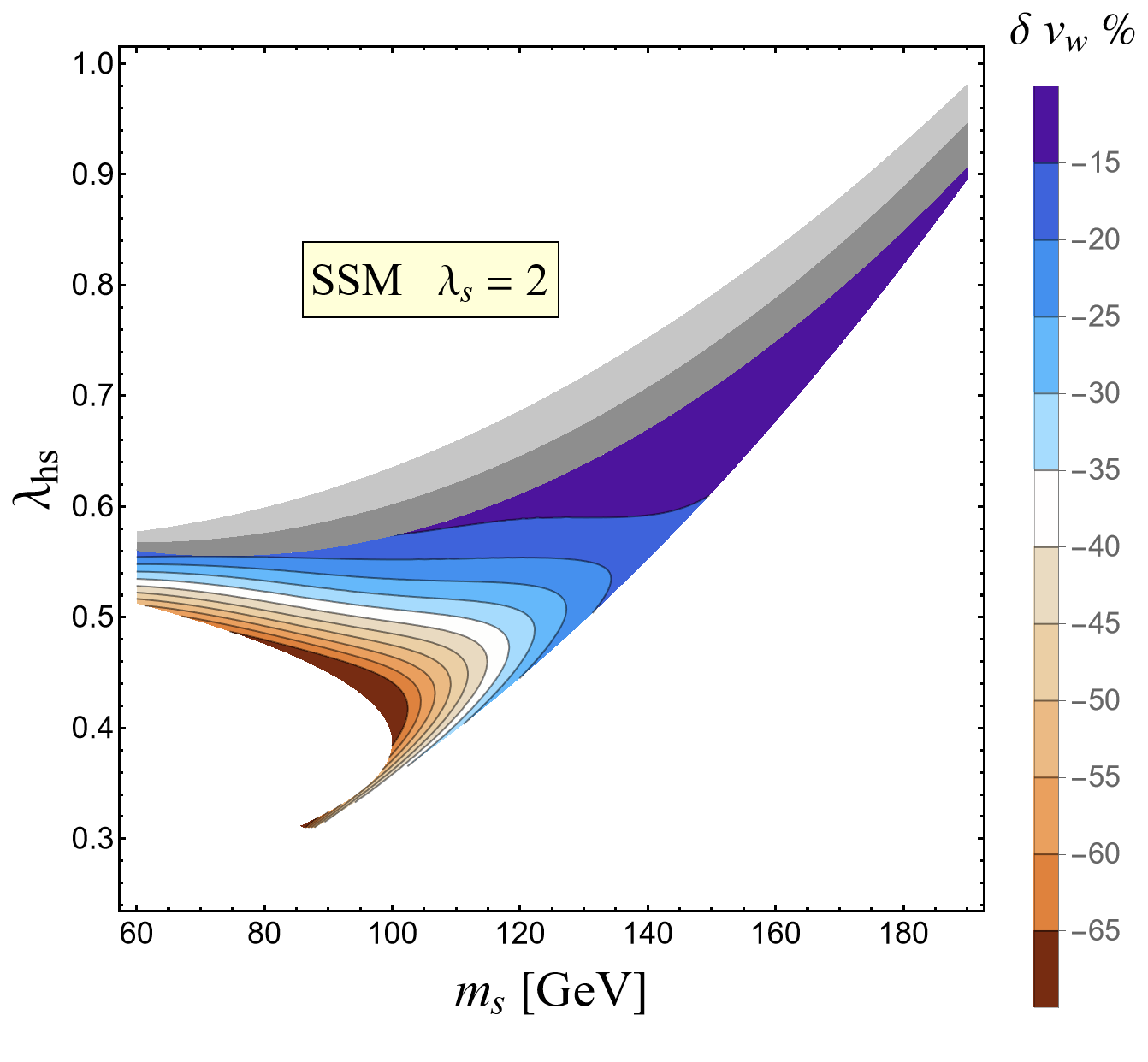}
    \caption{\textit{Left panel}. Contour plot of the wall velocity $v_w$ in the parameter space for $\lambda_s=2$. {\it Right panel}. Contour plot of the relative correction $\delta v_w$ to the LTE wall velocity.}
    \label{fig: paramspace ls2}
\end{figure}

Once the equations are solved, the total chemical potential for left-handed quarks can be written in terms of $\mu_{o,t_L},\,\mu_{o,b_L}$ and $\mu_{o,t_R}$ using \eqref{eq: chem pot light quarks} as 
\begin{equation}
    \mu_{o,B_L}=\frac12\sum_q\mu_{o,q_L}=\frac12 \left(1+4 D_0^{(t)}\right)\mu_{t_L}+\frac 12\left(1+4 D_0^{(b)}\right)\mu_{b_L}-2D_0^{(t)}\mu_{t_R}.
\end{equation}
The generated baryon asymmetry $\eta_B$ is then evaluated
\begin{equation}
    \eta_B=\frac{405\, \Gamma_{\rm sph}}{4\pi^2v_w\gamma_w g_* T}\int dz \mu_{o, B_L} f_{\rm sph}e^{-45 \Gamma_{\rm sph}|z|/4v_w\gamma_w},
\end{equation}
where $g_*$ is the effective number of degrees of freedom in the plasma, the function $f_{\rm sph}(z)\equiv {\rm min} \left(1,2.4 (T/\Gamma_{\rm sph}) e^{-40 h(z)/T}\right)$ is inserted by hand to effectively interpolate between the unsuppressed sphaleron rate in the symmetric phase $(f_{\rm sph}\to 1)$ and the suppressed one in the broken phase $(f_{\rm sph}\to 0)$, and $\Gamma_{\rm sph}$ is fixed to $\Gamma_{\rm sph}=10^{-6} T$. In the text, we present the results for the ratio between $\eta_B$ and the observed baryon asymmetry $\eta_{\rm obs}\sim 8.7\cdot 10^{-11}$.

From the explicit expression of $\eta_B$ given above, one can get an idea of how intricate the wall velocity dependence of the BAU is, as $v_w$ appears in the combination $v_w\gamma_w$ both in the denominator and in the exponential inside the integral, the two terms being responsible for the vanishing of $\eta_B$ in the $v_w\to 0$ and $v_w\to 1$ limit. Concerning the first one of the limits, this is due to the fact that for slower walls sphaleron interactions have more time to re-equilibrate any localised asymmetry before it enters the wall. This effect is encoded in the $v_w$-dependence of the exponential. As for the limit where the wall becomes ultra-relativistic, the suppression of $\eta_B$ arises since sphalerons do not have enough time to convert CP-asymmetries into baryon asymmetries.

\section{Numerical results for $\lambda_s=2$}
\label{sec- lambdas=2}

\begin{figure}
\centering
\includegraphics[width=0.45\linewidth]{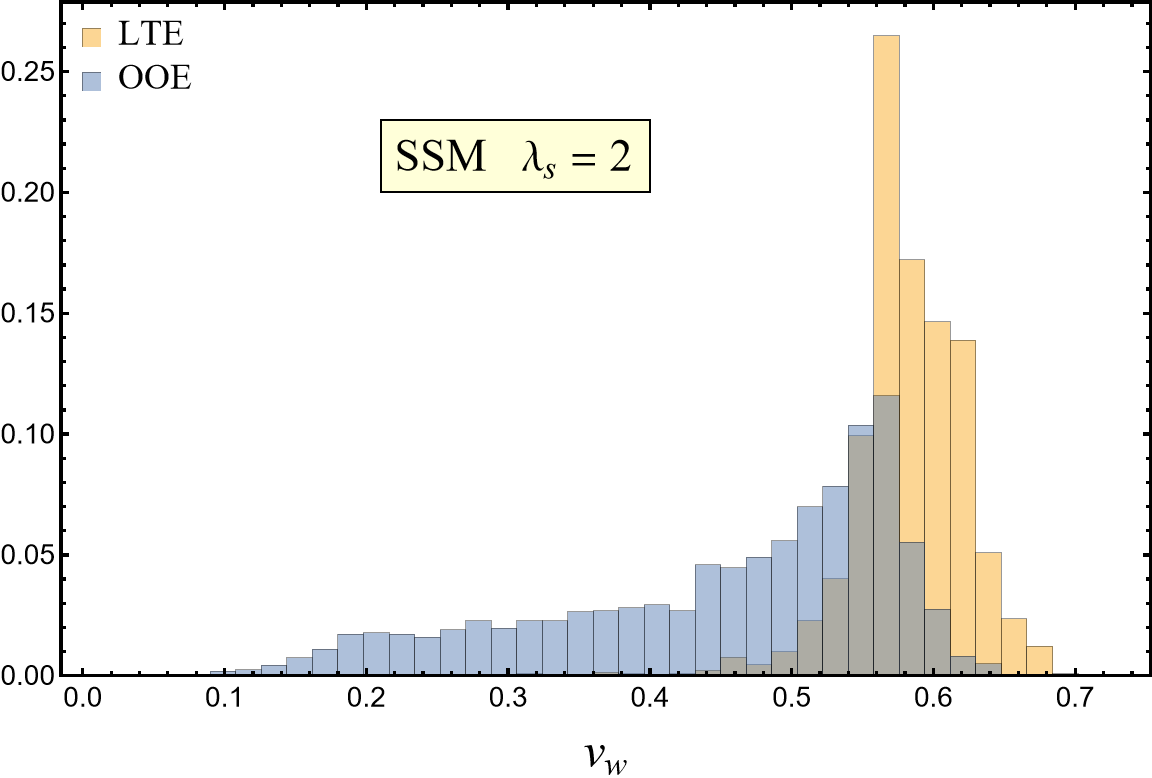}
\includegraphics[width=0.45\linewidth]{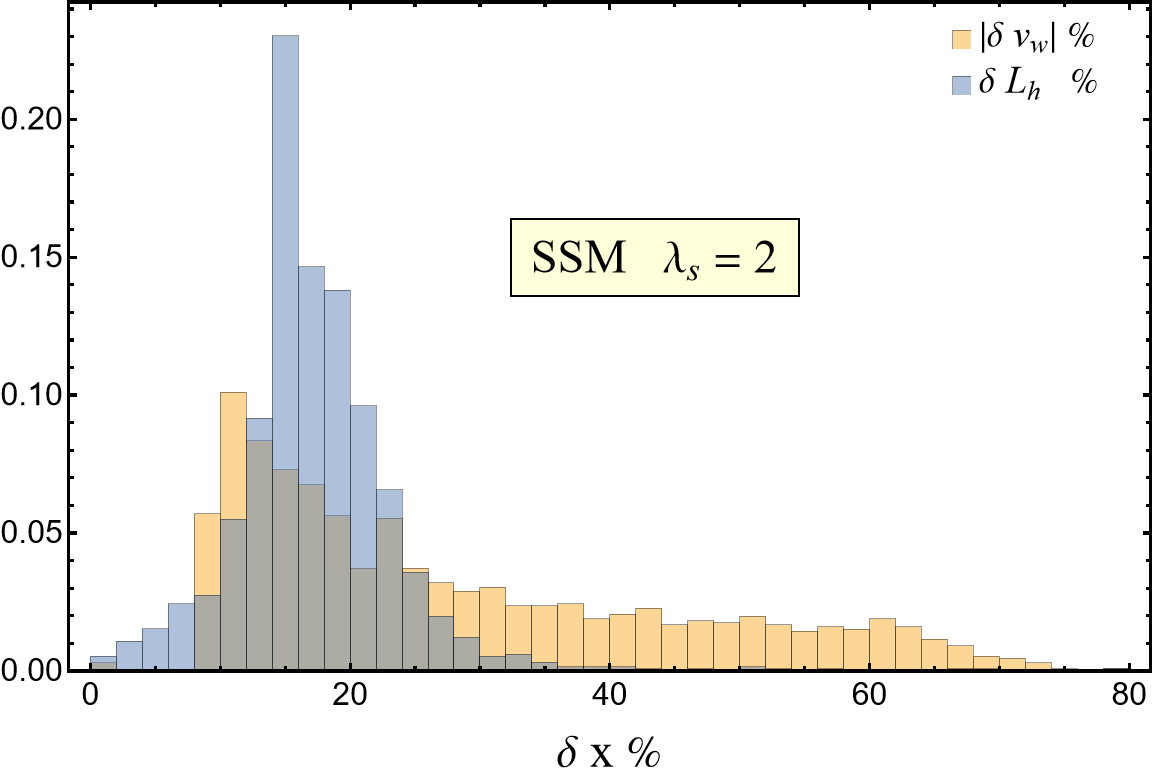}\\
\includegraphics[width=0.45\linewidth]{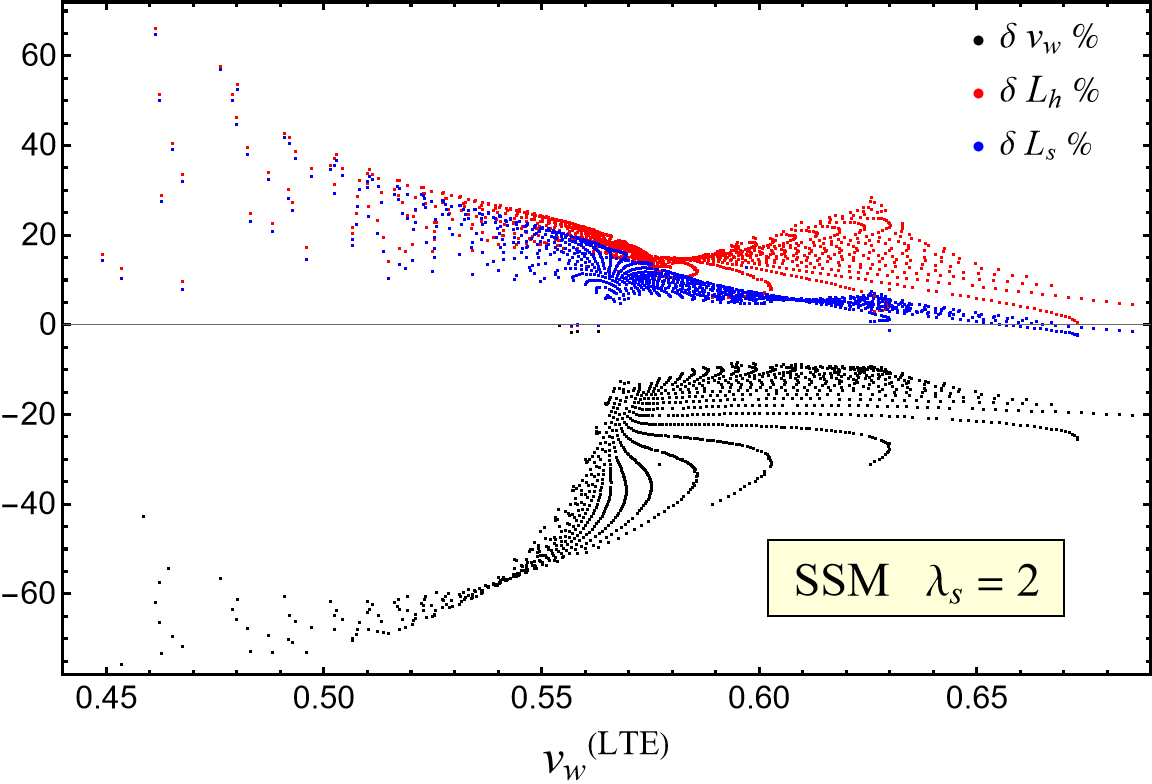}
    \caption{\textit{Upper left panel}. Histogram of the wall velocity $v_w$ for $\lambda_s=2$ within the LTE approximation (yellow bins) and with OOE contributions (blue bins). \textit{Upper right panel}. Histogram of the $|\delta v_w|$ (yellow bins) and $\delta L_h$ relative corrections. \textit{Lower panel}. Scatter plot of the relative corrections $\delta v_w$, $\delta L_h$ and $\delta L_s$ versus the LTE wall velocity $v_w^{(\rm LTE)}$. }
    \label{fig: histo ls2}
\end{figure}

We present in this appendix some results for the survey of the SSM with $\lambda_s=2$. The wall velocity $v_w$ with out-of-equilibrium contributions from the top quark, and the relative correction $\delta v_w$ with respect to the LTE solution are shown in Fig.\,\ref{fig: paramspace ls2}, left and right panel respectively. The light and dark grey bands have the same meaning as in the $\lambda_s=1$ plots. Comparison with Fig.\,\ref{fig:vw} shows the qualitative agreement between the two cases, with slightly larger corrections in the $\lambda_s=2$ scenario.

We collect in Fig.\,\ref{fig: histo ls2} histograms showing the distribution of $v_w$ and of the relative corrections $\delta v_w$ and $\delta L_h$ (upper left and right panel, respectively). In the lower panel, a scatter plot of $\delta v_w$, $\delta L_h$ and $\delta L_s$ versus the LTE wall velocity $v_w^{\rm (LTE)}$ is presented. Again, the agreement with the corresponding plots for $\lambda_s=1$ is evident.

\bibliographystyle{JHEP}

\bibliography{biblio.bib}

\end{document}